\documentclass[final,3p,times,twocolumn]{elsarticle/elsarticle}

\usepackage{graphicx}
\usepackage{amssymb}

\usepackage{amsmath}
\usepackage{amsfonts}
\usepackage{bm}
\usepackage{verbatim}
\usepackage{multirow}
\usepackage{booktabs} 
\usepackage[normalem]{ulem} 
\usepackage{units} 
\usepackage{float} 
\usepackage{hyperref}

\usepackage{color}
\definecolor{gray}{rgb}{0.5,0.5,0.5}
\definecolor{dred}{rgb}{0.5,0.0,0.0}
\definecolor{dgreen}{rgb}{0.0,0.5,0.0}
\definecolor{dblue}{rgb}{0.0,0.0,0.5}
\definecolor{violet}{rgb}{0.7,0.0,0.5}

\definecolor{lred}{rgb}{1.0,0.5,0.5}
\definecolor{lgreen}{rgb}{0.5,1.0,0.5}
\definecolor{lblue}{rgb}{0.5,0.5,1.0}

\def\O{\mathcal{O}}
\def\C{\mathcal{C}}
\def\V{\mathcal{V}}

\def\L{\mathcal{L}}
\def\F{\mathcal{F}}
\def\J{\mathcal{J}}
\def\P{\mathcal{P}}
\def\H{\mathcal{H}}

\def\G{\mathcal{G}}

\def\etot{\mathcal{E}}
\def\ekin{K}
\def\uGC{u}

\def\rhoPar{\varrho_\parallel}

\def\nablab{{\bm \nabla}}

\newcommand{\cb}{\color{blue} }
\newcommand{\cg}{\color{dgreen} }
\newcommand{\cred}{\color{red} }

\AtBeginDocument{\mathcode`v=\varv} 

\def\figures{.}

\journal{arXiv}

\begin{document}

\begin{frontmatter}
\title{Construction and analysis of guiding center distributions for tokamak plasmas with ambient radial electric field\vspace{-0.25cm}}

\author[rokk,naka]{Andreas~Bierwage\corref{cor1}}
\author[ipp]{Philipp~Lauber}
\author[rokk]{Noriyoshi~Nakajima}
\author[uot,naka]{Kouji~Shinohara}
\author[iter]{Guillaume~Brochard}
\author[kaist]{Young-chul~Ghim}
\author[kaist]{Wonjun~Lee}
\author[kyotou]{Akinobu~Matsuyama}
\author[naka]{Shuhei~Sumida}
\author[rokk,m2p2]{Hao~Yang}
\author[rokk]{Masatoshi~Yagi}

\cortext[cor1]{{\it Email address:} {\tt bierwage.andreas@qst.go.jp}}

\address[rokk]{QST, Rokkasho Institute for Fusion Energy, Aomori 039-3212, Japan}
\address[naka]{QST, Naka Institute for Fusion Science and Technology, Naka, Ibaraki 311-0193, Japan}
\address[ipp]{Max-Planck-Institut f\"{u}r Plasmaphysik, D-85748 Garching, Germany}
\address[uot]{Department of Complexity Science and Engineering, The University of Tokyo, Kashiwa, Chiba 277-8561, Japan}
\address[iter]{ITER Organisation, St.~Paul-lez-Durance 13115, France}
\address[kaist]{Department of Nuclear and Quantum Engineering, Korea Advanced Institute of Science and Technology (KAIST), Daejeon 34141, South Korea}
\address[kyotou]{Graduate School of Energy Science, Kyoto University, Uji 611-0011, Japan}
\address[m2p2]{CNRS, Centrale Marseille, M2P2, Marseille 13451, France\vspace{-1cm}}

\begin{abstract}
The contribution of a time-independent toroidally-symmetric radial electric field $E_r$ is implemented in {\tt VisualStart} [{\it Comp.\ Phys.\ Comm}.\ {\bf 275} (2022) 108305], a code whose purposes include the construction of guiding center (GC) drift orbit databases for the study of plasma instabilities in tokamaks.
$E_r$ is important for the thermal part of the velocity distribution and for fast particle resonances in the kHz frequency range. KSTAR, JT-60U and ITER tokamak cases are used as working examples to test our methods and discuss practical issues connected with $E_r$. Two points are worth noting: First, the GC orbit space is sampled in the magnetic midplane as before, and we find that in the presence of $E_r$, midplane-based coordinates are not only equivalent but superior to conventional constants of motion, allowing to attain high numerical accuracy and efficiency with a relatively simple mesh. Second, the periodic parallel acceleration and deceleration of GCs via the mirror force is modulated by $E_r$. Although this parallel electric acceleration averages to zero during a poloidal transit (or bounce) period, it has important consequences, one being the known shift of the trapped-passing boundary. Another consequence is that electric frequency shifts depend on the chosen reference point, so that some care is required when evaluating the $E_r$-dependence of transit frequencies for resonance analyses.
\end{abstract}

\end{frontmatter}


\tableofcontents


\section{General background and motivation}
\label{sec:motivation}

A weak but significant electric field ${\bm E}$ can form spontaneously in quasi-neutral plasmas, and it can be sustained as an ambient field through nonuniformities in the particle distributions, especially when different species have different charge-mass ratios. Such nonuniformities, in turn, can be sustained by sources and sinks of charged particles, their flows, and nonuniform magnetization. We begin our discussion with the latter.

In a magnetic torus, such as a tokamak, stellarator or a strong dipole field, light charged particles such as electrons, hydrogen and helium ions follow trajectories that are bent on two spatio-temporal scales, which we choose to distinguish as follows:
\begin{itemize}
	\item {\it local magnetization:} the small and rapid scale of gyration, spiraling along the magnetic field;
	\item {\it global magnetization:} cyclic parallel reflection off magnetic mirrors and drift across magnetic surfaces on the system transit scale.
\end{itemize}

\noindent The latter arises from the nonuniformity of the ambient magnetic field ${\bm B} = \nablab\times{\bm A}({\bm x})$ that we assume to be stationary (time-independent). Strictly speaking, the global magnetization effects and their modification by an ambient electric field ${\bm E} = -\nablab\Phi({\bm x})$ are already captured by the distortion of microscopic gyro-orbits \cite{SpitzerBook62, Qin00}. However, in strongly magnetized plasmas, the effect of these distortions becomes distinctly noticeable only after many gyrations, so that the above distinction between two scales is useful for at least two reasons:\vspace{-0.05cm}
\begin{enumerate}
	\item[(i)]   Local magnetization in a plasma with nonuniform density and temperature gives rise to a mean diamagnetic flow ${\bm V}_*$ that does not involve actual particle motion beyond the gyroradius scale.\footnote{We use the attribute ``diamagnetic'' in a narrow sense, referring only to the flow that emerges when local gyromotion is uncovered by gradients in the particle distribution. Some texts use this term more broadly to collectively refer to all magnetization flows, including electric drift. For instance, see Section ``4.2 Lorentz plasma'' of Ref.~\protect\cite{HelanderSigmarBook}.}
	In contrast, the mean flows associated with global magnetization (electric and magnetic drifts) constitute actual mass flows on large scales.
	\item[(ii)]  The collision time is usually much longer than a gyroperiod but can be comparable to or even shorter than the period of a particle's global transit, especially in cooler regions of the plasma. Global magnetization effects are then modified or lost, while local magnetization remains largely intact.
\end{enumerate}\vspace{-0.05cm}

The competition and interplay between the global magnetization effects and collisions in (ii) can have a strong effect on the plasma dynamics. For instance, they cause what is known as {\it neoclassical} ($\nablab B$- and curvature-enhanced) fluxes of particles, momentum and heat \cite{HelanderSigmarBook}. These fluxes are involved in the formation of the intrinsic electric field and plasma rotation that, in turn, affect wave-particle interactions, various instabilities on multiple scales, and the associated {\it anomalous} (fluctuation-enhanced) fluxes.

\begin{figure}
	[tbp]\vspace{-0.3cm}
	\centering
	\includegraphics[width=0.48\textwidth]{\figures/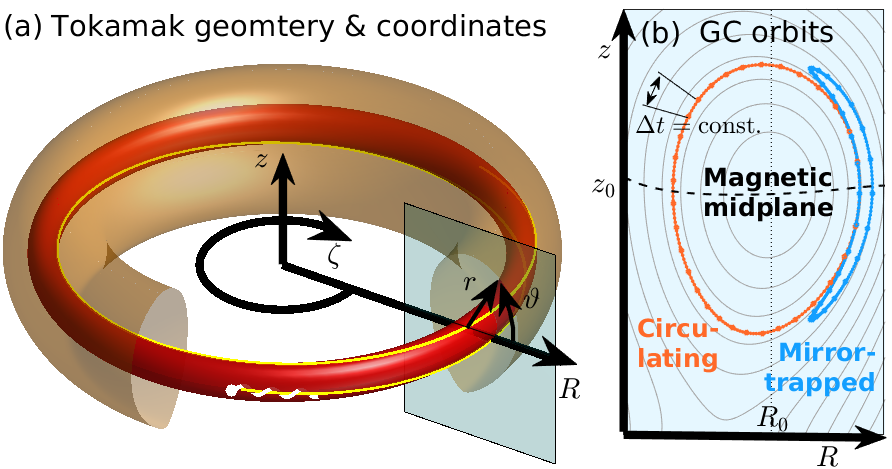}\vspace{-0.15cm}
	\caption{Tokamak geometry, coordinates, and orbit surfaces. Panel (a) shows the 3D structure of a magnetic surface (brown), a gyro-orbit (white), its guiding center (GC) trajectory (yellow) as well as the orbit surface that it covers (red). Panel (b) shows poloidal projections of two GC orbits, one circulating (orange) and one mirror-trapped (blue), whose modification in the presence of an ambient radial electric field $E_r$ will be analyzed in detail later (cf.~Figs.~\protect\ref{fig:13_jt60_deep-trapped}, \protect\ref{fig:14_jt60_deep-pass}). The dots indicate distances traveled during constant time intervals $\Delta t$. The magnetic midplane (dashed black curve) is defined in Eq.~(\protect\ref{eq:method_mid}) of Section~\protect\ref{sec:method_mid}.}
	\label{fig:01_geom}\vspace{-0.3cm}
\end{figure}

For the modeling and study of such transport phenomena and instabilities, it is often useful to split the distribution function $f({\bm v},{\bm x},t)$ of a given particle species into a reference $f_{\rm ref}$ and a perturbation $\delta f$ \cite{Aydemir94}:\vspace{-0.15cm}
\begin{equation}
	f({\bm v},{\bm x},t) = f_{\rm ref}({\bm v},{\bm x}) + \delta f({\bm v},{\bm x},t),
	\label{eq:intro_df}\vspace{-0.1cm}
\end{equation}

\noindent where $f_{\rm ref}$ is assumed to be stationary on the time scale on which $\delta f$ varies. The present work is concerned with the effect of an ambient electric field ${\bm E}({\bm x})$ on $f_{\rm ref}$ in regimes where the mean free path is sufficiently long for the particles to complete their poloidal transit around the torus without being significantly perturbed by collisions. This implies the existence of magnetic mirror-trapped orbits, and since the poloidal contour of deeply mirror-trapped orbits in a toroidal plasma has a banana-like shape in the guiding center approximation \cite{Littlejohn83}, this weakly collisional regime is known as ``banana regime'' in neoclassical theory \cite{Hinton76}. In this regime, the reference distribution $f_{\rm ref}$ can be adequately expressed as a function of a few so-called {\it constants of motion} (CoM, here denoted by ${\bm C}$) \cite{Morozov66b}, whose values identify toroidal surfaces that are covered by the orbits of unperturbed charged particles (see Fig.~\ref{fig:01_geom}, discussed in more detail in Section~\ref{sec:intro}). Formally, we write $f_{\rm ref}({\bm v},{\bm x}) \rightarrow f_{\rm com}({\bm C})$ with the Klimonotovitch representation\vspace{-0.2cm}
\begin{equation}
	f_{\rm com} = \sum_l W_l \delta({\bm C}_l),
	\label{eq:intro_fcom}\vspace{-0.3cm}
\end{equation}

\noindent where the index $l$ identifies unperturbed guiding center orbits that serve as (Dirac delta) basis functions $\delta({\bm C}_l)$ weighted by factors $W_l$.

The CoM distribution function $f_{\rm com}$ may be viewed as a {\it topological equilibrium}\footnote{In the terminology of classical thermodynamics, our $f_{\rm com}$ constitutes a special case of a non-equilibrium called {\it steady state} (e.g. see \href{https://sdkang.org/generalized-force-and-flux/}{\tt https://sdkang.org/generalized-force-and-flux/}).}
that forms in two stages. First, on the poloidal transit time scale $\tau_{\rm pol}$ (typically $0.01...1\,{\rm ms}$), the particles are distributed uniformly in time but nonuniformly in space on each orbit surface $\delta({\bm C}_l)$. The spatial modulation of the particles' density and velocity along ${\bm B}$ is determined by the magnetic mirror force and, as will be shown later, by the ambient electric field ${\bm E}$. Second, on a longer time scale $\tau \gg \tau_{\rm pol}$ (say, $1...100\,{\rm ms}$), the balance between collisions, sources and sinks with respect to each orbit surface determines the number of physical particles represented by each orbit. This determines the values of the weights $W_l$ in Eq.~(\ref{eq:intro_fcom}). Clearly, such a topological equilibrium $f_{\rm com}$ is an idealization that is valid on time scales over which the particle energy (Hamiltonian) can be considered to be invariant, and there will always be some portions of phase space $({\bm v},{\bm x})$, where this condition is violated. Nevertheless, we will work here under the assumption that $f_{\rm com}({\bm C})$ can be a useful approximation for burning-plasma-relevant conditions.

In tokamak-based magnetically confined fusion research, the use of CoM distribution functions $f_{\rm com}$ is common when dealing with particles that move at high velocities, such as runaway electrons, beam ions and fusion products in the $100\,{\rm keV}$ to multi-MeV energy range. Of course, due to their gradual collisional slow-down, a steady-state distribution does not only contain fast ions but usually extends all the way down to thermal energies. This, and the fact that plasma temperatures around $10\,{\rm keV}$ and higher must routinely be produced in burning fusion plasma cores, makes the use of CoM meaningful also for the description of thermal bulk ions.

Under burning-plasma-relevant conditions, thermal ions may drive Alfv\'{e}nic instabilities just like fast ions, albeit usually at low (near diamagnetic) frequencies of a few to tens of ${\rm kHz}$. This has been predicted theoretically \cite{Zonca99, Chen16} and confirmed experimentally \cite{Heidbrink21a, Heidbrink21b, Du21}. When dealing with dynamics at such low frequencies, (sheared) plasma rotation must be taken into account because the electric field, which is associated with much of the rotational flow \cite{Hinton76, Hirshman81, Ida98, Callen09}, alters the particles' transit frequencies and their reference distribution $f_{\rm ref}$.

Thus motivated, we included the effect of an ambient electric field with predominantly radial component $E_r = {\bm E}\cdot\hat{\bm e}_r$ (transverse to magnetic surfaces) in the code {\tt VisualStart} \cite{Bierwage22a}. This code has recently served as one of the prototypes for the development of an integrated workflow for the representation and transformation of charged particle distributions \cite{Bierwage22a, Benjamin23} in ITER's Integrated Modelling and Analysis Suite (IMAS) \cite{Imbeaux15}.

The purpose of this paper is to describe the numerical methods used in the upgraded {\tt VisualStart} code, discuss technical challenges and their solutions, characterize the impact of $E_r$ on a simple model distribution and on individual orbits, and compare the results with analytical estimates. Our methods and insights may, for instance, find use in the ongoing development of transport simulation workflows based on the theoretical concept of phase space zonal structures \cite{Zonca15b, Falessi19a, Falessi23, Lauber24, Meng24}, where CoM distributions are employed as well.

The following Section~\ref{sec:intro} revisits the theoretical concepts of charged particle motion, briefly describes our understanding of rotational flows in tokamaks, and outlines the workflow implemented in {\tt VisualStart} as well as the contents of the main part of this paper.

\vspace{-0.2cm}
\section{Theoretical concepts, formalism \& scope}
\label{sec:intro}

\vspace{-0.15cm}
\subsection{Charged particle motion in tokamak geometry}
\label{sec:intro_integrals}

Tokamaks and similar devices are used to confine quasi-neutral hydrogen isotope plasmas in a donut-shaped toroidal volume as illustrated in Fig.~\ref{fig:01_geom} by means of an ambient magnetic field ${\bm B}$. For our purposes, ${\bm B}$ can be assumed to be stationary (independent of time $t$) and axisymmetric with respect to rotation along the toroidal angle $\zeta$ around the $z$ axis, and is written as\vspace{-0.15cm}
\begin{equation}
	{\bm B} = B({\bm x})\hat{\bm b}({\bm x}) = \underbrace{\nablab\zeta\times\nablab\Psi_{\rm P}(r)}\limits_{\text{poloidal field } {\bm B}_{\rm pol}} + \underbrace{B_\zeta(r)\nablab\zeta}\limits_{\text{toroidal field } {\bm B}_{\rm tor}}\vspace{-0.25cm}
\end{equation}

\noindent with magnitude $B$, unit vector $\hat{\bm b}$, poloidal flux $2\pi\Psi_{\rm P}(r)$ and toroidal field component $B_{\rm tor} = B_\zeta(r)/R \approx B_0 R_0/R$, where $B_{\rm tor}$ and $B_\zeta$ are, respectively, the physical and covariant toroidal components of ${\bm B}$. The magnetic axis with field strength $B_0$ is located at the major radius $R_0$ from the vertical axis $z$ at height $z_0$. The argument $r$ of $\Psi_{\rm P}(r)$ implies that the magnetic axis is assumed to be surrounded by nested toroidal magnetic surfaces (the gray contours in Fig.~\ref{fig:01_geom}(b)) that can be labeled by a suitable (volume-averaged) minor radial coordinate $r$. As mentioned in Section~\ref{sec:motivation}, tokamaks also contain a (weak) ambient electric field ${\bm E} = -\nablab\Phi$ with electrostatic potential $\Phi(r)$ that we also assume to depend on $r$ only.

Nonrelativistic particles with electric charge $Ze$ and mass $M$ gyrate with frequency $\omega_{\rm B} = Ze B/M$. The field strength $B$ and the device's dimensions $2\pi R_0$ (toroidal circumference) and $2\pi a$ (poloidal circumference) are usually large enough for most particles to perform many gyrations during one poloidal or toroidal transit period. This condition can be expressed as\vspace{-0.15cm}
\begin{equation}
	\omega_{\rm pol}/\omega_{\rm B} \ll 1, \quad \omega_{\rm tor}/\omega_{\rm B} \ll 1,
	\label{eq:intro_t}\vspace{-0.15cm}
\end{equation}

\noindent where $\omega_{\rm pol} \equiv 2\pi/\tau_{\rm pol} \approx 2\pi\left<\smash{v_{\rm pol}}\right>/L_{\rm pol}$ and $\omega_{\rm tor} \equiv 2\pi/\tau_{\rm tor} \approx \left<\smash{v_{\rm tor}}\right>/R_0$ are the particle's angular frequencies along $\vartheta$ and $\zeta$ in Fig.~\ref{fig:01_geom}(a), with the respective mean velocities $\left<\smash{v_{\rm pol}}\right>$ and $\left<\smash{v_{\rm tor}}\right>$, and the drift orbit circumference $0 \leq L_{\rm pol} \lesssim 2\pi a$ in Fig.~\ref{fig:01_geom}(b). A particle's velocity vector ${\bm v}$ can then be usefully decomposed into perpendicular + parallel, or poloidal + toroidal components as
\begin{equation}
	{\bm v} = {\bm v}_\perp + v_\parallel\tilde{\hat{\bm b}} = {\bm v}_{\rm pol} + v_{\rm tor}R\widetilde{\nablab\zeta} = {\bm v}_{\rm pol} + v_\zeta\widetilde{\nablab\zeta},
	\label{eq:com_vfull}
\end{equation}

\noindent with physical and covariant components $v_{\rm tor}$ and $v_\zeta$. The tilde indicates that a quantity is evaluated at the coordinates $\{{\bm v}(t),{\bm x}(t)\}$ of a particle following a Lorentz orbit, which is indicated by the white helix in Fig.~\ref{fig:01_geom}(a). It is worth noting that, in a tokamak, condition (\ref{eq:intro_t}) implies that the gyroradius $\varrho_{\rm L} = v_\perp/\omega_{\rm B}$ is small compared to the length scale on which the ambient magnetic field varies:\vspace{-0.2cm}
\begin{equation}
	\varrho_{\rm L}\nablab \ln B \sim \frac{\varrho_{\rm L}}{R_0} \sim \underbrace{\frac{v_\perp}{v_\parallel}\frac{\omega_{\rm tor}}{\omega_{\rm B}}}\limits_{\text{circulating}} \sim \underbrace{\frac{v_\perp}{\left<\smash{v_{\rm pol}}\right>} \frac{L_{\rm pol}}{2\pi R_0} \frac{\omega_{\rm pol}}{\omega_{\rm B}}}\limits_{\text{mirror trapped}} \ll 1.
	\label{eq:com_b}\vspace{-0.2cm}
\end{equation}

We shall assume that\vspace{-0.2cm}
\begin{itemize}
	\item  any kind of scattering, be it due to radiative losses, Coulomb collisions or electromagnetic field fluctuations, is sufficiently weak to cause significant secular perturbations of a particle trajectory $\{{\bm v}(t),{\bm x}(t)\}$ only on time scales much longer than $\tau_{\rm pol}$.
\end{itemize}\vspace{-0.1cm}

\noindent Weak resonant wave-particle interactions on the transit time scale ($\tau_{\rm pol} \partial_t \sim \tau_{\rm tor} \partial_t \sim 1$) are allowed, including so-called adiabatic chirping via phase-space waterbags \cite{Hezaveh22}, whereas cyclotron resonances must generally be ruled out by requiring $\partial_t \ll \omega_{\rm B}$. These assumptions can be viewed as an extended definition of the ``banana regime'' that was mentioned in Section~\ref{sec:motivation}. Under these conditions, a Lorentz orbit's guiding center (GC),\vspace{-0.25cm}
\begin{equation}
	{\bm X}_{\rm gc}(t) \equiv {\bm x}(t) - {\bm \varrho}_{\rm L}(t) \quad \text{with} \quad {\bm \varrho}_{\rm L} \equiv {\bm v}\times \tilde{\hat{\bm b}}/\tilde{\omega}_{\rm B},
	\label{eq:com_gc}\vspace{-0.15cm}
\end{equation}

\noindent whose trajectory is indicated by the yellow line in Fig.~\ref{fig:01_geom}(a), will stay near and depart only slowly ($\gg \tau_{\rm pol}$) from an axisymmetric time-invariant toroidal surface, such as the one drawn red in Fig.~\ref{fig:01_geom}(a), which we call {\it orbit surface}. The particle trajectory can then be reasonably approximated by such an orbit surface, plus a set of initial conditions $(\vartheta_{\rm I},\zeta_{\rm I})$ that identify a particular unperturbed GC trajectory on that surface.

\subsection{Absolute and relative constants of motion (CoM)}
\label{sec:intro_com}

Each unperturbed orbit surface can be uniquely identified by triplets of exact constants of motion (CoM) \cite{Morozov66b}. Many choices of CoM exist, which differ only in their degree of intuitive appeal and practical utility, and are otherwise equivalent and can be transformed exactly from one to another. Following the examples shown in Fig.~1 of Ref.~\cite{Bierwage22a}, we distinguish between two classes, which shall here be called {\it absolute} and {\it relative} CoM.

{\it Absolute CoM} are based on the particle Hamiltonian and canonical actions that, for a tokamak-like configuration, are defined in terms of the gyrational, poloidal and toroidal angle integrals of the canonical momentum $\tilde{\bm p} \equiv Ze\tilde{\bm A} + M{\bm v}$ as $2\pi\tilde{J}_1 \approx \int_{\rm gyro}{\rm d}{\bm x}\cdot\tilde{\bm p}$, $2\pi\tilde{J}_2 \approx \int_{\rm pol}{\rm d}{\bm x}\cdot\tilde{\bm p}$, and $2\pi\tilde{J}_3 \approx \int_{\rm tor}{\rm d}{\bm x}\cdot\tilde{\bm p}$.\footnote{For instance, see Section 2.3 of Ref.~\protect\cite{BrochardThesis}. The reasons for our use of $\approx$ symbols when defining $\tilde{J}_{1,2,3}$ are explained in \protect\ref{apdx:can}.}
$\tilde{J}_3$ becomes algebraic and an exact invariant under the present assumption of perfect axisymmetry. The assumed explicit time-independence of the Hamiltonian $\tilde{\H}$ implies that the particle's total (kinetic + potential) energy $\tilde{\etot}$ is conserved exactly. Attaching the initial state of the parallel velocity's sign $\sigma = v_\parallel/|v_\parallel|$, we then use $\sigma\tilde{\etot}$ \cite{Bierwage22a} to take the place of the most cumbersome of the above integrals: $\tilde{J}_2$. Finally, it remains to deal with $\tilde{J}_1$. In the GC model that will be used in the main part of this paper, $\tilde{J}_1$ is going to be replaced by a magnetic moment $\mu$ that will be an exact invariant by construction (by eliminating any gyrophase-dependence) \cite{Littlejohn83}. In the present section, we are still considering full Lorentz orbits, where $\tilde{J}_1$ can often be adequately approximated by a series expansion (e.g., Eq.~(31) in Ref.~\cite{Littlejohn83}), unless the considered (time-independent) fields vary too much on the $\varrho_{\rm L}$ scale \cite{Carlsson01, Escande22}. As a first approximation, one assumes that $B$ is constant across the small gyroradius. This yields $\tilde{J}_1 \rightarrow \varrho_L M v_\perp/2 = M\mu_{(0)}/(Ze)$, where $\tilde{\mu}_{(0)} = M v_\perp^2/(2\tilde{B})$ is known as the leading-order magnetic moment. Higher-order corrections account for the nearby nonuniformity of ${\bm B}$ and ${\bm E}$, and the series should converge to a practically constant $\tilde{J}_1$ when condition (\ref{eq:com_b}) is satisfied. Since the expansion of $\tilde{J}_1$ needs to be truncated in practical implementations, some time-dependence remains. However, at least in the regime of adiabatic invariance, where $\tilde{J}_1$ oscillates periodically rather than chaotically \cite{Carlsson01, Escande22}, $\tilde{J}_1$ could in principle be replaced by a mean (bounce-averaged)  magnetic moment $\tilde{\mu} \propto \tau_{\rm pol}^{-1}\oint_{\rm pol}{\rm d}t\,\tilde{J}_1$ that is practically a CoM even for full Lorentz orbits. We then obtain the widely used set
\begin{subequations}
	\begin{align}
		\tilde{\mu} =&\; M v_\perp^2/(2\tilde{B}) + ... \propto \tilde{J}_1 \quad \text{magnetic moment},
		\label{eq:com_full_mu} \\
		\tilde{\etot} =&\; \tilde{\ekin} + Ze\tilde{\Phi} \qquad\qquad\quad\;\, \text{total energy},
		\label{eq:com_full_etot} \\
		\tilde{P}_\zeta =&\; -Ze\tilde{\Psi}_{\rm P} + M v_\zeta \propto \tilde{J}_3 \;\;\;  \parbox{2.9cm}{canonical toroidal \\ angular momentum,}
		\label{eq:com_full_pzeta}
	\end{align}
	\label{eq:com_full}\vspace{-0.2cm}
\end{subequations}

\noindent where $\tilde{\ekin} = Mv^2/2 = Mv_\perp^2/2 + Mv_\parallel^2/2$ is the kinetic energy. Recent application of absolute CoM for GC orbits include the Advanced Transport model for Energetic Particles (ATEP) \cite{Lauber24, Meng24} and fast ion orbit tomography \cite{Rud24}. The attribute ``absolute'' refers to the fact that the values of $\{\tilde{\mu},\sigma\tilde{\etot},\tilde{P}_\zeta\}$ are the same anywhere along the orbit, so that a reference point is not required.

{\it Relative CoM}, on the other hand, bear their name because they are defined by the value of a time-dependent quantity like $\tilde{K}(t)$ or $v_\parallel(t)$ at a well-defined reference point on an orbit. The relative CoM used in the present work are defined at the point where an orbit crosses the magnetic midplane in Fig.~\ref{fig:01_geom}(b) either from above or from below. Such coordinates are used, for instance, when sampling the phase space of guiding centers in our code {\tt VisualStart} for quiet starts of global particle-in-cell (PIC) simulations \cite{Bierwage22a, Bierwage12a}, for the numerical analysis of nonlinear fast particle dynamics \cite{Bierwage16a, Bierwage16b}, and for orbit tomography of fast ion phase space \cite{Stagner17, Stagner22} in the Orbit Weight Computational Framework (OWCF) \cite{Jaerleblad24}.

It must be noted that mathematically elegant canonical coordinates $\{\tilde{J}_1,\tilde{J}_2,\tilde{J}_3\}$ or our absolute CoM $\{\tilde{\mu},\sigma\tilde{\etot},\tilde{P}_\zeta\}$ in Eq.~(\ref{eq:com_full}) are often numerically troublesome, and the present work is no exception. We will find that it is advantageous, for reasons of accuracy, performance and simplicity, to utilize relative CoM that are measured at suitable reference points.

\subsection{Remarks on parallel circulation and radial drifts}
\label{sec:intro_circ_drift}

In the case of absolute CoM, ${\bm C} = \{\tilde{\mu},\sigma\tilde{\etot},\tilde{P}_\zeta\}$, the direction of circulation must also be specified in order to uniquely identify an orbit surface containing circulating particles, like the orange contour in Fig.~\ref{fig:01_geom}(b). It is customary to use the sign $\sigma \equiv v_\parallel/|v_\parallel|$ of the parallel velocity for this purpose (see Section 3 of Ref.~\cite{Bierwage22a} for notes of caution). When ${\bm E} = 0$, all mirror-trapped orbits, like the blue contour in Fig.~\ref{fig:01_geom}(b), consist of portions with positive and negative $v_\parallel$. The situation becomes more complicated in the presence of an ambient radial electric field $E_r$, which adds an offset to an orbit's toroidal precession in the laboratory frame of reference. This has the consequence that even a subset of the mirror-trapped particle population has parallel velocities $v_\parallel$ that oscillate without ever changing sign in the lab frame. We will inspect this effect in some detail later (Fig.~\ref{fig:13_jt60_deep-trapped}).

The mathematical form of absolute CoM in Eq.~(\ref{eq:com_full}) tells us that $v_\zeta$ must vary in time for $\tilde{\mu}$ and $\tilde{\etot}$ to remain constant in a field with nonuniform $B$. The form of $\tilde{P}_\zeta$ in Eq.~(\ref{eq:com_full_pzeta}) then implies that orbit surfaces (${\bm C} = {\rm const}$.), like the orange contour in Fig.~\ref{fig:01_geom}(b), deviate from a magnetic flux surfaces ($\Psi_{\rm P} = {\rm const}$., gray contours). This is physically realized by periodic drifts across magnetic surfaces due to the gradient $\nablab B$ and curvature $\hat{\bm b}\cdot\nablab\hat{\bm b}$, whose vectors in a tokamak are mainly radial and satisfy ${\bm B}\cdot\nablab\hat{\bm b} \approx \nablab B \approx -\hat{\bm e}_R B_0/R$. Moreover, the same field nonuniformity $\nablab B$, together with conservation of total energy $\tilde{\etot} = M v_\perp^2/2 + M v_\parallel^2/2 + Ze\tilde{\Phi}$ and magnetic moment $\tilde{\mu}$, causes some particles to be mirror-trapped on the outboard (low-field) side of the torus, where $B \approx B_0 R_0/R$ is weaker, giving rise to orbits with banana-shaped contours like the blue curve in Fig.~\ref{fig:01_geom}(b). These features of what we referred to in Section~\ref{sec:motivation} as {\it global magnetization} depend on the charge-mass-ratio $Ze/M$, the velocity ${\bm v}$, and are modified by the ambient electric field ${\bm E}$.

\subsection{Local Maxwellian versus CoM distribution}
\label{sec:intro_distrib}

In regions where the plasma is sufficiently cool and dense for the characteristic collision time $\tau_{\rm coll}$ to be much shorter than the poloidal transit time $\tau_{\rm pol}$, such that $\tau_{\rm coll} \ll \tau_{\rm pol}$, the particle distribution tends to relax towards a state that may be approximated by an isotropic {\it local} Maxwellian of the form\vspace{-0.1cm}
\begin{equation}
	f_{\rm loc}(r,\tilde{\ekin}) = \left(\frac{M}{2\pi}\right)^{3/2} \frac{N(r)}{T^{3/2}(r)}\exp\left(-\frac{\tilde{\ekin}}{T(r)}\right).
	\label{eq:intro_fMloc}\vspace{-0.1cm}
\end{equation}

\noindent The attribute ``local'' refers to the fact that global magnetization effects are overriden by collisions when $\tau_{\rm coll} \ll \tau_{\rm pol}$. For the spatial dependence this means that the distribution can be expressed in terms of a local number density $N(r)$ and temperature $T(r)$ that are magnetic flux functions, here expressed in terms of a minor radial coordinate $r(\Psi_{\rm P})$, whose unit vector $\hat{\bm e}_r \equiv \nablab r/|\nablab r|$ is transverse to magnetic surfaces. Likewise, the velocity-dependence is captured by the local kinetic energy $\tilde{K}$.

In contrast, in the core of a tokamak that has been heated to conditions where a deuterium-tritium plasma is expected to enter a controlled burning state ($T \gtrsim 10\,{\rm keV}$, $N \lesssim 10^{20}\,{\rm m}^{-3}$ \cite{WessonBook}), the poloidal transit time of ions is expected to be shorter than their collision time: $\tau_{\rm pol} \ll \tau_{\rm coll}$. The form of their steady-state distribution $f_{\rm eq}({\bm v},{\bm x})$ can then be significantly influenced by the variation of $v_\parallel$ and the radial drifts discussed in Section~\ref{sec:intro_circ_drift} above. The consequence is that the steady state $f_{\rm eq}$ may deviate significantly from the collision-dominated state that would be characterized by the local Maxwellian (\ref{eq:intro_fMloc}). Instead, $f_{\rm eq}$ approaches the topological equilibrium distribution of collisionless confined orbits, $f_{\rm com}$, that depends only on the CoM identifying unperturbed orbit surfaces, such as the set $\{\tilde{\mu}, \sigma\tilde{\etot},\tilde{P}_\zeta\}$ in Eq.~(\ref{eq:com_full}).

The mathematical form of the CoM in Eq.~(\ref{eq:com_full}) highlights the fact that the nonuniform magnetization intertwines the configuration space ${\bm x}$ with the velocity space ${\bm v}$, and this causes $f_{\rm eq}$ to deviate from $f_{\rm loc}$ when the characteristic collision time $\tau_{\rm coll}$ exceeds the relevant global transit time, here $\tau_{\rm pol}$. Since the ratio $\tau_{\rm coll}/\tau_{\rm pol}$ is also a function of both ${\bm x}$ and ${\bm v}$, we anticipate that a `perfect' model of $f_{\rm eq}$ would have to be more complicated, with the local Maxwellian $f_{\rm loc}$ and the CoM distribution $f_{\rm com}$ being more accurate at low and high velocities, respectively. If a choice needs to be made, $f_{\rm com}$ appears to be more broadly applicable because the radial drifts are ignorable at low velocities, so that $\tilde{P}_\zeta \rightarrow -Ze\tilde{\Psi}_{\rm P}(r)$.

If one were to use the local Maxwellian $f_{\rm loc}$ as a reference state $f_{\rm ref}$ for instability analyses and transport models in a sparse hot plasma in domains where the difference between $\tilde{P}_\zeta$ and $-Ze\tilde{\Psi}_{\rm P}$ plays a non-negligible role, the resulting inconsistencies can be handled as follows:
\begin{enumerate}
	\item[(a)] maintain the deviation $\Delta f_{\rm eq} \equiv f_{\rm eq} - f_{\rm ref} \approx f_{\rm com} - f_{\rm loc}$ artificially in a {\it pseudo-equilibrium}; or
	\item[(b)] permit a so-called {\it prompt relaxation} during which the discrepancy is absorbed by the perturbation in Eq.~(\ref{eq:intro_df}) as $\delta f(t) = f(t) - f_{\rm ref} = f(t) - f_{\rm eq} + \Delta f_{\rm eq}$. This is essentially a form of a ``non-quiet start''.
\end{enumerate}	

Pseudo-equilibria (a) are often used to freeze the reference state of the bulk plasma, for instance, when estimating steady state fluxes of particles and heat, the ambient electric field and plasma rotation. They are also used when the true magnetohydrodynamic (MHD), drift-kinetic or gyrokinetic ``equilibrium'' is not accurately known or subject to numerical inaccuracies.

Prompt relaxation (b) is sometimes employed (or tolerated) when modeling minority species such as fast ions, whose populations are sufficiently small to have (at leading order) a negligible effect on the bulk plasma's reference state. The initial fast ion distribution may, for instance, be based on a measured radial density profile $N(r)$, which then relaxes to a near-CoM distribution.

Both of the above methods are convenient tools for testing the plasma response on short time scales. However, prompt relaxation may result in poor control over the initial conditions of a simulation, and its transients may have undesirable side effects, including signal-noise correlations (e.g., Fig.~1 of Ref.~\cite{Bierwage12a}). Both methods tend to come with a lack of self-consistency and potentially limit our ability to study the connection and interplay between processes on multiple time scales, including the slow evolution of the reference state. When collisions are rare, the use of CoM can help to overcome these issues as it allows to construct a reference state $f_{\rm ref}$ that more closely resembles the true steady state $f_{\rm eq}$ by accounting at least for the global effects of magnetization (drifts and mirror force, and electric modifications thereof). As was expressed in Eq.~(\ref{eq:intro_fcom}), each drift orbit surface can be seen as a unique basis function in 3D CoM space. The long-term effects of weak sources, collisions, sinks and even instabilities can be taken into account via the orbits' weights.

\vspace{-0.3cm}
\subsection{Considerations regarding full Lorentz orbits}
\label{sec:intro_fo}

From the point of view of particle simulations, the use of full Lorentz orbits is attractive due to the formal simplicity of their equations of motion and their high degree of realism: Full orbit motion implicitly accounts for all global magnetization and gyroaveraging effects, and it gives rise to diamagnetic drift. Difficulties are however encountered when one tries to construct a discrete CoM mesh for the representation of distribution functions $f({\bm v},{\bm x})$ in terms of full Lorentz orbits in 3D, because the reduction from 6D phase space to a 3D CoM space is not entirely straightforward: Only dimensions six and five --- the toroidal angle $\zeta$ and poloidal orbit time $\tau$ --- are readily eliminated via the assumption of axisymmetry and time-independence. In order to eliminate the fourth dimension --- the gyrophase $\xi$ --- and sample orbits in the magnetic midplane (the dashed line in Fig.~\ref{fig:01_geom}(b)), we need to specify the initial gyrophase $\xi_{\rm I} \equiv \xi(t=0)$ at the particle's launch point, which poses the following dilemma: On each transit, a gyrating particle crosses the midplane multiple times (cf.~Fig.~\ref{fig:d01_lorentz2pi}), but in order to perform the required mapping onto the midplane, one needs a {\it unique crossing point}. As discussed in \ref{apdx:can}, one way to deal with this (and with related issues for all angles) is to utilize the concept of a guiding center, whose instantaneous location may be defined as in Eq.~(\ref{eq:com_gc}): ${\bm X}_{\rm gc}(t) \equiv {\bm x}(t) - {\bm \varrho}_{\rm L}(t)$.

It is desirable to initialize both ${\bm x}(t)$ and ${\bm X}_{\rm gc}(t)$ in the midplane at $t=0$. The gyrophase $\xi_{\rm I}$ must then be such that the gyroradius vector ${\bm \varrho}_{\rm L}(t) = {\bm v}\times\tilde{\hat{\bm b}}/\tilde{\omega}_{\rm B}$ is aligned with the midplane; that is, approximately horizontal:\vspace{-0.2cm}
\begin{equation}
	{\bm \varrho}_{\rm L}(\xi_{\rm I}) \approx \pm\rho_{\rm L}\hat{\bm e}_R \quad \Leftrightarrow \quad |{\bm v}(\xi_{\rm I})| \approx |v_z|, \quad v_R(\xi_{\rm I}) \approx 0.
	\label{eq:fo_xc_horiz}\vspace{-0.0cm}
\end{equation}

\noindent However, Eq.~(\ref{eq:fo_xc_horiz}) complicates the computation of the Jacobians for transformations like	$\{\tilde{\mu},\sigma\tilde{\etot},\tilde{P}_\zeta\} \rightarrow \{\tilde{\ekin},v_\parallel,R\}$ because\vspace{-0.15cm}
\begin{equation}
	v_\parallel \tilde{B} = {\bm v}\cdot\tilde{\bm B} = v_\zeta \tilde{B}^\zeta + v_{\rm pol} \tilde{\bm B}_{\rm pol} \approx v_\zeta \tilde{B}^\zeta + v_z \tilde{B}^z,
	\label{eq:fo_vparB}\vspace{-0.15cm}
\end{equation}

\noindent contains not only $v_\zeta$ but also $v_{\rm pol}\approx v_z(\tilde{\ekin},v_\parallel,R) \neq 0$.

Many particles have at least one gyrophase $\xi = \xi_{\rm I}$ for which $v_{\rm pol}(\xi_{\rm I}) = 0$, so that Eq.~(\ref{eq:fo_vparB}) reduces to $v_\parallel \tilde{B} = v_\zeta \tilde{B}^\zeta$. For some particles, $v_{\rm pol}$ never vanishes in the midplane, but even then the approximation $v_\parallel \tilde{B} \approx v_\zeta \tilde{B}^\zeta$ is usually reasonably accurate in tokamaks. The canonical angular toroidal momentum then acquires the convenient form $\tilde{P}_\zeta = -Ze\tilde{\Psi}_{\rm P} + M v_\zeta \rightarrow -Ze\tilde{\Psi}_{\rm P} + M v_\parallel \tilde{B}/\tilde{B}^\zeta$ \cite{BrochardEPS24}. What is less convenient though is the fact that ${\bm \varrho}_{\rm L}$ is now approximately vertical:\vspace{-0.2cm}
\begin{equation}
	{\bm \varrho}_{\rm L}(\xi_{\rm I}) \approx \pm\rho_{\rm L}\hat{\bm e}_z \quad \Leftrightarrow \quad |{\bm v}(\xi_{\rm I})| \approx |v_R|, \quad v_z(\xi_{\rm I}) \approx 0.
	\label{eq:fo_xc_vert}\vspace{-0.0cm}
\end{equation}

\noindent Thus, initializing full orbits with gyrophases $\xi_{\rm I}$ satisfying Eq.~(\ref{eq:fo_xc_vert}) puts their center points ${\bm X}_{\rm gc}(\xi_{\rm I})$ above or below the midplane, while $\dot{\bm x} = {\bm v}$ is tangential to the midplane. This necessitates further minor approximations when converting full orbits to midplane-based relative CoM coordinates, gyrocenters or guiding centers.
	
\vspace{-0.25cm}
\subsection{GC approximation capturing global magnetization}
\label{sec:intro_gc}

The above issues arose because $\tilde{P}_\zeta$ and $\tilde{\mu}$ are expressed in terms of $v_\zeta$ and $v_\perp$, respectively, which constitutes a mixture of the two projections of the particle velocity ${\bm v}$ in Eq.~(\ref{eq:com_vfull}). This problem disappears if one eliminates the fast gyration and performs a reduction to gyrocenters (gyrokinetic model with gyroaveraged fields) or guiding centers (drift-kinetic model with local fields). These models rely on the slow variation (in space and time) of ${\bm B}$ and ${\bm E}$ \cite{Littlejohn83, ChenH24} as in Eq.~(\ref{eq:com_b}). $\mu$ can then be defined to be conserved exactly at the desired order, and $P_\zeta$ contains the parallel velocity $u$ (instead of $v_\zeta$) by default. This facilitates straightforward transformations between different coordinates, including our absolute and relative CoM in 3D. The drifts associated with global magnetization are captured in a physically transparent fashion. The diamagnetic drift associated with local magnetization arises via the gradient of the distribution function's (perpendicular) pressure moment.

In this work, the reference distribution $f_{\rm ref}$ is described as a function that depends only on the CoM of unperturbed guiding center (GC) drift orbits. The usual set of absolute CoM $\{\mu,\sigma_{\uGC}\etot,P_\zeta\}$ is given, for instance, by Eqs.~(3.60)--(3.61) of Ref.~\cite{Cary09}. Omitting the negligible ponderomotive potential from $\etot$ as discussed in the last paragraph of Appendix C.2 of Ref.~\protect\cite{Bierwage22d}, we have
\begin{subequations}
	\begin{align}
		\mu \equiv&\; Mv_\perp^2/(2B) = \text{\rm constant by construction},
		\label{eq:intro_gc_com_mu} \\
		\etot =&\; K + Ze\Phi = M\uGC^2/2 + \mu B + Ze\Phi,
		\label{eq:intro_gc_com_etot} \\
		P_\zeta =&\; -Ze\Psi_{\rm P} + M\uGC B_\zeta/B,
		\label{eq:intro_gc_com_pzeta}
	\end{align}
\label{eq:intro_gc_com}\vspace{-0.3cm}
\end{subequations}

\noindent where $\uGC = \sigma_{\uGC}\sqrt{2(K - \mu B)/M}$ is the parallel GC velocity with sign $\sigma_{\uGC} \equiv \uGC/|\uGC|$. In contrast to $\{\tilde{\mu},\sigma\tilde{\etot},\tilde{P}_\zeta\}$ in Eq.~(\ref{eq:com_full}) for the full Lorentz orbit, the fields appearing in the GC energy $\etot$, magnetic moment $\mu$, and canonical toroidal angular momentum $P_\zeta$ are evaluated at the GC position ${\bm X}_{\rm gc}(t)$. Unlike the full orbit's $\tilde{\mu}$, the GC's lowest-order magnetic moment $\mu$ is defined to be an exact constant \cite{Littlejohn83}. The difference between a full orbit's $\tilde{\ekin}$ and the GC kinetic energy $\ekin = Mu^2/2 + \mu B$ is perhaps most easily seen when the conservation of kinetic energy is broken via the presence of an electric field: $\hat{\ekin}$ and $\ekin$ evolve differently because $\Phi({\bm x}(t)) \neq \Phi({\bm X}_{\rm gc}(t)$).

\subsection{Plasma rotation}
\label{sec:intro_rot}

As was noted in Section~\ref{sec:motivation}, when dealing with dynamics in the diamagnetic drift range of frequencies (a few to tens of ${\rm kHz}$), the effect of (sheared) plasma rotation must be taken into account. In tokamaks, plasma rotation, or mean flows in general, can arise as follows:\vspace{-0.15cm}
\begin{itemize}
	\item[(i)]  {\it A stationary ambient radial electric field} $E_r$ causes flows that are composed of the individual particles' electric drifts, ${\bm v}_{\rm E} = {\bm E}\times{\bm B}/B^2$. If $E_r$ is effectively uniform on the gyroradius scale, the mean electric flow ${\bm V}_{\rm E}$ becomes independent of the charge-mass ratio $Ze/M$, so it is identical for all species.\footnote{${\bm V}_{\rm E}$ then takes the form of a bulk mass flow whose contribution to the electric current $\mu_0{\bm J} \approx \nablab\times{\bm B}$ is negligible. Under this condition, ${\bm V}_{\rm E}$ does not perturb the magnetic field at leading order, but its inertia can shift the plasma relative to magnetic surfaces. This, in turn, creates an intrinsic parallel electric field $E_\parallel = \hat{\bm b}\cdot{\bm E}$ that accelerates and decelerates electrons in such a way as to maintain quasi-neutrality. By the same token the radial shift of ion orbits due to magnetic drift that is evident from $P_\zeta$ in Eq.~(\protect\ref{eq:intro_gc_com_pzeta}) contributes to $E_\parallel$. In the present work, we assume that the centrifugal shift and $E_\parallel$ are ignorable.}
	
	\item[(ii)]  {\it Gradients in density or pressure} can ``uncover'' (= make macroscopically measurable due to incomplete cancellation) the motion of charged particles. Uncovered local gyration yields diamagnetic flows ${\bm V}_*$, which are global but do not carry mass. Uncovered drift-orbits yield global mass flows \cite{SpitzerBook62}.
	
	\item[(iii)]  {\it Parallel ion mass flows} along the magnetic field ${\bm B}$ can be produced by direct momentum input (e.g., beam injection). In addition, the magnetization flows (i) and (ii), which are locally transverse to ${\bm B}$ at leading order, have parallel components due to the nonuniformity of ${\bm B}$ in a torus.
	Moreover, elastic Coulomb collisions can generate a parallel flow that tends to cancel the poloidal component of the ion mass flow (i), effectively converting (or redirecting) it into toroidal flow.
\end{itemize}

In the present work, the ambient $E_r$ field (i) is treated as a free input parameter. Density and pressure gradients (ii), including those produced by loss boundaries, contribute via global magnetization flows. Diamagnetic flow (local magnetization) is not present in the moments taken and will be estimated from the pressure gradient using Eq.~(\ref{eq:vdiamag}) if needed. The collisionless intrinsic part of the parallel flow in item (iii) is closely connected with the mirror force and modulated magnetic drift, so it is also captured by our GC orbit-based description. 

\begin{figure*}
	[tbp]\vspace{-2.4cm}
	\centering
	\includegraphics[width=0.9\textwidth]{\figures/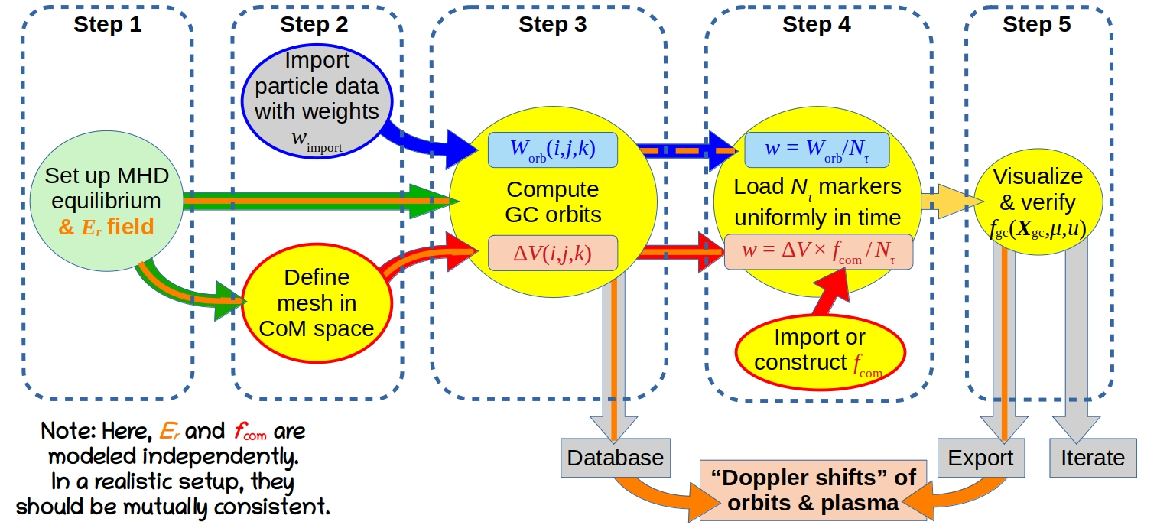}\vspace{-0.2cm}
	\caption{Relevant part of the extended {\tt VisualStart} workflow in five steps as described in the text. (Adapted from Fig.~2 of Ref.~\protect\cite{Bierwage22a}.) The present paper focuses on the tasks highlighted yellow. The orange lines indicate the flow of information associated with the $E_r$ field model.}
	\label{fig:02_vstart}\vspace{-0.3cm}
\end{figure*}

Meanwhile, the collisional redirection or damping of poloidal flow and the associated generation of additional parallel flow that we mentioned in item (iii) must be modeled by other means. For strong and moderate collisionality, this collisional flow conversion is usually described in terms of a Fermi-like acceleration process known as ``magnetic pumping'' \cite{Berger58, Rosenbluth72, Stix73}. In the present context, this may be most readily realized via transit-time heating, where collisions in a nonuniform ${\bm B}$ field irreversibly generate a parallel flow whose poloidal component tends to cancel the original poloidal mass flow. The result of such a calculation could be incorporated here by weighting the GC orbits in Eq.~(\ref{eq:intro_fcom}) in accordance in with a suitable distribution function, such as a shifted Maxwellian model $\propto \exp(-M|\uGC - U_\parallel|^2/[2 T_\parallel])$ with precomputed mean parallel flow $U_\parallel$. However, as discussed in Ref.~\cite{Novakovskii97}, poloidal flow dynamics can be rather complicated in weakly collisional plasmas, possibly involving zonal oscillations known as geodesic acoustic modes (GAM) \cite{Winsor68}; i.e., a nonstationary state.

Here, we ignore the effects of flow relaxation processes and work with an unshifted CoM Maxwellian as a simple test case, which, albeit not realistic, has merits on its own. For instance, this choice simplifies benchmarks and can make certain physical aspects more transparent than a realistic distribution function would.

\subsection{Workflow and outline}
\label{sec:intro_outline}

In Section~\ref{sec:method}, we describe how the effect of an ambient radial electric field $E_r$ is incorporated in the code {\tt VisualStart} \cite{Bierwage22a} and explain how our method of slicing the GC orbit space along the lines of midplane-based relative CoM coordinates unifies the merits of simplicity, accuracy and efficiency.

The extended workflow comprising {\tt VisualStart} is shown in Fig.~\ref{fig:02_vstart}. The field configuration is prepared in Step 1 (Sec.~\ref{sec:method_equil}), which now includes the preparation of a model for the radial electric field $E_r$ (Sec.~\ref{sec:method_fld}). As indicated by the bold orange lines overlaying the green arrows, the $E_r$ field is first used in Step 2 when slicing the CoM space into volume elements $\Delta\V$ (Sec.~\ref{sec:method_mid}--\ref{sec:method_slice}). In Step 3, the same $E_r$ field is used when we compute GC orbits using the equations in Section~\ref{sec:method_gc}. Continuing along the red arrows in Fig.~\ref{fig:02_vstart}, the GC orbits are then weighted by a model distribution function $f_{\rm com}$ and sampled by marker particles in Step 4 (Sec.~\ref{sec:method_weight}).

The methods described in the present paper focus primarily on the workflow branch indicated by red arrows between the yellow ellipses in Fig.~\ref{fig:02_vstart}, where one constructs a GC distribution from scratch on a user-defined CoM mesh (Step 2) and a user-defined model distribution function $f_{\rm com}$ (Step 4). Of course, {\tt VisualStart} can also process precomputed particle data and turn it into a CoM distribution, which is the blue branch of the workflow in Fig.~\ref{fig:02_vstart}. Provided that the imported particle data and $E_r$ are mutually consistent, this is straightforward and hence not elaborated further in this paper.

In the final Step 5 in Fig.~\ref{fig:02_vstart}, we visualize and analyze the constructed GC distribution $f_{\rm gc}$. This is the subject of Section~\ref{sec:example}, where the extended code is applied to examples with realistic geometry based on KSTAR, JT-60U and ITER tokamak plasmas.

In Section~\ref{sec:gc}, we verify our numerical results against analytical estimates at the level of individual GC orbits. We also analyze the effect of $E_r$ on orbit contours and their transit frequencies, and discuss implications for resonance analyses in rotating plasmas. For this purpose, the results of Step 3 (orbit database) and Step 5 (moments of the distribution function) can be used to evaluate the respective electric ``Doppler shifts'' as indicated below the main workflow in Fig.~\ref{fig:02_vstart}. In doing so, one must account for ambiguities (``reference point bias'') in an orbit's electric frequency shift, which are caused by the electric modulation of the mirror force.

Section~\ref{sec:summary} contains a summary and concluding remarks concerning the possible application of our methods for the realistic modeling of particle distributions.

The Appendices contain more information about our test cases, numerical benchmarks and mathematical derivations. We emphasize that the GC orbit weights and the electric field are treated as free input parameters, so their physical consistency must be enforced by other means (e.g., via iteration with neoclassical codes).

\section{Methods for phase space sampling with $E_r$}
\label{sec:method}

\subsection{Background equilibrium}
\label{sec:method_equil}

The inclusion of $E_r$ field effects is important for problems such as the realistic modeling of tokamak equilibria, the analysis of low-frequency resonances, and the estimation of transport and losses. At the next level, the effect of $E_r$ on the redistribution of particles and instabilities may also feed back on the magnetohydrodynamic (MHD) equilibrium. However, here we ignore such feedback effects and assume that the magnetic configuration satisfies the conventional Grad-Shafranov equation with scalar pressure $P_{\rm mhd}(r)$ and without plasma rotation effects. The MHD equilibria and other parameters of the working examples we use are described in detail in \ref{apdx:examples}.

\subsection{Fields, spatial coordinates, normalizations}
\label{sec:method_fld}

We consider axisymmetric ambient magnetic and electric fields, ${\bm B} = \nablab\times{\bm A}$ and ${\bm E} = -\nablab\Phi$, of the form
\begin{subequations}
	\begin{align}
		{\bm B} =&\; \nablab\zeta\times\nablab\Psi_{\rm P} + I(\Psi_{\rm P})\nablab\zeta,
		\label{eq:method_B} \\
		{\bm E} =&\; E_{\Psi_{\rm P}} \nablab\Psi_{\rm P} = -\frac{{\rm d}\Phi(\Psi_{\rm P})}{{\rm d}\Psi_{\rm P}}\nablab\Psi_{\rm P},
		\label{eq:method_E}
	\end{align}
\label{eq:method_BE}\vspace{-0.4cm}
\end{subequations}

\noindent where $2\pi\Psi_{\rm P}$ is the poloidal magnetic flux, $I \equiv B_\zeta = \partial_\zeta{\bm r}\cdot{\bm B} = R B_{\rm tor}$ is the covariant component of the magnetic field along the toroidal angle $\zeta$, and $\Phi$ is the electrostatic potential. We basically work with right-handed cylinder coordinates $(R,z,\zeta)$ as shown in Fig.~\ref{fig:01_geom}(a), although, in many cases, we map the poloidal plane using the auxiliary radial and vertical coordinates $X \equiv R - R_0$ and $Y \equiv z - z_0$ relative to the magnetic axis $(R_0,z_0)$. The values of the magnetic field strength and poloidal flux function at the axis are denoted by $B_0$ and $\Psi_{\rm P0}$. Another pair of coordinates that we use within the last closed flux surface is the volume-averaged minor radius $r(\Psi_{\rm P}) \in [0,a]$ and the geometric poloidal angle $\vartheta$ measured counter-clockwise starting from $z = z_0$ on the low-field side ($X>0$) as also shown in Fig.~\ref{fig:01_geom}(a). Furthermore, we make use of the normalized poloidal flux $\psi_{\rm P} \equiv (\Psi_{\rm P} - \Psi_{\rm P0}) / (\Psi_{{\rm P}a} - \Psi_{\rm P0})$, where $\Psi_{{\rm P}a} \equiv \Psi_{\rm P}(r=a)$ is the value at the last closed flux surface.

\begin{figure}
	[tb]
	\centering
	\includegraphics[width=0.48\textwidth]{\figures/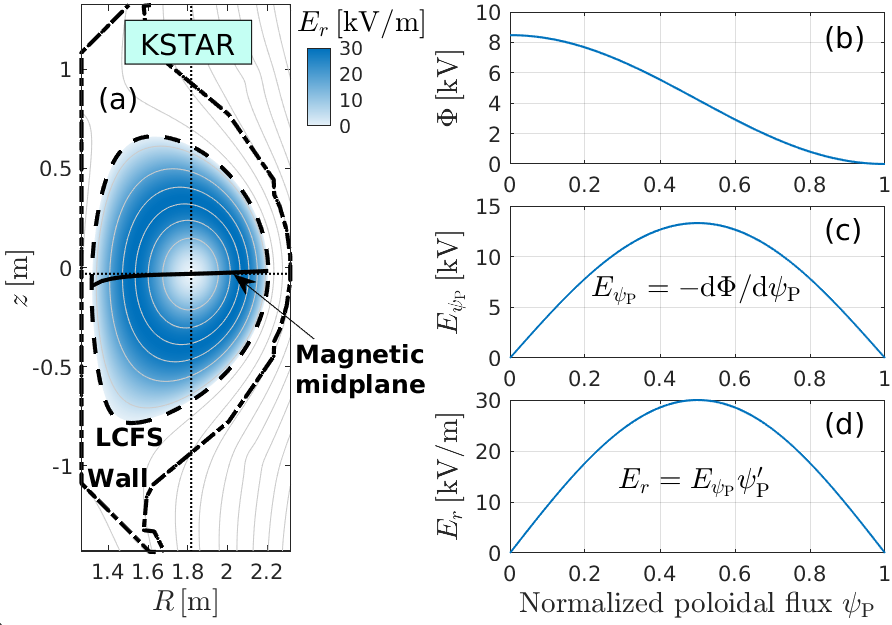}
	\caption{Analytical model of the radial electric field $E_r$ that is used as a working example in this paper. The blue-shaded contours in panel (a) show $E_r(R,z)$ in the poloidal plane of an MHD equilibrium based on KSTAR shot 18567 \protect\cite{Lee23}. Panel (a) also shows some magnetic flux surface contours (gray), the last closed flux surface (LCFS, dashed), the wall (dash-dotted), and the magnetic midplane (solid black). The magnetic axis is located at $R_0 = 1.82\,{\rm m}$, $z_0 = -0.03\,{\rm m}$ (dotted black). Panels (b)--(d) show the radial profiles of the electrostatic potential $\Phi$ and the covariant electric field components $E_{\psi_{\rm P}}$ and $E_r$ from Eq.~(\protect\ref{eq:er_model}).}
	\label{fig:03_kstar_E30_model}%
\end{figure}

In the present work, we use a simple analytical model of the electrostatic potential and radial electric field:
\begin{subequations}
	\begin{align}
		\Phi(\psi_{\rm P}) =&\; \frac{|E_{r0}| + E_{r0}\cos(\pi\psi_{\rm P})}{\pi \psi_{\rm P}'(0.5)},
		\label{eq:er_model_phi} \\
		\Rightarrow \quad E_{\Psi_{\rm P}}(\psi_{\rm P}) =&\; -\frac{{\rm d}\Phi}{{\rm d}\Psi_{\rm P}} = E_{r0}\frac{\sin(\pi\psi_{\rm P})}{\Psi_{\rm P}'(0.5)}, \\
		\Rightarrow \quad E_r(\psi_{\rm P}) =&\; \Psi_{\rm P}' E_{\Psi_{\rm P}} = E_{r0} \frac{\sin(\pi\psi_{\rm P})}{\psi_{\rm P}'(0.5) / \psi_{\rm P}'(\psi_{\rm P})},
	\end{align}
	\label{eq:er_model}\vspace{-0.3cm}
\end{subequations}

\noindent where $\Psi_{\rm P}' \equiv {\rm d}\Psi_{\rm P}/{\rm d}r$. We thus have $E_r(0) = E_r(1) = 0$ and $E_r(\psi_{\rm P} = 0.5) = E_{r0}$. The profiles for the case $E_{r0} = 30\,{\rm kV/m}$ are shown in Fig.~\ref{fig:03_kstar_E30_model}. The CoM space sampling procedure described in Section~\ref{sec:method_slice} below assumes that $\Phi$ includes a constant offset such that\vspace{-0.1cm}
\begin{equation}
	{\rm min}\{\Phi\} = 0\vspace{-0.1cm}
\end{equation}

\noindent in the domain of interest, which is realized here by the term $|E_{r0}|$ in the numerator of Eq.~(\ref{eq:er_model_phi}) for $\psi_{\rm P} \in [0,1]$.

We normalize the magnetic field by the on-axis value $B_0$, velocities by a reference velocity $v_0$, and energies by $M v_0^2$, where $M$ is the particle mass. Lengths remain unnormalized unless stated otherwise. This yields\vspace{-0.1cm}
\begin{gather}
	\hat{u} = \frac{u}{v_0}, \;\; \hat{\mu} = \frac{\mu B_0}{M v_0^2}, \;\; \hat{\ekin} = \frac{\ekin}{M v_0^2}, \;\; \hat{\bm B} = \frac{\bm B}{B_0}, \;\; \hat{\Psi}_{\rm P} = \frac{\Psi_{\rm P}}{B_0}, \nonumber \\
	\hat{E}_{\hat{\Psi}_{\rm P}} = \frac{E_{\Psi_{\rm P}}}{v_0} = -\frac{{\rm d}\hat{\Phi}}{{\rm d}\hat{\Psi}_{\rm P}}, \quad \hat{\Phi} = \frac{\Phi}{v_0 B_0}, \quad \hat{\bm E} = \frac{\bm E}{v_0 B_0}.
	\label{eq:norm}
\end{gather}

\subsection{Guiding center equations of motion}
\label{sec:method_gc}

The motion of guiding centers (GC) with electric charge $Ze$ and mass $M$ is described in terms of Hamiltonian equations of motion that govern the evolution of the GC position ${\bm X}_{\rm gc}(t)$ and the GC parallel velocity $u(t)$ with constant magnetic moment $\mu$ as described by Littlejohn \cite{Littlejohn83}. Adopting the notation of Eq.~(C13) in Ref.~\cite{Bierwage22d}, and ignoring terms originating from the ponderomotive potential $-M v_{\rm E}^2/2$ in Eq.~(C.3) of Ref.~\cite{Bierwage22d}, which is negligibly small in our cases,\footnote{See the last paragraph of Appendix C.2 of Ref.~\protect\cite{Bierwage22d}}
we have
\begin{subequations}
	\begin{align}
		\dot{\bm X}_{\rm gc} =&\; {\bm v}_\parallel^* + {\bm v}_{\rm d}^* + {\bm v}_{\rm E}^*,
		\label{eq:mdl_dXgc_dt}
		\\
		M\dot{\uGC} =&\; {\bm B}^* \cdot (Ze{\bm E} - \mu\nablab B) / B_\parallel^*, \nonumber \\
		=&\; (ZeB E_\parallel + Z e B E_\nabla - \mu{\bm B}^*\cdot\nablab B) / B_\parallel^*,
		\label{eq:mdl_du_dt}
	\end{align}\vspace{-0.4cm}
\end{subequations}

\noindent with $\uGC = \hat{\bm b}\cdot\dot{\bm X}_{\rm gc}$, and with the definitions\vspace{-0.1cm}
\begin{subequations}
	\begin{align}
		&{\bm B}^* \equiv {\bm B} + \frac{M \uGC}{Ze} \nablab\times\hat{\bm b}, \quad B_\parallel^* \equiv {\bm B}^*\cdot\hat{\bm b},
		\label{eq:gc_def_b_star}
		\\
		&{\bm v}^*_\parallel \equiv \uGC\frac{\bm B}{B_\parallel^*}, \quad {\bm v}^*_{\rm d} \equiv \frac{\mu}{Ze} \frac{\hat{\bm b}\times\nablab B}{B_\parallel^*} + \frac{M\uGC^2}{Ze } \frac{\nablab\times\hat{\bm b}}{B_\parallel^*},
		\label{eq:gc_def_uPar_uD_star}
		\\
		&{\bm v}_E^* \equiv \frac{{\bm E}\times {\bm B}}{B B_\parallel^*}, \; E_\parallel \equiv \hat{\bm b}\cdot{\bm E}, \;\, E_\nabla \equiv \uGC\frac{{\bm E}_\perp\cdot(\nablab\times\hat{\bm b})}{\omega_{\rm B}}.
		\label{eq:gc_def_vE_star}
	\end{align}\vspace{-0.3cm}
\end{subequations}

\noindent Note that the magnetic field gradients are related as\vspace{-0.15cm}
\begin{equation}
	B\nablab\times\hat{\bm b} = \hat{\bm b}\times\nablab B + \nablab\times{\bm B} = \hat{\bm b}\times\nablab B + \mu_0{\bm J},\vspace{-0.2cm}
\end{equation}

\noindent with plasma current density ${\bm J} = \mu_0^{-1}\nablab\times{\bm B}$ and vacuum permeability $\mu_0$. The ${\bm E}$ field in the form of Eq.~(\ref{eq:method_E}) has $E_\parallel = 0$ and $E_r \neq 0$, and contributes via the terms\vspace{-0.15cm}
\begin{subequations}
	\begin{align}
		{\bm E} \times {\bm B} \stackrel{E_r}{=}&\; E_{\Psi_{\rm P}} \left(\nablab\Psi_{\rm P}\times(\nablab\zeta\times\nablab\Psi_{\rm P}) + I\nablab\Psi_{\rm P}\times\nablab\zeta\right) \nonumber \\
		=&\; E_{\Psi_{\rm P}} (R B_{\rm pol}^2 \hat{\bm e}_\zeta - R B_{\rm tor} {\bm B}_{\rm pol}), \\
		E_\nabla\frac{B}{B^*_\parallel} \stackrel{E_r}{=}&\; \uGC \frac{{\bm v}_{\rm E}^*\cdot\nablab B}{\omega_{\rm B}} \qquad \parbox{2.85cm}{(electric modulation of the mirror force),}
	\end{align}\vspace{-0.2cm}
\end{subequations}

\noindent where ${\bm B}_{\rm pol} = \nablab\zeta\times\nablab\Psi_{\rm P}$ and $B_{\rm pol}^2 = |\nablab\Psi_{\rm P}|^2 |\nablab\zeta|^2 = |\nablab\Psi_{\rm P}|^2/R^2$, and where ${\bm J}\cdot{\bm E} = 0$ holds, because we assume that the electric field has the form ${\bm E} = E_{\Psi_{\rm P}}(\Psi_{\rm P})\nablab\Psi_{\rm P}$ and because the Grad-Shafranov equilibrium current satisfies ${\bm J}\cdot\nablab\Psi_{\rm P} = 0$ owing to $\nablab P_{\rm mhd} = {\bm J}\times{\bm B}$ and pressure $P_{\rm mhd}(\Psi_{\rm P})$ being a flux function. 

Applying the normalizations in Eq.~(\ref{eq:norm}) yields
\begin{subequations}\vspace{-0.2cm}
	\begin{align}
		\hat{\bm B}^* =&\; \hat{\bm B} + \varrho_0 \hat{\uGC} \frac{\hat{\bm b}\times\nablab\hat{B} + \nablab\times\hat{\bm B}}{\hat{B}}, \\
		\hat{\bm v}_{\rm E}^* =&\; \frac{\hat{\bm E} \times \hat{\bm B}}{\hat{B} \hat{B}^*_\parallel} \stackrel{E_r}{=} \frac{\hat{E}_{\hat{\Psi}_{\rm P}}}{\hat{B} \hat{B}^*_\parallel} R (\hat{B}_{\rm pol}^2 \hat{\bm e}_\zeta - \hat{B}_{\rm tor} \hat{\bm B}_{\rm pol}),
		\label{eq:gc_vE_norm}
		\\
		\hat{E}_\nabla \frac{\hat{B}}{\hat{B}_\parallel^*} =&\; \varrho_0 \hat{u}\frac{\hat{\bm E}_\perp\cdot(\nablab\times\hat{\bm b})}{\hat{B}_\parallel^*} \stackrel{E_r}{=} \varrho_0 \hat{\uGC} \frac{\hat{\bm v}_{\rm E}^*\cdot\nablab \hat{B}}{\hat{B}},
	\end{align}\vspace{-0.35cm}
\end{subequations}

\noindent with gyroradius parameter $\varrho_0 \equiv v_0/\omega_{\rm B0} = M v_0/(Z e B_0)$. The discretized equations of motion become
\begin{subequations}\vspace{-0.2cm}
	\begin{align}
		\Delta {\bm X}_{\rm gc} =&\; \Delta\hat{t} \; (\hat{\bm v}_\parallel^* + \hat{\bm v}_{\rm d}^* + \hat{\bm v}_{\rm E}^*),
		\label{eq:gc_norm_X} \\
		\Delta\hat{\uGC} =&\; \Delta\hat{t}\left(\frac{1}{\varrho_0} \frac{{\bm E}\cdot\hat{\bm B}^*}{\hat{B}\hat{B}^*_\parallel} - \hat{\mu}\frac{\hat{\bm B}^* \cdot \nablab \hat{B}}{\hat{B}_\parallel^*}\right) \nonumber \\
		\stackrel{E_r}{=}&\;\Delta\hat{t}\left(\frac{\hat{\uGC} \hat{\bm v}_{\rm E}^*}{\hat{B}} - \frac{\hat{\mu} \hat{\bm B}^*}{\hat{B}_\parallel^*}\right) \cdot \nablab \hat{B},
		\label{eq:gc_norm_u}
	\end{align}
	\label{eq:gc_norm}\vspace{-0.35cm}
\end{subequations}

\noindent where $\Delta\hat{t} \equiv \Delta t v_0$ is the time step in units of length.

It is interesting to note that using $\rhoPar \equiv \uGC/\omega_{\rm B}$ instead of $\uGC$ as an independent dynamic variable in the GC Lagrangian (cf.~\ref{apdx:gc_alt}) turns Eq.~(\ref{eq:mdl_du_dt}) into\vspace{-0.1cm}
\begin{equation}
	\dot{\varrho}_\parallel = -\left(\rhoPar^2\omega_{\rm B} + \frac{\mu B}{M\omega_{\rm B}}\right) \frac{{\bm B}^*\cdot\nablab B}{B B^*_\parallel}\vspace{-0.2cm}
\end{equation}

\noindent where the explicit ${\bm E}$-dependence in the form of a parallel acceleration term $Ze{\bm E}\cdot{\bm B}^*/M$ is replaced by a term proportional to $\rhoPar^2$, which depends on the electric field only implicitly via the relation $M\rhoPar^2\omega_{\rm B}^2 = 2(\ekin - \mu B) = 2(\etot - Ze\Phi - \mu B)$. Evidently, the parallel electric acceleration occurs only when the GC drifts across magnetic flux surfaces, and it vanishes (and switches sign) at points where $\dot{\bm X}_{\rm gc}\cdot\nablab\Psi_{\rm P} = 0$; namely, in the magnetic midplane that we introduce in Section~\ref{sec:method_mid} below. We will show in Section~\ref{sec:gc_eaccel} that this behavior causes systematic biases that must be taken into account when examining the $E_r$-dependence of GC transit frequencies.

\subsection{Magnetic midplane}
\label{sec:method_mid}

The magnetic midplane, which is drawn as a solid black line in Fig.~\ref{fig:03_kstar_E30_model}(a), can be defined by the condition\vspace{-0.1cm}
\begin{equation}
	[{\bm B}\cdot\nablab B](R_{\rm mid},z_{\rm mid}) = 0.
	\label{eq:method_mid}\vspace{-0.2cm}
\end{equation}

\noindent In up-down symmetric plasmas, the magnetic midplane coincides with the horizontal plane at the height of the magnetic axis at $z=z_0$ (dotted in Fig.~\ref{fig:03_kstar_E30_model}(a)). In general, Eq.~(\ref{eq:method_mid}) constitutes the set of all GC stagnation points (``orbit axes''), which can be understood as follows.

Equation~(\ref{eq:method_mid}) identifies the set of points where $B$ is minimized or maximized with respect to motion along $\hat{\bm b}$; that is, where $B$ is locally uniform at leading order. This local uniformity implies that parallel acceleration and radial drifts vanish ($\dot{\uGC} = 0$ and $\dot{\bm X}_{\rm gc}\cdot\nablab\Psi_{\rm P} = 0$), which explains why Eq.~(\ref{eq:method_mid}) is independent of the charge-mass ratio $Ze/M$, velocity $\dot{\bm X}_{\rm gc}$ and electric field ${\bm E}$.\footnote{In a Hamiltonian gyrokinetic model, the gyroaverage of the electric field around a gyrocenter generally introduces a dependence on charge, mass and velocity when ${\bm E}$ is nonuniform. The uniqueness of the magnetic midplane is, however, preserved by the requirement that gyroaveraging is performed only on the fluctuating components of ${\bm E}$ and ${\bm B}$, not the ambient fields. This is also important for GC models with {\it ad hoc} gyroaveraging \protect\cite{Bierwage22d}. In the case of full Lorentz orbits, the midplane's uniqueness seems evident to us only in the up-down symmetric case but remains obscure in general, although we have not seen evidence to the contrary so far. This adds to the issues in Section~\protect\ref{sec:intro_fo}.}

This leads us to conjecture that the existence of a magnetic midplane is closely connected to the condition for the existence of an idealized topological equilibrium (MHD or otherwise) in the sense that {\it all collisionless charged particle orbits are confined}.\footnote{Landreman \& Catto state in Section I and Appendix A of Ref.~\protect\cite{Landreman12} that such {\it confinement} is defined more precisely as the condition that ``radial'' drifts $\dot{\bm X}_{\rm gc}\cdot\nablab\Psi_{\rm P}$ must vanish on average.}
This is also known as {\it omnigenity}. Indeed, ${\bm B}\cdot\nablab B$ appears explicitly in the condition for omnigenity in Eq.~(11) of Ref.~\cite{Landreman12} by Landreman \& Catto, who generalized the quasi-symmetry condition by Helander \& Simakov \cite{Helander08,Simakov11}. As stated in Section~II~A of Ref.~\cite{Landreman12}, $({\bm B}\times\nablab\Psi_{\rm P})\cdot\nabla B$ vanishes whenever ${\bm B}\cdot\nablab B$ does, so the former constitutes an alternative expression for Eq.~(\ref{eq:method_mid}).

During one poloidal turn along a single field line, $B$ must have an equal number of minima and maxima. Usually there is one each in axi- or quasi-axisymmetric systems (with a degeneracy at the magnetic axis), so by virtue of Eq.~(\ref{eq:method_mid}) we may assume that there exists only one unique magnetic midplane within each magnetic separatrix of a tokamak plasma. Each GC orbit crosses the midplane exactly twice (with degeneracies for point-like stagnation orbits and at the trapped-passing boundary), so the entire GC orbit space can be sampled on a grid that is aligned with $(R_{\rm mid},z_{\rm mid})$. Moreover, we assume here that for each radial position $R$ there exists exactly one midplane point $z_{\rm mid}(R)$, a condition that seems to be satisfied in many cases of practical interest but may be relaxed if necessary. Additional midplanes may exist in magnetic islands and the scrape-off-layer. For instance, the JT-60U case that we will use as one of our working examples has an external midplane that supports mirror-trapped GC orbits near the inner divertor (see Fig.~\ref{fig:a02_jt60u_profs} of \ref{apdx:examples}). When pressure anisotropy is included, a different treatment may be required to capture more exotic orbits \cite{CooperGA07} (and the validity of MHD equilibria may then need to be reassessed).

In our numerical implementation, we approximate the midplane curve $z_{\rm mid}(R)$ by a polynomial. In the following,  we will use the short-hand notation\vspace{-0.1cm}
\begin{equation}
	g(X) \equiv g(R,z_{\rm mid}(R))
	\label{eq:method_fct_mid}\vspace{-0.1cm}
\end{equation}

\noindent to express the spatial dependence of any field $g$ with respect to the auxiliary radial coordinate $X \equiv R - R_0$ in the midplane $z_{\rm mid}(R)$.

\subsection{GC phase space mapping with relative \& absolute constants of motion}
\label{sec:method_com}

The motion of a GC in the ambient fields given by Eq.~(\ref{eq:method_BE}) conserves the total energy $\etot$ and the canonical toroidal angular momentum $P_\zeta$ in Eq.~(\ref{eq:intro_gc_com}),\vspace{-0.1cm}
\begin{equation}
	\etot = \ekin + Ze\Phi, \quad \quad P_\zeta = -Ze\Psi_{\rm P} + M u I/B,
	\label{eq:etot_pzeta}\vspace{-0.1cm}
\end{equation}

\noindent where one can interchangeably use the parallel velocity $\uGC(t)$ or the kinetic energy\vspace{-0.1cm}
\begin{equation}
	\ekin(t) = \mu B_{\rm gc}(t) + Mu^2(t)/2,\vspace{-0.05cm}
\end{equation}

\noindent with $B_{\rm gc}(t) \equiv B(R_{\rm gc}(t),z_{\rm gc}(t))$ and $\mu = {\rm const}$. Evidently, in the absence of an electric field ($\Phi={\rm const}.$), $\ekin$ becomes an absolute constant of motion and $\uGC$ an algebraic function of $B$. In the following, $\overline{\bm C}$ and ${\bm C}$ will represent sets of absolute and relative CoM, respectively.

The relation between Cartesian coordinates $({\bm v},{\bm x})$ and our set of absolute CoM $\overline{\bm C} = \{\sigma_\uGC\etot,\mu,P_\zeta\}$ is given, for instance, by Eq.~(75) of Ref.~\cite{Porcelli94}. In our notation,\vspace{-0.05cm}
\begin{equation}
	{\rm d}^3{\bm v}{\rm d}^3{\bm x} = \sum\limits_{\sigma_\uGC} \frac{{\rm d}\etot}{M} \frac{{\rm d}\mu B_0}{M} \frac{{\rm d}P_\zeta}{Ze B_0} {\rm d}\tau {\rm d}\xi {\rm d}\zeta.
	\label{eq:model_dv3dx3_can}\vspace{-0.15cm}
\end{equation}

\noindent Since all variables are independent of the gyrophase $\xi$ and toroidal angle $\zeta$, these two angle coordinates may be readily integrated over to give ${\rm d}\xi{\rm d}\zeta \rightarrow (2\pi)^2$. As in Ref.~\cite{Bierwage22a}, we initialize GC orbits in the magnetic midplane given by Eq.~(\ref{eq:method_mid}), and the remainder of the poloidal plane is sampled by the angle $\tau \in [0,\tau_{\rm pol})$, which has a time-like metric in the GC model and effectively labels a GC's position along the poloidal $(R,z)$ contour of its drift orbit that is identified by the values of $\overline{\bm C}$ and has a poloidal transit period $\tau_{\rm pol}$. Integration over $\tau$ introduces a factor $\tau_{\rm pol}(\overline{\bm C})$ on the right-hand side of Eq.~(\ref{eq:model_dv3dx3_can}), and it is this $\overline{\bm C}$-dependent factor that makes the coordinates $\{\sigma_\uGC\etot,\mu,P_\zeta,\tau,\xi,\zeta\}$ non-canonical.

In practice, absolute CoM are often problematic. When representing the GC phase space in terms of marker particles, it must be sliced up into cells (volume elements) along the chosen coordinate lines, but in the presence of an ambient electric field ${\bm E} = -\nablab\Phi$, the slicing along lines of constant $\etot = \ekin - Ze\Phi$ becomes awkward (see Section~\ref{sec:method_slice} below). Thus, we will usually work with midplane-based relative CoM, and use absolute CoM only for verification (\ref{apdx:benchmark}). Here, we perform the transformation\vspace{-0.05cm}
\begin{equation}
	\overline{g}(\etot,P_\zeta) \rightarrow g(\ekin_{\rm I},X_{\rm I}),\vspace{-0.05cm}
\end{equation}

\noindent where $g$ is an arbitrary field, and $(\ekin_{\rm I},X_{\rm I})$ are the initial coordinates of GC orbits in the midplane as in Eq.~(\ref{eq:method_fct_mid}). Omitting the subscript ``I'' for brevity, Eq.~(\ref{eq:model_dv3dx3_can}) becomes\vspace{-0.05cm}
\begin{equation}
	{\rm d}^3{\bm v}{\rm d}^3{\bm x} = (2\pi)^2 \sum\limits_{\sigma_\uGC} \frac{{\rm d}\ekin}{M} \frac{{\rm d}\mu B_0}{M} {\rm d}X {\rm d}\tau \times\J,
	\label{eq:model_dV_mid}\vspace{-0.05cm}
\end{equation}

\noindent where $\J$ denotes the Jacobian for the transformation $(\ekin,X)\rightarrow(\etot,P_\zeta)$. Littlejohn's factor $B_\parallel^*/B$ in Eqs.~(6) and (A.7) of Ref.~\cite{Bierwage22a} does not appear in Eq.~(\ref{eq:model_dV_mid}) and subsequent equations here because the latter are all evaluated in the magnetic midplane, where $B_\parallel^* \rightarrow B$. Expressing $\overline{\bm C} = \{ \sigma_\uGC\etot,\mu,P_\zeta\}$ in terms of ${\bm C} = \{\sigma_\uGC\ekin,\mu,X\}$, and applying the normalizations of Eq.~(\ref{eq:norm}), we have
\begin{subequations}
	\begin{align}
		\hat{\etot}(\ekin,\mu,X) =&\; \hat{\ekin} + \frac{\hat{\Phi}(X)}{\varrho_0},
		\label{eq:model_com_etot} \\
		\hat{P}_\zeta(\ekin,\mu,X) =&\; -\hat{\Psi}_{\rm P}(X) + \varrho_0 \frac{\hat{I}(X)}{\hat{B}(X)}\hat{u}(\ekin,\mu,X).
		\label{eq:model_com_pzeta}
	\end{align}
	\label{eq:model_com}\vspace{-0.2cm}
\end{subequations}

\noindent The parallel velocity has the functional form
\begin{equation}
	\hat{u}^2(\ekin,\mu,X) \equiv 2\hat{\ekin} - 2\hat{\mu}\hat{B}(X).
	\label{eq:model_u2}
\end{equation}

\noindent The partial derivatives
\begin{gather}
	\left.\frac{\partial\hat{\etot}}{\partial\hat{\ekin}}\right|_{\mu,X} = 1,\quad
	\left.\frac{\partial\hat{\etot}}{\partial\hat{\mu}}\right|_{\ekin,X} = 0,\quad
	\left.\frac{\partial\hat{\etot}}{\partial X}\right|_{\ekin,\mu} = \frac{1}{\varrho_0}\frac{{\rm d}\hat{\Phi}}{{\rm d}X}, \nonumber
	\\
	\left.\frac{\partial\hat{\mu}}{\partial\hat{\ekin}}\right|_{\mu,X} = 0,\quad
	\left.\frac{\partial\hat{\mu}}{\partial\hat{\mu}}\right|_{\ekin,X} = 1,\quad
	\left.\frac{\partial\hat{\mu}}{\partial X}\right|_{\ekin,\mu} = 0,
	\label{eq:model_jac_elements}
	\\
	\left.\frac{\partial\hat{P}_\zeta}{\partial\hat{\ekin}}\right|_{\mu,X} = \frac{\varrho_0\hat{I}}{\hat{u}\hat{B}},\quad
	\left.\frac{\partial\hat{P}_\zeta}{\partial\hat{\mu}}\right|_{\ekin,X} = -\frac{\varrho_0\hat{I}}{\hat{u}},\quad
	\left.\frac{\partial\hat{P}_\zeta}{\partial X}\right|_{\ekin,\mu} \equiv \F, \nonumber
\end{gather}

\noindent with
\begin{equation}
	\F \equiv \left.\frac{\partial\hat{P}_\zeta}{\partial X}\right|_{\ekin,\mu} = -\frac{{\rm d}\hat{\Psi}_{\rm P}}{{\rm d}X} + \varrho_0\left[\frac{\hat{u}}{\hat{B}}\frac{{\rm d}\hat{I}}{{\rm d}X} - \left(\hat{\ekin} + \frac{\hat{u}^2}{2}\right) \frac{\hat{I}}{\hat{u}\hat{B}^2} \frac{{\rm d}\hat{B}}{{\rm d}X}\right],
	\label{eq:method_jac_f}
\end{equation}

\noindent yield the Jacobian determinant $\J \equiv |\partial\overline{\bm C}/\partial{\bm C}| = |\G|$ with
\begin{align}
	\G \equiv&\; \left.\frac{\partial\hat{P}_\zeta}{\partial X}\right|_{\etot,\mu} = \F - \frac{\hat{I}}{\hat{u}\hat{B}} \frac{{\rm d}\hat{\Phi}}{{\rm d}X} = \F + \frac{\hat{I}\hat{E}_{\hat{\Psi}_{\rm P}}}{\hat{u}\hat{B}} \frac{{\rm d}\hat{\Psi}_{\rm P}}{{\rm d}X}
	\label{eq:method_jac_g} \\
	=&\; \left(\frac{\hat{I}\hat{E}_{\hat{\Psi}_{\rm P}}}{\hat{u}\hat{B}} - 1\right)\frac{{\rm d}\hat{\Psi}_{\rm P}}{{\rm d}X} + \frac{\varrho_0}{\hat{u}} \left[\frac{\hat{u}^2}{\hat{B}}\frac{{\rm d}\hat{I}}{{\rm d}X} - \left(\hat{\ekin} + \frac{\hat{u}^2}{2}\right) \frac{\hat{I}}{\hat{B}^2} \frac{{\rm d}\hat{B}}{{\rm d}X}\right] \nonumber \\
	=&\; \left(\frac{\hat{I}\hat{E}_{\hat{\Psi}_{\rm P}}}{\hat{u}\hat{B}} - 1 + \frac{\varrho_0\hat{u}}{\hat{B}} \frac{{\rm d}\hat{I}}{{\rm d}\hat{\Psi}_{\rm P}}\right) \frac{{\rm d}\hat{\Psi}_{\rm P}}{{\rm d}X} - \frac{\varrho_0}{\hat{u}} \left(\hat{\ekin} + \frac{\hat{u}^2}{2}\right) \frac{\hat{I}}{\hat{B}^2} \frac{{\rm d}\hat{B}}{{\rm d}X}. \nonumber
\end{align}

The GC orbit space spanned by the absolute CoM $\overline{\bm C} = \{\sigma_\uGC\etot,\mu,P_\zeta\}$ or the relative CoM ${\bm C} = \{\sigma_\uGC\ekin,\mu,X\}$ is discretized on respective grids with indices $(i,j,k)$. Moreover, as in Section~6.1 of Ref.~\cite{Bierwage22a}, each GC orbit $(i,j,k)$ is then represented by $N_\tau$ marker particles that are labeled by the index $l = 1,...,N_\tau$. Being distributed uniformly in time, these markers represent identical volume elements proportional to $\Delta\tau_l = \tau_{\rm pol}/N_\tau$. Equation~(33) of Ref.~\cite{Bierwage22a} can then be generalized as
\begin{subequations}
	\begin{align}
		{\rm d}^3\hat{\bm v}{\rm d}^3{\bm x} \approx&\; \frac{1}{2}(2\pi)^2 \Delta\hat{\etot}_i \Delta\hat{\mu}_j\left|\Delta\hat{P}_{\zeta,k}\right|_{\etot,\mu} v_0 \Delta\tau_{ijkl}
		\label{eq:model_dVol_transf_can} \\
		=&\; \frac{1}{2}(2\pi)^2 \Delta\hat{\ekin}_i \Delta\hat{\mu}_j\left|\Delta\hat{P}_{\zeta,k}\right|_{\ekin,\mu} v_0 \Delta\tau_{ijkl} \left|\frac{\G}{\F}\right|
		\label{eq:model_dVol_transf_P} \\
		=&\; \frac{1}{2}(2\pi)^2 \Delta\hat{\ekin}_i \Delta\hat{\mu}_j\left|\Delta X_k\right|_{\ekin,\mu} v_0 \Delta\tau_{ijkl} |\G|
		\label{eq:model_dVol_transf_X} \\
		=&\; \frac{1}{2}(2\pi)^2 \Delta\hat{\etot}_i \Delta\hat{\mu}_j\left|\Delta X_k\right|_{\etot,\mu} v_0 \Delta\tau_{ijkl} |\G|,
		\label{eq:model_dVol_transf_canX}
	\end{align}	
	\label{eq:model_dVol_transf}\vspace{-0.4cm}
\end{subequations}

\noindent where the subscripts of $|...|_{\etot,\mu}$ or $|...|_{\ekin,\mu}$ indicate the coordinates that are kept constant when slicing the GC phase space into cells. Since we will later replace $\mu$ by a signed pitch angle coordinate $\alpha$, the index $j$ may here be taken to capture both signs of $\sigma_\uGC$. The factor $1/2$ corrects for the double-counting of orbits on the midplane's low- and high-field side as the index $k$ runs along the midplane from the inner to the outer boundary.

In the code, we replace the magnetic moment $\mu$ with the auxiliary pitch coordinate $\Lambda \equiv \mu B_0/\ekin = \hat{\mu}/\hat{\ekin}$, so that the functional forms of $\hat{\mu}$ and $\hat{u}^2$ in Eq.~(\ref{eq:model_u2}) become\vspace{-0.5cm}
\begin{subequations}
	\begin{align}
		\hat{\mu}(\ekin,\Lambda,P_\zeta) =&\; \Lambda \hat{\ekin}, \\
		\hat{u}^2(\ekin,\Lambda,X) =&\; 2\hat{\ekin}(1 - \Lambda\hat{B}(X)).
	\end{align}
\end{subequations}

\noindent This is done for convenience in order to make the integration limits independent of the kinetic energy: $\Lambda \in [0,1/\hat{B}]$ instead of $\hat{\mu} \in [0,\hat{\ekin}/\hat{B}]$. The partial derivatives in the lower-left part of Eq.~(\ref{eq:model_jac_elements}) are then replaced by\vspace{-0.1cm}
\begin{gather}
	\left.\frac{\partial\hat{\mu}}{\partial\hat{\ekin}}\right|_{\Lambda,X} = \Lambda,\quad
	\left.\frac{\partial\hat{\mu}}{\partial\Lambda}\right|_{\ekin,X} = \hat{\ekin},
	\\
	\left.\frac{\partial\hat{P}_\zeta}{\partial\hat{\ekin}}\right|_{\Lambda,X} = \frac{\varrho_0\hat{I}}{\hat{u}\hat{B}}(1 - \Lambda \hat{B}),\quad
	\left.\frac{\partial\hat{P}_\zeta}{\partial\Lambda}\right|_{\ekin,X} = -\frac{\varrho_0\hat{I}}{\hat{u}}\hat{\ekin}. \nonumber
\end{gather}

\noindent After some cancellations (or using chain rules), the Jacobian determinant acquires a factor $\hat{\ekin}$, so one substitutes $\Delta\hat{\mu}_j \rightarrow \hat{\ekin}_i\Delta\Lambda_j$ in Eq.~(\ref{eq:model_dVol_transf}) for relative CoM. Similarly, in the case of absolute CoM we can define the conserved pitch coordinate $\overline{\Lambda} \equiv \hat{\mu}/\hat{\etot}$ and substitute $\Delta\hat{\mu}_j \rightarrow \hat{\etot}_i\Delta\overline{\Lambda}_j$, although here the energy-dependence is inevitable in the presence of an electric field: $\overline{\Lambda} \in [0,\hat{\ekin}/(\hat{B}\hat{\etot})]$.

As in Ref.~\cite{Bierwage22a}, we also utilize the pitch angle $\alpha$ and define a conserved counterpart $\overline{\alpha}$:\vspace{-0.1cm}
\begin{subequations}
	\begin{align}
		\sin\alpha \equiv&\; \sigma_{\uGC}\sqrt{1 - \Lambda\hat{B}} = \uGC/v_{\rm gc}, \\
		\sin\overline{\alpha} \equiv&\; \sigma_{\uGC}\sqrt{1 - \overline{\Lambda}\hat{B}},
	\end{align}
	\label{eq:method_alpha}\vspace{-0.4cm}
\end{subequations}

\noindent with $v_{\rm gc} \equiv \sqrt{\ekin/(2M)}$ and $\sigma_{\uGC} \equiv \uGC/|\uGC|$ (see Section 3 of Ref.~\cite{Bierwage22a} for details concerning the sign of $\uGC$). Using the pitch angle $\alpha$ (or $\overline{\alpha}$) to control the nonuniformity of the $\Lambda$ (or $\overline{\Lambda}$) mesh allows us to take advantage of the fact that at least $\alpha \in [-\pi/2,-\pi/2]$ has unique limits that do not depend on other variables.

\begin{figure}
	[tbp]
	\centering
	\includegraphics[width=0.48\textwidth]{\figures/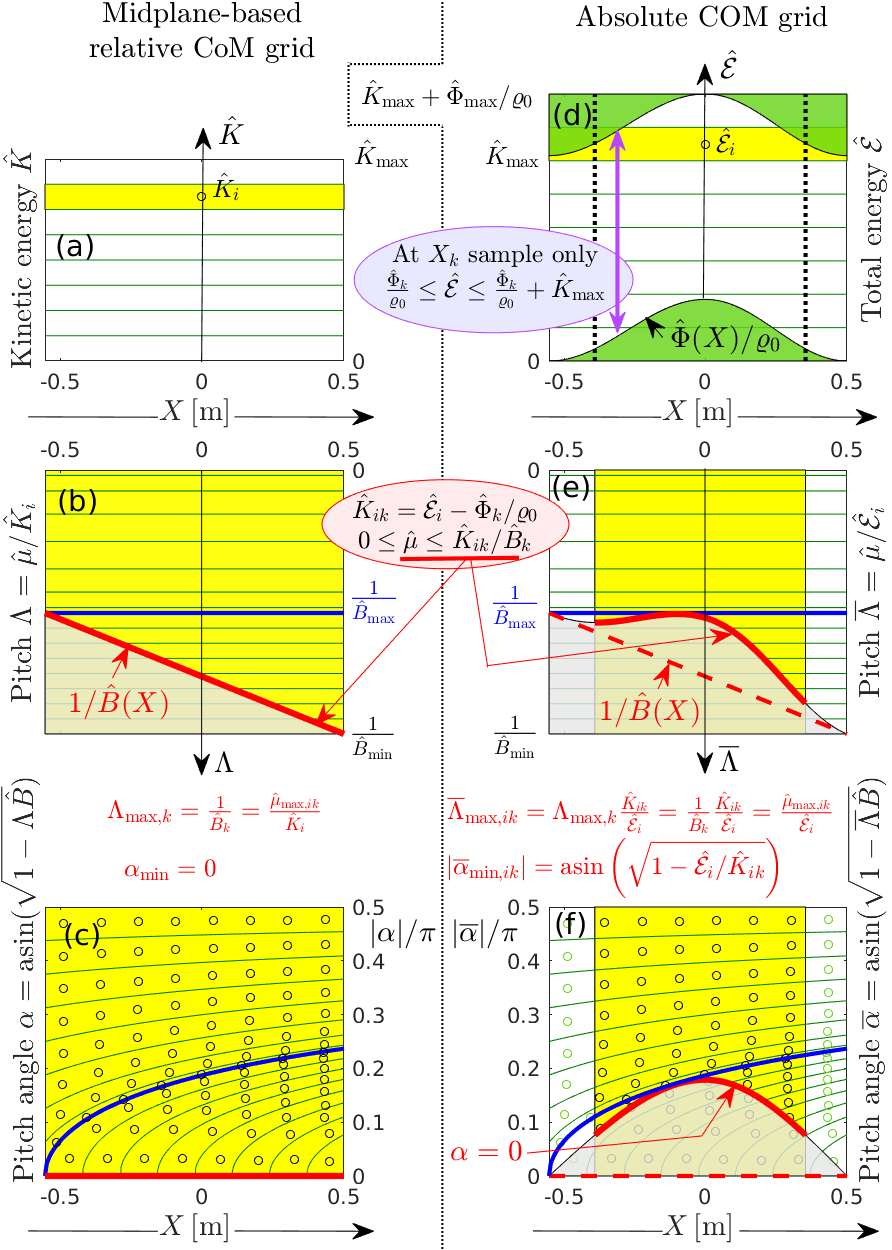}
	\caption{Comparison of midplane-base relative (left) and absolute (right) CoM space grid options in the extended {\tt VisualStart} code that includes an ambient $E_r$ field. The spatial dimensions are those of a KSTAR plasma with major radius $R_0 \approx 1.8\,{\rm m}$, whose midplane extends over the range $-0.55\,{\rm m} \leq X \leq 0.50\,{\rm m}$ from the inner to the outer wall (see Fig.~\protect\ref{fig:03_kstar_E30_model}). Given the magnetic geometry, the required input parameters are the electrostatic potential $\Phi(r)$, the kinetic energy range $[\ekin_{\rm min},\ekin_{\rm max}]$, and the numbers of grid points in energy, midplane pitch angle (magnetic moment), and midplane radius (canonical toroidal angular momentum). The model profile $\Phi(r)$ of Fig.~\protect\ref{fig:03_kstar_E30_model} becomes $\Phi(X)$ as shown in panel (d). In the present illustrative example, we let $\hat{\ekin}_{\rm min} = 0$ and $\hat{\Phi}_{\rm max} = 0.3\times\varrho_0\hat{\ekin}_{\rm max}$ with $\varrho_0 = 6.5\,{\rm mm}$, and cut the respective energy axes in (a) and (d) uniformly into $N_{\ekin} = N_{\etot} = 8$ cells. Panels (b) and (e) effectively show the grids of the magnetic moment $\mu$, using the pitch coordinates $\Lambda \equiv \hat{\mu}/\hat{\ekin}_i \in [0,1/B]$ and $\overline{\Lambda} \equiv \hat{\mu}/\hat{\etot}_i \in [0,\hat{\ekin}_{ik}/(\hat{\etot}_iB)]$ with $\hat{\ekin}_{ik} = \hat{\etot}_i-\hat{\Phi}_k/\varrho_0$. The equivalent grids in pitch angles $\alpha$ and $\overline{\alpha}$ defined in Eq.~(\protect\ref{eq:method_alpha}) are shown in panels (c) and (f) (only positive values are shown here). As in Fig.~6 of Ref.~\protect\cite{Bierwage22a}, we choose to sample the subdomain $[0,1/B_{\rm max}]$ of $\Lambda$ and $\overline{\Lambda}$ (above the blue line) uniformly in pitch angle $\alpha(X_{\rm min})$ at the midplane's high-field side boundary $X_{\rm min}$, and the remaining domain $[1/B_{\rm max},1/B_{\rm min}]$ (below the blue line) is sampled uniformly in $X$, here using $N_\alpha = N_X = 8$ cells. Finally, the $(\alpha,X)$ and $(\overline{\alpha},X)$ domains are sampled as described in Section~4 of Ref.~\protect\cite{Bierwage22a}, and the resulting cell centers are indicated by open circles in panels (c) and (f). Note that $\{\ekin, \alpha, X\}$ in the present context represent the initial (I) positions $\{\ekin_{\rm I}, \alpha_{\rm I}, X_{\rm I}\}$ of a GC orbit in the midplane, so they constitute relative CoM. The yellow-shaded area and bold red line indicate the domains of valid samples satisfying the constraints $\hat{\ekin}_{ik} = \hat{\etot}_i - \hat{\Phi}_k \in [\hat{\ekin}_{\rm min},\hat{\ekin}_{\rm max}]$ and $\Lambda\hat{B} = \overline{\Lambda}\hat{B}\hat{\etot}/\hat{\ekin} \in [0,1]$, respectively. For the relative CoM space, Fig.~\protect\ref{fig:05_kstar_orbtypes} shows the distribution of GC orbit types for thermal deuterons.}
	\label{fig:04_kstar_grid}%
\end{figure}

\subsection{Procedure for slicing the CoM space}
\label{sec:method_slice}

Our CoM space slicing procedure, which is a generalization of Fig.~6 of Ref.~\cite{Bierwage22a}, is illustrated in Fig.~\ref{fig:04_kstar_grid}, showing a side-by-side comparison between the methods of slicing along lines of constant $(\ekin,\Lambda)$ (relative CoM, Fig.~\ref{fig:04_kstar_grid} left column) and slicing along lines of constant $(\etot,\overline{\Lambda})$ (absolute CoM, Fig.~\ref{fig:04_kstar_grid} right column). CoM spaces are first sliced into cells $\Delta\ekin_i$ or $\Delta\etot_i$ (Fig.~\ref{fig:04_kstar_grid} top row), which are then further divided into cells $\Delta\Lambda_j$ or $\Delta\overline{\Lambda}_j$ (Fig.~\ref{fig:04_kstar_grid} middle row). These, in turn, are divided into cells $\Delta X$ or $\Delta P_\zeta$ (Fig.~\ref{fig:04_kstar_grid} bottom row), which are sliced along lines of constant $\ekin$ and $\Lambda$  or constant $\etot$ and $\overline{\Lambda}$.

The crucial advantage of slicing the orbit space in relative CoM coordinates is that, for any kinetic energy $\hat{\ekin}_i \geq 0$, the entire $(\alpha,X)$ plane can be filled with valid\footnote{A valid sample satisfies $\ekin\geq 0$ and $\mu B = \ekin - M\uGC^2/2 \geq 0$. One may also impose additional constraints such as $0 \leq \ekin_{\rm min} \leq \ekin \leq \ekin_{\rm max}$.}
samples as indicated by the yellow-shaded area. The simplicity of the domain boundaries translates into simple boundary cell shapes, allowing to achieve high numerical accuracy with relatively simple algorithms.

In contrast, in the case of absolute CoM, slicing along lines of constant $\etot$ imposes variable validity limits on the samples in the $(\overline{\alpha},X)$ plane. First, the range of $X$ is limited by the constraint $\hat{\ekin}_{ik} = \hat{\etot}_i - \hat{\Phi}_k \in [\hat{\ekin}_{\rm min},\hat{\ekin}_{\rm max}]$ as indicated by the green-shaded areas in Fig.~\ref{fig:04_kstar_grid}(d). Second, $\overline{\alpha}$ is limited by the constraint $1 - \overline{\Lambda}\hat{B} \in [0,1]$ as indicated by the gray-shaded area in panels (e) and (f). The nonuniformity of $\Phi \in [0,\Phi_{\rm max}]$ implies that the lower limit of the constraint $1 - \overline{\Lambda}\hat{B} \geq 0$, which corresponds to $\alpha \approx 0$ (or $\uGC \approx 0$), generally has a curved form that cuts our simple $(\overline{\alpha},X)$ cells into complicated shapes as can be inferred from the intersection of the bold red line with the horizontal green lines in Fig.~\ref{fig:04_kstar_grid}(e). In order to obtain accurate results with reasonable computational effort, one would have to develop a more sophisticated meshing algorithm that accumulates samples on both sides of the bold red curve and accurately computes the effective sizes of the adjacent boundary cells in dependence on the $\Phi(X)$ profile.

The two options shown in Fig.~\ref{fig:04_kstar_grid} have both been implemented in {\tt VisualStart} for testing purposes. The results of these tests can be found in \ref{apdx:benchmark}, where we demonstrate in detail the above-mentioned difficulties associated with absolute CoM space slicing and the merits of midplane-based relative CoM, which will be used in the remainder of this paper.

\begin{figure}
	[tb]\vspace{-0.3cm}
	\centering
	\includegraphics[width=0.48\textwidth]{\figures/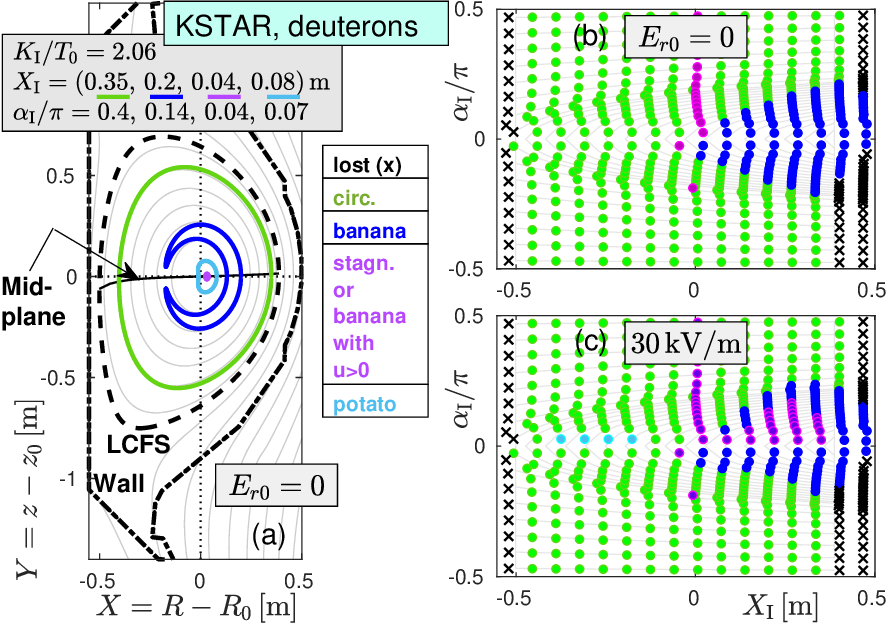}\vspace{-0.2cm}
	\caption{GC orbit space grid and distribution of orbit classes in the case based on KSTAR shot 18567 \protect\cite{Lee23}, whose parameters are summarized in Table~\protect\ref{tab:tok} and Fig.~\protect\ref{fig:a01_kstar_profs}. As in Fig.~\ref{fig:03_kstar_E30_model}, panel (a) shows the poloidal cross-section, here with one example for each of the four orbit classes: circulating (green), banana (blue), stagnation (magenta), and potato (cyan). For $E_{r0} = 0$ and $30\,{\rm kV/m}$, which is considered to be realistic for this plasma, panels (b) and (c) show the distribution of orbit classes in the space of initial radii $X_{\rm I}$ and pitch angles $\alpha_{\rm I}$ on the magnetic midplane (the solid black line in panel (a)). The chosen initial kinetic energy, $\ekin_{\rm I} = 2.06\times T_0 \approx 6.7\,{\rm keV}$, is roughly twice the thermal energy on axis. Stagnation and potato orbits are rare at this low energy and only a few of them are captured here since we use a sparse grid, which consists of only $N_\alpha \times N_X = 16\times 16$ cells to facilitate the visibility of individual samples. Here, our $X_{\rm I}$-mesh extends into the scrape-off layer (outside the LCFS), where only some banana orbits remain confined, while others are lost (black crosses).}\vspace{-0.2cm}
	\label{fig:05_kstar_orbtypes}%
\end{figure}

For the midplane-based relative CoM grid in Fig.~\ref{fig:04_kstar_grid}(c), which is based on a KSTAR plasma that was studied in Ref.~\cite{Lee23}, Fig.~\ref{fig:05_kstar_orbtypes} shows the distribution of GC orbit types for thermal deuterons with initial (I) kinetic energy $\ekin_{\rm I} = 2.06 \times T_0 \approx 6.7\,{\rm keV}$ at the midplane. Fig.~\ref{fig:05_kstar_orbtypes}(b) shows the situation for $E_{r0} = 0$, and panel (c) for a radial electric field with peak value $E_{r0} = 30\,{\rm kV/m}$. One can see that the boundaries of the domain of mirror-trapped ``banana'' (blue) and ``potato'' orbits (cyan)\footnote{In Fig.~\protect\ref{fig:05_kstar_orbtypes}, potato orbits exist only in very narrow layers around the rim of the trapped-passing boundary on the low-field side ($X>0$) and near $\alpha \approx 0$ on the high-field side ($X < 0$). Our coarse mesh has not captured any potato orbits in Fig.~\protect\ref{fig:05_kstar_orbtypes}(b) and only a few in Fig.~\protect\ref{fig:05_kstar_orbtypes}(c).}
is visibly shifted towards positive initial pitch angles $\alpha_{\rm I}$. For given $E_r$, this shift is larger at lower energies and smaller at higher energies.

Our orbit classification algorithm is based on the procedure described in Section~4 of Ref.~\cite{Bierwage22a}. For $E_{\rm r0} = 0$ in panels (a) and (b) of Fig.~\ref{fig:05_kstar_orbtypes}, circulating and stagnation orbits are both passing orbits, and banana and potato orbits are both mirror-trapped orbits, the difference being whether or not they enclose the magnetic axis. In the presence of an electric field, our group of ``stagnation orbits'', which are colored magenta in Fig.~\ref{fig:05_kstar_orbtypes}, includes
\begin{itemize}
	\item both passing and mirror-trapped orbits that do not encircle the magnetic axis, and whose parallel velocity $\uGC$ does not have an alternating sign.
\end{itemize}

\noindent In Fig.~\ref{fig:05_kstar_orbtypes}(c) with $E_{r0} = 30\,{\rm kV/m}$, this criterion is satisfied by some of the deeply trapped banana orbits, whose electric precession is fast enough to fully offset the sign reversal of $\uGC$. A proper distinction between such electrically ``Doppler-shifted'' banana orbits and true stagnation orbits would require an accurate estimation of the electric modification of the precession frequency and mirror force. This is easier said than done (and not done here), in part because it requires a transformation to an inertial ``$E_r = 0$'' frame of reference, which does not strictly exist. Further details will be discussed in Section~\ref{sec:gc} when we analyze individual orbits (cf.~Fig.~\ref{fig:13_jt60_deep-trapped}).

\subsection{GC distribution and weighting}
\label{sec:method_weight}

Each GC orbit in Fig.~\ref{fig:05_kstar_orbtypes} can be given an arbitrary weight $W_{ijk}$, which has the form
\begin{equation}
	W_{ijk} = f_{\rm com}(\etot_i, \mu_j, P_{\zeta,k}) \Delta\V_{ijk}.
\end{equation}

\noindent $j$ covers both signs of $\sigma_\uGC$ as mentioned below Eq.~(\ref{eq:model_dVol_transf}). The normalized volume element $\Delta\hat{\V}_{ijk} = \Delta\V_{ijk}/v_0^3$,\vspace{-0.2cm}
\begin{equation}
	\Delta\hat{\V}_{ijk} = \sum\limits_{l=1}^{N_{\tau,ijk}}[{\rm d}^3\hat{\bm v}{\rm d}^3{\bm x}]_{ijkl} = \frac{1}{2}(2\pi)^2 \Delta\hat{\etot}_i \Delta\hat{\mu}_j\Delta \hat{\P}_{\zeta,k} v_0 \tau_{{\rm pol},ijk},
	\label{eq:model_dVol_orb}
\end{equation}

\noindent is obtained by summing Eq.~(\ref{eq:model_dVol_transf}) over the orbit's $N_{\tau,ijk}$ markers, whose identical weights $w_{ijkl}$ are given by
\begin{equation}
	w_{ijkl} = \frac{W_{ijk}}{N_{\tau,ijk}} = f_{\rm com}(\etot_i, \mu_j, P_{\zeta,k}) \frac{\Delta\V_{ijk}}{N_{\tau,ijk}}, \quad l = 1,...,N_\tau.
	\label{eq:method_w_mk}
\end{equation}

\noindent The value $f_{\rm com}(\etot_i, \mu_j, P_{\zeta,k})$ of the GC CoM distribution function determines the number of physical particles that are represented by the GC CoM space volume element $\Delta\V_{ijk}$.

In order to illustrate how to construct models for $f_{\rm com}$ let us consider the isotropic local Maxwellian
\begin{equation}
	f_{\rm loc}(r,\ekin) = \left(\frac{M}{2\pi}\right)^{3/2}\frac{N_{\rm ref}(r)}{T_{\rm ref}^{3/2}(r)} \exp\left(\frac{-\ekin}{T_{\rm ref}(r)}\right),
	\label{eq:fM_loc}
\end{equation}

\noindent where $N_{\rm ref}(r)$ and $T_{\rm ref}(r)$ are the reference profiles for the number density and temperature. Eq.~(\ref{eq:fM_loc}) differs from Eq.~(\ref{eq:intro_fMloc}) only in that we replaced the Lorentz orbit kinetic energy $\tilde{\ekin}$ by the GC kinetic energy $\ekin$, and $r$ is now a GC position, not that of a particle. In the absence of an ambient electric field, $\ekin$ is constant and only $r(t)$ varies along a GC orbit. One may then assign GC orbit weights by replacing $r(t)$ with a suitable CoM, such as
\begin{itemize}
	\item  the radius $r_{\rm lfs}$ or $r_{\rm hfs}$ of the orbit's low-field-side (LFS) or high-field-side (HFS) midplane crossings,
	\item  the turning-point radius $r_{\uGC=0}$ of a banana orbit, or
	\item  the mean (orbit-time-averaged) minor radius $\left<r\right>$,
\end{itemize}
\begin{equation}
	\left<r\right> = \frac{1}{\tau_{\rm pol}}\oint{\rm d}t\; r(t).
\end{equation}

\noindent Any of these (or any other equivalent) choices are valid provided that they are used consistently and in a way that satisfies our double-counting assumption, which is expressed by the factor $1/2$ in Eqs.~(\ref{eq:model_dVol_transf}) and (\ref{eq:model_dVol_orb}). For instance, if one chooses to use $r_{\rm lfs}$, this coordinate must then always be used to evaluate $f_{\rm com}$, regardless of whether the initial position $r_{\rm I}$ of an orbit sample is located on the LFS or HFS.

In the presence of an ambient radial electric field, the kinetic energy $\ekin(t) = \etot - Ze\Phi(r(t))$ also becomes a function of time. In order to obtain a model for $f_{\rm com}$ that has a well-defined value for each orbit contour $(i,j,k)$, we substitute $\ekin(t)$ with some suitable CoM in the same way as $r(t)$ above, such as $\ekin_{\rm lfs}$, $\ekin_{\rm hfs}$, or $\ekin_{\uGC=0}$, or $\left<\ekin\right>$.\footnote{One might be tempted to replace $\ekin$ with the total energy $\etot$ and write $\exp(-|\etot|/T)$. However, this would introduce an explicit dependence of the orbit weight $W_{ijk}$ on the electrostatic potential $\Phi$, which does not seem physically meaningful: a Maxwellian is the result of a collisional equilibration process that depends only on the relative velocities of the colliding particles, not their potential energies. Similarly, $N_{\rm ref}$ and $T_{\rm ref}$ could be expressed as explicit functions of $|P_\zeta|$, but it is easy to find more meaningful spatial coordinates that are CoM.}
In this paper, we chose to weight our GC orbits with an isotropic CoM Maxwellian model $f_{\rm mdl}$ that is expressed in terms of orbit-averaged coordinates, $\left<r\right>$ and $\left<\ekin\right>$, as\vspace{-0.3cm}
\begin{equation}
	\hspace{-0.2cm} f_{\rm mdl}(\left<r\right>,\left<\ekin\right>) = \left(\frac{M}{2\pi}\right)^{3/2}\frac{N_{\rm ref}(\left<r\right>)}{T_{\rm ref}^{3/2}(\left<r\right>)} \exp\left(\frac{-\left<\ekin\right>}{T_{\rm ref}(\left<r\right>)}\right).
	\label{eq:fM}\vspace{-0.25cm}
\end{equation}

The choice of orbit-averaged coordinates $(\left<r\right>,\left<\ekin\right>)$ appears not only physically meaningful; it also seems to be the most straightforward way to minimize biases that alternative choices such as $(r_{\rm lfs},\ekin_{\rm lfs})$ or $(r_{\rm hfs},\ekin_{\rm hfs})$ would introduce via the parallel electric acceleration term $\propto E_\nabla = \varrho_\parallel {\bm v}_{\bm E}\cdot\nabla B$ that modulates the mirror force in Eqs.~(\ref{eq:mdl_du_dt}) and (\ref{eq:gc_norm_u}) (see Section~\ref{sec:gc_eaccel} for details).

\vspace{-0.2cm}
\subsection{Summary of the model, its scope, and test cases}
\label{sec:method_summary}

We model charged particle distributions by factorizing them as in Eq.~(\ref{eq:intro_fcom}) and using as basis functions $\delta({\bm C})$ the orbit surfaces defined by the Hamiltonian motion of guiding centers (GC) in stationary ambient fields ${\bm B}$ and ${\bm E} = -\nablab\Phi$. The form of the latter was shown in Fig.~\ref{fig:03_kstar_E30_model}. The GC orbit space is sampled on the magnetic midplane as shown in Figs.~\ref{fig:04_kstar_grid} and \ref{fig:05_kstar_orbtypes}, and the trajectory of each orbit is sampled uniformly in time to give an exact equilibrium distribution in the sense that it is only a function of three constants of motion (CoM).

Due to the reasons outlined in Section~\ref{sec:method_slice}, we use a set of midplane-based relative CoM, $\{\ekin_{\rm I},\alpha_{\rm I},X_{\rm I}\}$ for sampling the GC orbit space. It is important to remember that a nonzero $E_r$ field causes double-counted orbits --- sampled once on the high-field side (HFS) and once on the low-field side (LFS) midplane --- to appear in {\it different} cells of the kinetic energy grid $\ekin_i$. It is thus important to choose the range $\ekin \in [\ekin_{\rm min}, K_{\rm max}]$ wide enough to capture all relevant orbits; that is, all orbits with significant weight $W_{ijk} = f_{\rm com}\Delta\V_{ijk}$.

The use of monoenergetic test distributions as in our previous paper \cite{Bierwage22a} is no longer applicable here because kinetic energy is not conserved in the presence of $E_r$. Instead, as described in Section~\ref{sec:method_weight}, we use the isotropic Maxwellian-type model distribution $f_{\rm mdl}$ in Eq.~(\ref{eq:fM}) with orbit-averaged radius $\left<r\right>$ and kinetic energy $\left<\ekin\right>$. In other words, we assign exponentially less weight to orbits with higher kinetic energies $\left<\ekin\right>$. Of course, the resulting overall GC distribution $f_{\rm gc}(\ekin,\alpha,R,z)$ will not be isotropic locally, especially in cases where density and temperature gradients uncover the effects of radial magnetic drifts (e.g., see Fig.~13 of Ref.~\cite{Bierwage22a}), which are here modified by electric acceleration and drift, one consequence being the shift of the trapped-passing boundaries in Fig.~\ref{fig:05_kstar_orbtypes}.

We have developed and tested our methods only for cases with relatively smooth $\Phi$ and $E_r$, satisfying Eq.~(\ref{eq:com_b}) in the form $\varrho_{\rm L}\nabla\ln\Phi \sim \varrho_{\rm L}/a \ll 1$. Cases with steeply varying electric fields must be treated with additional care. Although the physics of so-called {\it orbit squeezing} \cite{Hinton95} is captured, the neglect of gyroaveraging implies that the GC model itself becomes invalid in regions with steep ambient field gradients.
 
We close this summary with a few technical remarks that are relevant for building a GC orbit database:\vspace{-0.2cm}
\begin{itemize}
	\item  Since orbits are followed only for a single poloidal transit, one may be tempted to sacrifice some accuracy to save calculation time by increasing the time step size; in our case, $\Delta t_{\rm RK}$ of a 4th-order Runge-Kutta solver. Individual GC orbits and (purely additive) {\it even} velocity moments ($0,v^2,...$) of their distribution function may indeed remain fairly accurate. In contrast, {\it odd} moments like the flow density ${\bm \Gamma}$ in Eq.~(\ref{eq:example_mom_gamma}) below, rely on the cancellation of large positive and negative flows in cases with nearly isotropic velocity distributions like our Eq.~(\ref{eq:fM}). We found that the residual mean flow can easily be obscured by unphysical artifacts (e.g., non-zero divergence $\nablab\cdot{\bm \Gamma} \neq 0$) if the chosen time step is too large (see Section~\ref{sec:example_dt} below).\vspace{-0.15cm}
	
	\item  We use a polynomial to represent the magnetic midplane curve, with the risk that some stagnation orbits and some potato orbits very close to trapped-passing boundary may be missed or corrupted. In {\tt VisualStart}, some safeguards have been implemented to at least eliminate the corrupted orbits.\vspace{-0.2cm}
\end{itemize}

\subsection{Remarks concerning grid aliasing noise}
\label{sec:method_grid_alias}

If one uses a method that samples the CoM space on a regular mesh as described in Section~\ref{sec:method_slice} and illustrated in Figs.~\ref{fig:04_kstar_grid} and \ref{fig:05_kstar_orbtypes}, subsequent binning of the marker particles on another diagnostic mesh can give rise to a phenomenon that we call ``grid aliasing noise'': The computed radial profiles or velocity distributions acquire spiky structures when the diagnostic mesh is similar but not identical to the CoM mesh that was used to sample the orbit space.

For instance, in Ref.~\cite{Bierwage22a}, our primitive binning procedure (described in Appendix~A of Ref.~\cite{Bierwage22a}) produced noise-like artifacts in plots of the velocity distribution at large values of the pitch angle $\alpha = {\rm asin}(u/v_{\rm gc})$ (see Figs.~13, 15 and 18 of \cite{Bierwage22a}). Meanwhile, radial profiles of the number density $N(r)$ (see Figs.~10 and 17 of \cite{Bierwage22a}) remained fairly smooth because we dealt mostly with energetic ions whose relatively large radial drifts spread their drift orbits (= CoM samples) across multiple cells along the minor radius $r$, which is not a CoM.

In the present paper, we utilize Maxwellian-like distributions (\ref{eq:fM}) that contain a large number of particles with low energies (few keV and less). Their GC orbits lie near surfaces of constant $r(\psi)$, so that profiles like $N(r)$ will sometimes contain significant noise-like artifacts due to the said grid aliasing effect. For instance, see the arrows in Figs.~\ref{fig:07_summary_n_t_mk}(w,x) and \ref{fig:10_jt60u_E30_n-vtor_K-scan}(e,n).

This type of noise increases with decreasing kinetic energy (cf.~Fig.~\ref{fig:10_jt60u_E30_n-vtor_K-scan}). Moreover, since the relative effect of radial drifts decreases with increasing distance from the plasma center, the magnitude of grid aliasing noise tends to increase with increasing radius (cf.~Figs.~\ref{fig:07_summary_n_t_mk} and \ref{fig:10_jt60u_E30_n-vtor_K-scan}). In addition, grid aliasing noise can be further enhanced by an ambient radial electric field $E_r$. For instance, in the configurations that we study here, a positive $E_r > 0$ enhances the radial drifts of co-passing ions (cf.~Fig.~\ref{fig:14_jt60_deep-pass}(b-7) and (c-7)) and reduces the radial drift of counter-passing ions (not shown graphically, but evident from Eq.~(\ref{eq:gc_dr_pass}) in \ref{apdx:gc_dr_du}). At the same time, $E_r > 0$ shifts the domain of mirror-trapped particles towards larger positive pitch angles $\alpha$, as we saw in Fig.~\ref{fig:05_kstar_orbtypes}(c). This effect will be particularly strong in our JT-60U case (Fig.~\ref{fig:b01_jt60_orbtypes}(c) of \ref{apdx:benchmark}). This reduces the number of co-passing orbits (with enhanced radial drifts) and increases the number of counter-passing orbits (with reduced radial drift), yielding a net reduction of the smoothing effect of radial drifts and, thus, larger grid aliasing noise (cf.~Fig.~\ref{fig:07_summary_n_t_mk}(w)).

Evidently, grid aliasing noise is an entirely systematic artifact that results from regular CoM space sampling and the use of a primitive marker binning scheme. In principle, it should be possible to accurately eliminate such kinds of artifacts with a more sophisticated marker binning scheme, provided that one possesses additional information about the form of the distribution function $f_{\rm com}$ between the samples taken. For instance, combining the assumption of $f_{\rm com}$ being smooth with knowledge about the grids used for sampling and binning, grid weights can be adjusted systematically to reconstruct a smooth binned distribution. We have not implemented such an advanced binning procedure yet, since grid aliasing noise is only an aesthetic issue here.

A simpler but less precise method to avoid grid aliasing noise is to introduce a small random shift in the locations where markers are loaded relative to the regular CoM mesh in Fig.~\ref{fig:04_kstar_grid}. The randomization comes at the expense of introducing errors in the volume elements $\Delta\hat{\V}_{ijk}$, which are usually computed from equations like (\ref{eq:model_dVol_orb}) based on the assumption of a regular CoM mesh. A randomization technique is currently used in {\tt EPCoM} \cite{BrochardEPS24}, because one of that code's purposes is to generate binned CoM distributions that (after further smoothing) can be used to compute gradients of $f_{\rm com}$. These gradients, in turn, can then be used in instability codes that employ the $\delta f$ method (c.f.~\cite{Aydemir94} and references therein).

\section{Application examples}
\label{sec:example}

\begin{table}
	\centering
	\begin{tabular}{@{\hspace{0.05cm}}c@{\hspace{0.1cm}}|@{\hspace{0.1cm}}c@{\hspace{0.1cm}}|@{\hspace{0.1cm}}c@{\hspace{0.1cm}}|@{\hspace{0.1cm}}c@{\hspace{0.05cm}}}
		\hline\hline
		& KSTAR & JT-60U & ITER \\
		\hline
		$R_0/a$ & $1.82\,{\rm m}/0.58\,{\rm m}$ & $3.39\,{\rm m}/0.97\,{\rm m}$ & $6.32\,{\rm m}/2.71\,{\rm m}$ \\
		$B_0/I_{\rm p}$ & $1.77\,{\rm T}/0.5\,{\rm MA}$ & $1.17\,{\rm T}/0.6\,{\rm MA}$ & $2.7\,{\rm T}/7.5\,{\rm MA}$ \\
		$T_0$ & $3.2\,{\rm keV}$ & $1.28,{\rm keV}$ & $5\,{\rm keV}$ \\
		$v_0^{({\rm D}^+)}$ & $550\,{\rm km/s}$ & $350\,{\rm km/s}$ & $700\,{\rm km/s}$ \\
		$\varrho_0^{({\rm D}^+)}$ & $6.5\,{\rm mm}$ & $6.3\,{\rm mm}$ & $5.3\,{\rm mm}$ \\
		\hline\hline
	\end{tabular}\vspace{-0.1cm}
	\caption{Plasma parameters: Aspect ratio $R_0/a$, on-axis field strength $B_0$, plasma current $I_{\rm p}$, on-axis temperature $T_0$, thermal velocity $v_0({\rm D}^+) \equiv \sqrt{2 T_0/M}$, thermal gyroradius $\varrho_0 \equiv v_0/\omega_{\rm B0}$ of deuterons. The KSTAR case is based on shot 18567 \protect\cite{Lee23}. The JT-60U case is based on shot E039672, similar to that studied in Refs.~\protect\cite{Shinohara02, Bierwage17a, Bierwage18}. The ITER case is a simplified up-down symmetric version of model shot 101006r50, with central safety factor near unity ($q_0 \sim 1$).}\vspace{-0.2cm}
	\label{tab:tok}
\end{table}

\subsection{Setup and diagnostics}
\label{sec:example_setup}

In this section, we discuss the results of numerical solutions obtained with the methods described in Section~\ref{sec:method}. The calculations were performed in realistic tokamak geometry for three examples, which are based on models of KSTAR, JT-60U and ITER plasmas. The main parameters are summarized in Table~\ref{tab:tok} and further details can be found in \ref{apdx:examples}. All cases have their toroidal field $B_{\rm tor} \equiv B_\zeta R$ and plasma current in the positive-$\zeta$ direction of right-handed cylinder coordinates $(R,z,\zeta)$. We assume pure deuterium plasmas.

In our KSTAR case \cite{Lee23}, rapid toroidal rotation with speeds up to $200\,{\rm km/s}$ suggest that $E_r$ may have reached $30\,{\rm kV/m}$ (cf.~Fig.~\ref{fig:a01_kstar_profs}). For comparison, this fairly high value will also be used in the JT-60U and ITER cases. For a 9 MA ITER scenario, an earlier numerical study estimated vales up to $\pm 10\,{\rm kV/m}$ globally and $-30\,{\rm kV/m}$ locally near the edge \cite{Tani15}, so our $E_{r0} = 30\,{\rm kV/m}$ constitutes a moderate exaggeration. For JT-60U, we also show results for a weaker $E_{r0} = 3\,{\rm kV/m}$ because the measured rotation speed in the reference experiment \cite{Shinohara02, Bierwage17a, Bierwage18} was $\lesssim 30\,{\rm km/s}$, suggesting that the radial electric field was a only few ${\rm kV/m}$ (cf.~Fig.~\ref{fig:a02_jt60u_profs}). The use of $E_{r0} = 30\,{\rm kV/m}$ in this JT-60U case should be regarded as a deliberate exaggeration with the purpose of emphasizing the trends and testing the accuracy of the numerical methods and algorithms at the extreme.

The kinetic energy axis was sampled uniformly in the range $0 < \ekin < 6T_0$ ($\hat{\ekin}_{\rm max} = 3$), which corresponds to initial velocities in the range $v_{\rm I} \in [0,\sqrt{6}v_0]$, where $v_0 \equiv \sqrt{2T_0/M}$ and $T_0 \equiv T_{\rm ref}(0)$. We sampled the entire range of possible initial pitch angles $\alpha_{\rm I} \equiv \sin^{-1}(\uGC_{\rm I}/v_{\rm I}) \in [-\pi/2,\pi/2]$. The initial midplane positions $X_{\rm I}$ of our orbit samples were constrained to the domain of closed flux surfaces. The number of cells was $N_\ekin \times N_\alpha \times N_X = 16\times32\times64$, giving about $10^5$ confined orbits.
 
After building databases of deuteron orbits for several values of the electric field amplitude $E_{r0}$, the orbits were weighted with the CoM Maxwellian $f_{\rm mdl}$ in Eq.~(\ref{eq:fM}). In the KSTAR and JT-60U cases, the reference profiles $N_{\rm ref}(r)$ and $T_{\rm ref}(r)$ for the number density and temperature resemble experimental measurements (cf.~Figs.~\ref{fig:a01_kstar_profs} and \ref{fig:a02_jt60u_profs}). In the ITER case (Fig.~\ref{fig:a03_iter_profs}), we used flat profiles $N_{\rm ref} = N_0$ and $T_{\rm ref} = T_0$ for test purposes.

For the following analysis, each orbit contour was sampled with a mean marker density of $50/(2\pi a)$ and a minimum of $5$ markers per orbit. With these choices, each case was represented by about $2\times 10^6$ markers. The resulting GC distribution $f_{\rm gc}$ was diagnosed by binning the marker weights on diagnostic grids as described in Appendix~A of Ref.~\cite{Bierwage22a}. In the following subsections, we will look at the number density $N$, flow density vector ${\bm \Gamma}$, and energy density (pressure) $P$,\vspace{-0.2cm}
\begin{subequations}
	\begin{align}
		N =&\; \int{\rm d}^3 v\, \frac{B}{B_\parallel^*} f_{\rm gc},
		\label{eq:example_mom_dens} \\
		{\bm \Gamma} =&\; \int{\rm d}^3 v\,\frac{B}{B_\parallel^*} \dot{\bm X}_{\rm gc} f_{\rm gc} \equiv \frac{{\bm J}}{Ze} \equiv N{\bm V},
		\label{eq:example_mom_gamma} \\
		P =&\; \int{\rm d}^3 v\,\frac{B}{B_\parallel^*}\frac{2\ekin}{3} f_{\rm gc} \equiv N T \stackrel{\rm Maxw.}{\longrightarrow} N_{\rm ref} T_{\rm ref},
		\label{eq:example_mom_pre}
	\end{align}\vspace{-0.3cm}
	\label{eq:example_mom}
\end{subequations}

\noindent normalized by $N_0 \equiv N_{\rm ref}(0)$, $N_0 v_0$, and $N_0 M v_0^2$, respectively. As indicated on the right-hand side of Eq.~(\ref{eq:example_mom_gamma}), the flow density ${\bm \Gamma}$ corresponds to the current density ${\bm J}$ divided by the species' charge $Ze$, and we define the mean flow velocity vector in regions of nonvanishing density as ${\bm V} \equiv {\bm \Gamma}/N$. Similarly, we define a temperature as $T \equiv P/N$. For a Maxwellian-like distribution, we expect to obtain $P \approx N_{\rm ref} T_{\rm ref}$ as indicated on the right-hand side of Eq.~(\ref{eq:example_mom_pre}). The major-radial, vertical, toroidal and poloidal flow density components are
\begin{subequations}
	\begin{gather}
		\Gamma_R \equiv {\bm \Gamma}\cdot\nablab R, \quad \Gamma_z \equiv {\bm \Gamma}\cdot\nablab z, \quad \Gamma_{\rm tor} \equiv R{\bm \Gamma}\cdot\nablab\zeta, \\
		\Gamma_{\rm pol} \equiv \frac{{\bm \Gamma}\cdot{\bm B}_{\rm pol}}{B_{\rm pol}} = \frac{\Gamma_R B_R + \Gamma_z B_z}{(B_R^2 + B_z^2)^{1/2}}.
	\end{gather}
\end{subequations}

\noindent We measured fields like $N(R,z)$, cross-sectional profiles $N(X)$ at $Y \equiv z-z_0 = 0$, and flux-surface-averaged profiles like $N(r)$ or $N(\psi_{\rm P})$. The number of markers per diagnostic cell, denoted by $N_{\rm mpc}$, was also measured.

\subsection{Convergence test for time step $\Delta t$}
\label{sec:example_dt}

In {\tt VisualStart}, a semi-empirical formula is used to estimate the optimal time step size $\Delta t_{\rm ref}$ for each marker particle to trace its GC orbit. A marker pushing time step satisfying $\Delta t/\Delta t_{\rm ref} \sim 1$ can then be expected to yield a reasonable compromise between accuracy and speed. However, that formula was conceived with only the accuracy of individual orbits in mind, and its adequacy for the moments of the modeled GC distribution must be checked case by case.

\begin{figure}
	[tb]\vspace{-0.1cm}
	\centering
	\includegraphics[width=0.48\textwidth]{\figures/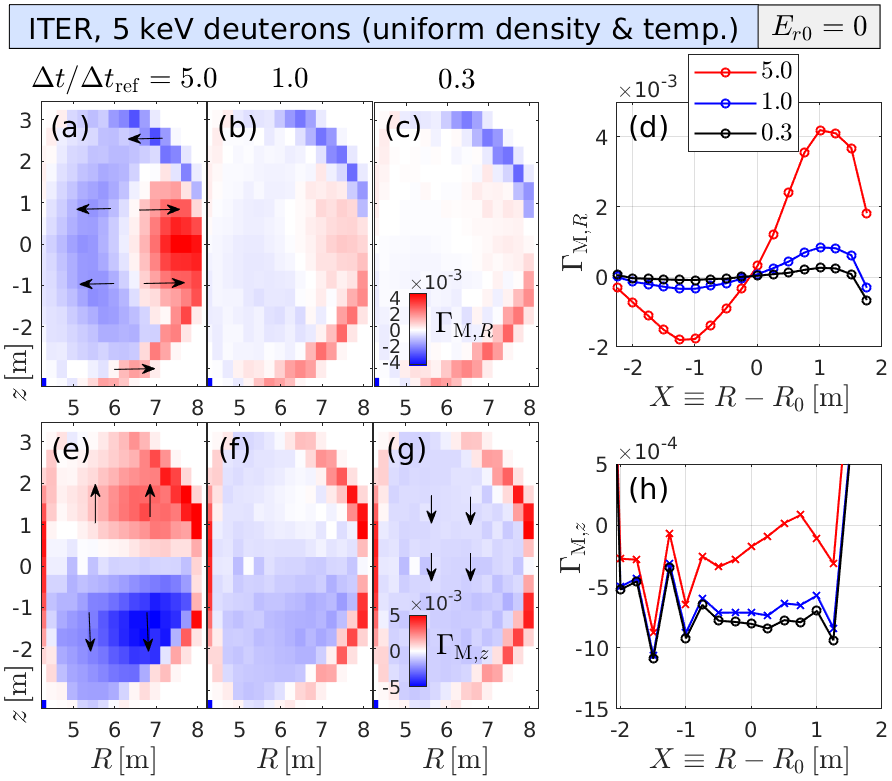}\vspace{-0.2cm}
	\caption{Convergence tests for residual (non-electric) magnetization flows with respect to the marker pushing time step $\Delta t$. The ITER test case with uniform number density $N_{\rm ref}=N_0$ and temperature $T_{\rm ref}=T_0$ is used. Panels (a)-(c) show the major radial component $\Gamma_{{\rm M},R}$ and panels (e)--(g) the vertical component $\Gamma_{{\rm M},z}$ of the non-electric flow density vector field ${\bm \Gamma}_{\rm M}(R,z)$ for $\Delta t = (5.0, 1.0, 0.3)\times \Delta t_{\rm ref}$. Panels (d) and (h) show the profiles of $\Gamma_{{\rm M},R}$ and $\Gamma_{{\rm M},z}$ at the height of the magnetic axis ($z=z_0$) as functions of $X \equiv R - R_0$.}\vspace{-0.2cm}
	\label{fig:06_iter_E0_residual-flow-dt}%
\end{figure}

At the end of Section~\ref{sec:method_summary}, we noted that representing distribution functions using GC orbits that are computed quickly with low accuracy can yield unphysical mean residual flows. This is demonstrated in Fig.~\ref{fig:06_iter_E0_residual-flow-dt}, where we plot the major radial and vertical components, $\Gamma_{{\rm M},R}$ and $\Gamma_{{\rm M},z}$, of the non-electric flow density ${\bm \Gamma}_{\rm M} \equiv {\bm \Gamma}(E_{r0}=0)$ in the ITER test case for three different time step sizes: $\Delta t = (5.0, 1.0, 0.3)\times \Delta t_{\rm ref}$.

Our model distribution $f_{\rm mdl}$ in Eq.~(\ref{eq:fM}) is isotropic and we let $E_{r0} = 0$, so the mean flows seen in Fig.~\ref{fig:06_iter_E0_residual-flow-dt} must be due to the effects of the mirror force and magnetic drifts (hence the subscript ``M''). Since we used a uniform density and temperature in this case, the expected result is a uniform flow field, except near the plasma boundary. Specifically, we expect $\Gamma_{{\rm M},R} \approx 0$ and $\Gamma_{{\rm M},z} \propto B^{-3} B_{\rm tor}\partial_R B \approx {\rm const}. < 0$ in the plasma interior.\footnote{The nonzero negative $\Gamma_{{\rm M},z}$ in Fig.~\protect\ref{fig:06_iter_E0_residual-flow-dt}(g) reflects the fact that the $\nablab B$ drift is directed downward everywhere. Individual particles remain confined due to the helical magnetic field, so that the vertical drift towards ($z > z_{\rm mid}$) and away from ($z < z_{\rm mid}$) the magnetic midplane cancel. Overall particle conservation is maintained via a positive return flow $\Gamma_{{\rm M},z} > 0$ at the boundary. Some deviation from $\Gamma_{{\rm M},z} \approx {\rm const}$.\ can be expected since the loss boundaries of more energetic orbits can lie deeper inside the plasma.}

Panels (b) and (f) of Fig.~\ref{fig:06_iter_E0_residual-flow-dt} show that our default time step $\Delta t = \Delta t_{\rm ref}$ yields unphysically divergent flows in the plasma core that are weak but still noticeable compared to the physical boundary flows. Panels (c) and (g) show that reducing the step size to $\Delta t = 0.3\times\Delta t_{\rm ref}$ reduces the unphysical flows to a barely visible level on the present scale, and yields $\Gamma_{{\rm M},z} \approx {\rm const}$.\ in panel (h) as expected. In contrast, panels (a) and (e) show that increasing the time step size to $\Delta t = 5\times\Delta t_{\rm ref}$ makes the unphysical residual core flow\footnote{Our 4th-order Runge-Kutta integrator conserves $\etot$ and $P_\zeta$ only approximately. We suspect that the resulting spiraling motion causes the unphysical residual flow structure seen in Fig.~\protect\ref{fig:06_iter_E0_residual-flow-dt}.}
comparable to the physical boundary flow. On the low-field side ($X>0$), the divergent radial velocity reaches $V_{{\rm M},R} = \Gamma_{{\rm M},R}/N_0 \approx 4.5 v_0 \approx 3\,{\rm km/s}$. This corresponds to the electric drift velocity that is produced by an electric field with strength $(3\,{\rm km/s})/2.7\,{\rm T} \approx 1\,{\rm kV/m}$. Thus, a large time step is justified in cases with strong radial electric field, such as $E_{r0} = \pm 30\,{\rm keV}$ in our examples. For weak $|E_r| < 10\,{\rm kV/m}$, shorter time steps are required.

\begin{figure*}
	[tb]\vspace{-2.5cm}
	\centering
	\includegraphics[width=0.96\textwidth]{\figures/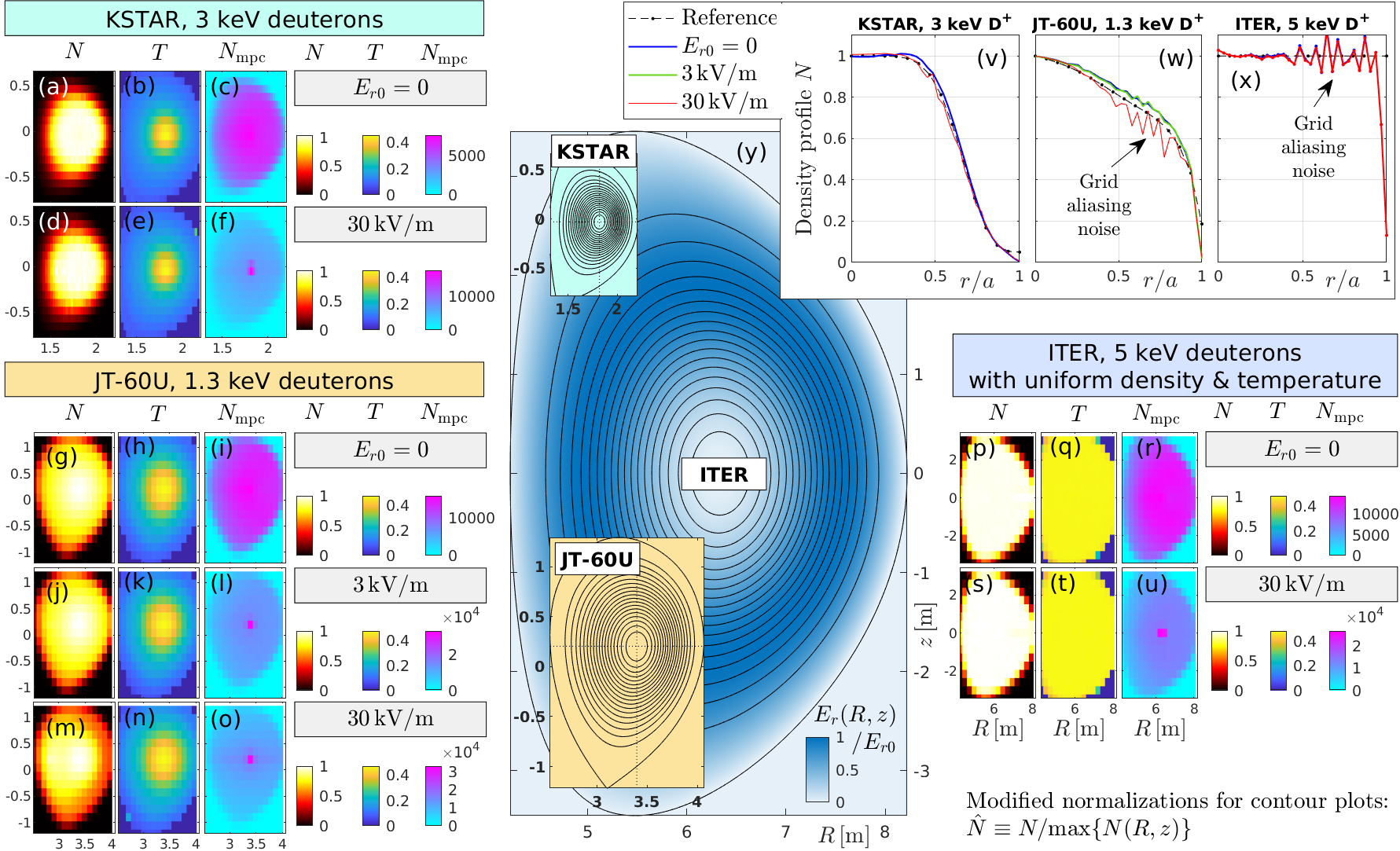}\vspace{-0.2cm}
	\caption{Overview of normalized number density ($N$) and temperature ($T \equiv P/N$) fields, as well as the distribution of markers per cell ($N_{\rm mpc}$) computed for the KSTAR (a--f), JT-60U (g--o) and ITER cases (p--u), with corresponding color bars next to them. In each set, we show results obtained without electric field ($E_{r0} = 0$), followed by results for $3\,{\rm kV/m}$ (JT-60U only) and $30\,{\rm kV/m}$ (all). Flux-surface-averaged profiles $N(r)$ are plotted in panels (v--x) in comparison with the respective reference profile $N_{\rm ref}(r)$ (dashed black) that enters our Maxwellian model in Eq.~(\protect\ref{eq:fM}). The cause and properties of the ``grid aliasing noise'' that is indicated by arrows in panels (w) and (x) were discussed in Section~\protect\ref{sec:method_grid_alias}. In the ITER case, we used constant $N_{\rm ref} = 1$ and $T_{\rm ref} = 0.5$ (normalized by $M v_0^2 = 2T_0$). Panel (y) and its insets show magnetic flux surfaces (black rings) and the relative size of the poloidal plasma cross-section of each device. The blue-shaded contours show the structure of our $E_r(R,z)$ model.}\vspace{-0.2cm}
	\label{fig:07_summary_n_t_mk}%
\end{figure*}

\subsection{Overview of results}
\label{sec:example_results}

For all three devices, Fig.~\ref{fig:07_summary_n_t_mk} summarizes the results for $N$, $T$ and $N_{\rm mpc}$. Panels (a)--(u) show 2D contour plots in the $(R,z)$-plane, and panels (v)--(x) show the radial profiles of the number density $N(r)$ for one or two values of the radial electric field in the range $0\leq E_{r0} \leq 30\,{\rm kV/m}$. At the level of detail shown, $E_r$ has no readily visible influence on the temperature field $T(R,z)$. For realistic values of $E_{r0} = 30\,{\rm kV/m}$ in KSTAR (d)--(f) and $E_{r0} = 3\,{\rm kV/m}$ in JT-60U (j)--(l), the density field $N(R,z)$ also remains nearly unchanged. However, after raising the value for the JT-60U case to $E_{r0} = 30\,{\rm kV/m}$, Fig.~\ref{fig:07_summary_n_t_mk}(m) shows a notable inward shift of the density contours $N(R,z)$. The cause of this will be discussed in Section~\ref{sec:example_dens} below.

The marker density $N_{\rm mpc}(R,z)$ is expected to be elevated near the midplane, especially near center, because we have chosen to load at least $5$ markers per orbit, regardless of how small its contour is. Interestingly, $N_{\rm mk}(R,z)$ appears smooth for $E_{r0} = 0$, and a distinct spike near the axis is observed only for nonzero $E_{r0}$ in Fig.~\ref{fig:07_summary_n_t_mk}(f,l,o,u).\footnote{We speculated that the marker accumulation may be enhanced by the fact that $E_r'$ in our model (\protect\ref{eq:er_model}) has a nonzero gradient at the axis. However, a test with an $E_r$ profile whose gradient approaches zero at the axis produced a similar result, so there must be a different reason.}
In any case, the smoothness of the physical density field $N(R,z)$ in Fig.~\ref{fig:07_summary_n_t_mk}(d,j,m,s) confirms that artificial structures in the marker density $N_{\rm mpc}(R,z)$ are properly balanced by the marker weight in Eq.~(\ref{eq:method_w_mk}), so the results appear to be consistent.

\begin{figure*}
	[tbp]\vspace{-1.35cm}
	\centering
	\includegraphics[width=0.48\textwidth]{\figures/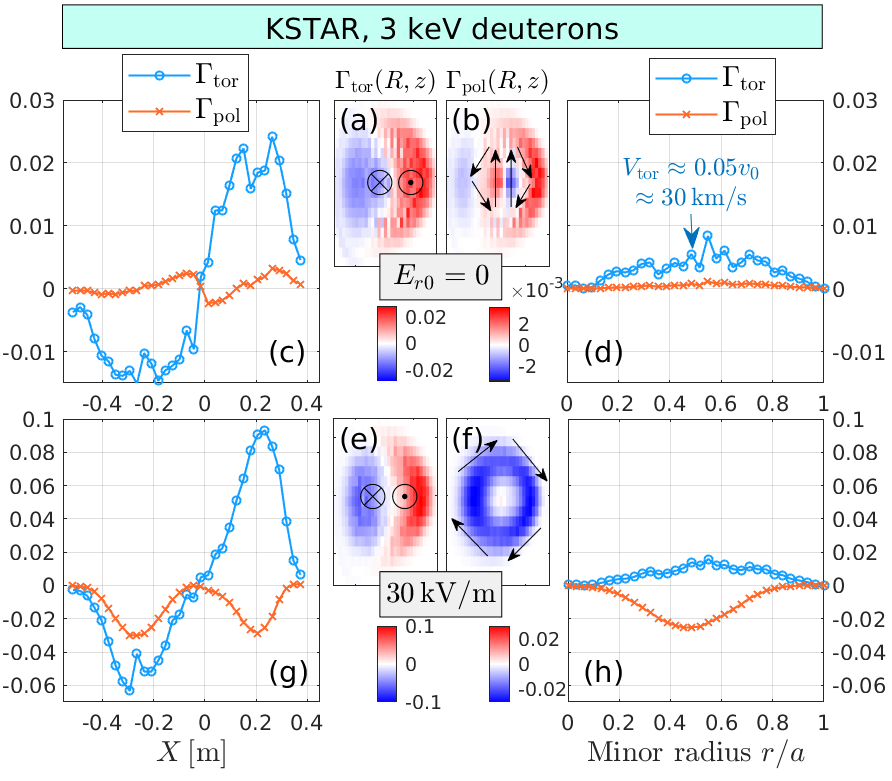} $\quad$
	\includegraphics[width=0.48\textwidth]{\figures/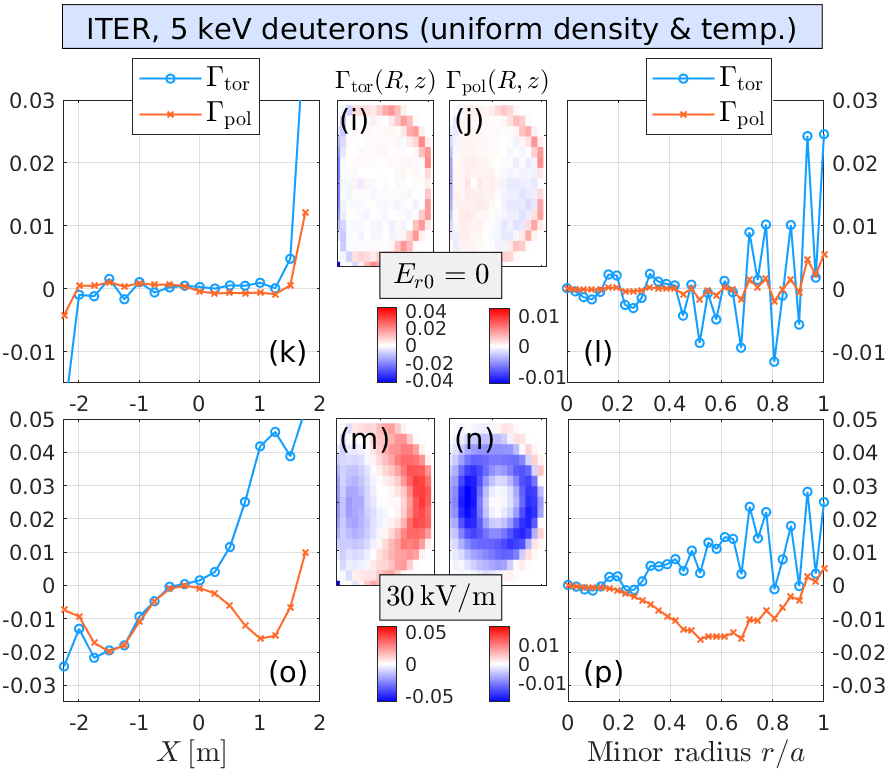}\vspace{-0.15cm}
	\caption{Overview of the toroidal and poloidal components of the flow density profiles, $\Gamma_{\rm tor}$ and $\Gamma_{\rm pol}$, in the KSTAR case (left) and the ITER case (right), for $E_{r0} = 0$ (top) and $30\,{\rm kV/m}$ (bottom). Panels (c), (g), (k) and (o) show the cross-sectional profiles $\Gamma_{\rm tor}(X)$ and $\Gamma_{\rm pol}(X)$ near the midplane ($Y\approx 0$). Panels (d), (h), (l) and (p) show the flux-surface-averaged profiles $\Gamma_{\rm tor}(r)$ and $\Gamma_{\rm pol}(r)$ as functions of the volume-averaged minor radius $r/a$. All radial profiles were smoothed by averaging over 3 grid points with equal weights. In the central column for each case, panels (a,b), (e,f), (i,j) and (m,n) show contour plots of the respective flow density fields in the poloidal $(R,z)$ plane, using the same axis limits as in Fig.~\protect\ref{fig:07_summary_n_t_mk}.}\vspace{-0.2cm}
	\label{fig:08_kstar-iter_flow-prof}%
\end{figure*}

\begin{figure}
	[tbp]
	\centering
	\includegraphics[width=0.48\textwidth]{\figures/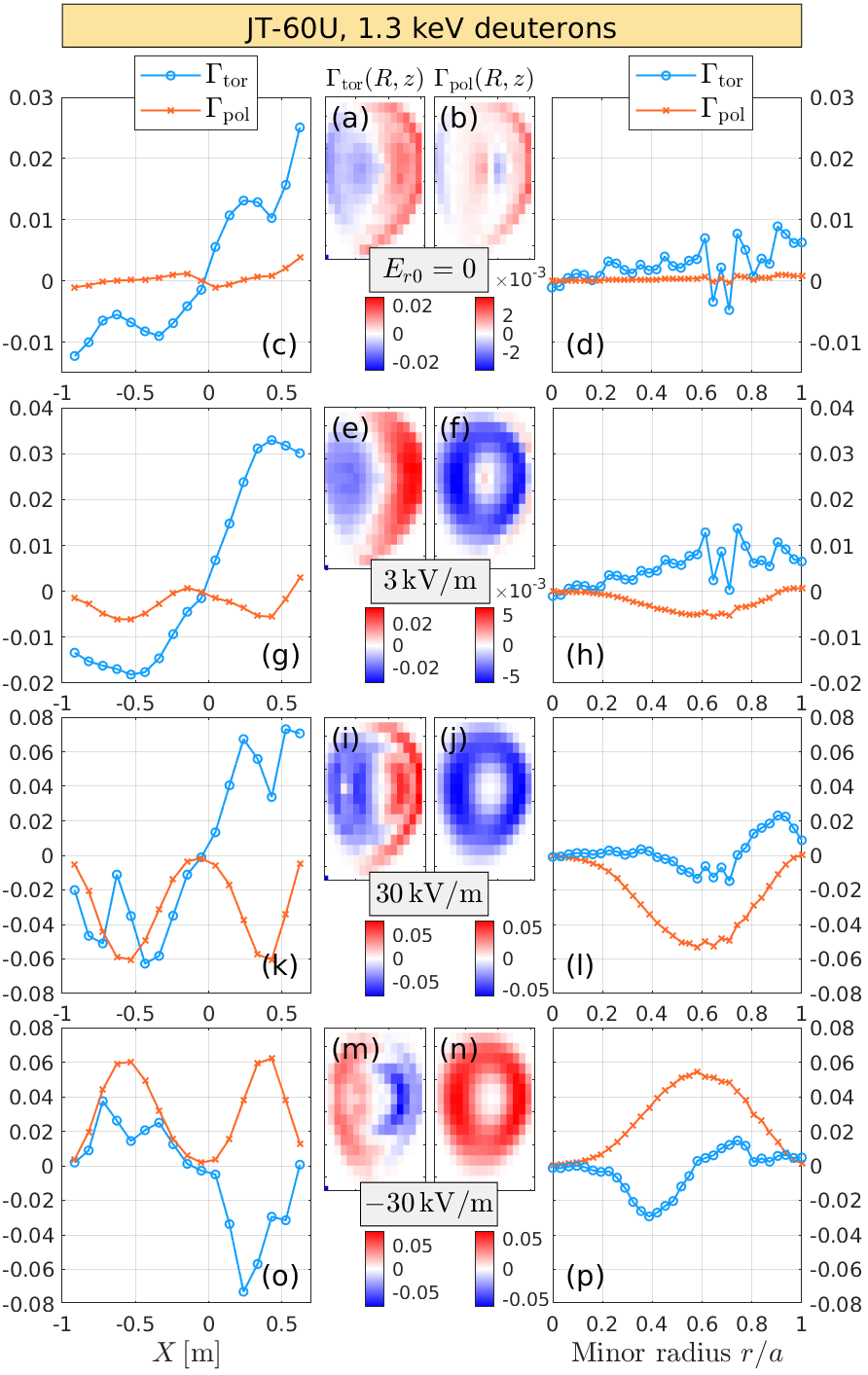}\vspace{-0.2cm}
	\caption{Toroidal and poloidal flow densities, $\Gamma_{\rm tor}$ and $\Gamma_{\rm pol}$, in the JT-60U case with $E_{r0} = 0$ (top), $3\,{\rm kV/m}$, and $\pm 30\,{\rm kV/m}$ (bottom). Arranged as Fig.~\protect\ref{fig:08_kstar-iter_flow-prof}. For results with uniform $N_{\rm ref}$ and $T_{\rm ref}$, see Fig.~\protect\ref{fig:b02_jt60_flat30_benchmark}.}\vspace{-0.25cm}
	\label{fig:09_jt60u_flow-prof-sign}%
\end{figure}

Contours and radial profiles of the toroidal and poloidal flow density fields, $\Gamma_{\rm tor}$ and $\Gamma_{\rm pol}$, are plotted in Fig.~\ref{fig:08_kstar-iter_flow-prof} for KSTAR and ITER, and in Fig.~\ref{fig:09_jt60u_flow-prof-sign} for the JT-60U case. For each case, the left column shows cross-sectional profiles at $Y=0$ as functions of $X$, the right column shows flux-surface-averaged profiles as functions of $\psi_{\rm P}$, and the central column shows the 2D contours of $\Gamma_{\rm tor}$ and $\Gamma_{\rm pol}$ in the same domain as in Fig.~\ref{fig:07_summary_n_t_mk}.

The upper row of Figs.~\ref{fig:08_kstar-iter_flow-prof} and \ref{fig:09_jt60u_flow-prof-sign} shows the non-electric flow density ${\bm \Gamma}_{\rm M} \equiv {\bm \Gamma}(E_{r0}=0)$. In Fig.~\ref{fig:08_kstar-iter_flow-prof}(a)--(d) for the KSTAR case and Fig.~\ref{fig:09_jt60u_flow-prof-sign}(a)--(d) for the JT-60U case, these residual magnetization flows are globally uncovered by density and temperature gradients. Counter-passing particles dominate on the high-field side (HFS) and co-passing particles dominate on the low-field side (LFS). This results in oppositely directed toroidal flows $\Gamma_{\rm M,tor}$ as shown in panels (a) and (c) of Fig.~\ref{fig:08_kstar-iter_flow-prof} for KSTAR. Meanwhile, the poloidal component $\Gamma_{\rm M,pol}$ consists of a pair of counter-rotating vortices as indicated by the black arrows in Fig.~\ref{fig:08_kstar-iter_flow-prof}(b). (A more complete picture was shown in Fig.~11 of Ref.~\cite{Bierwage22a}.) In the ITER case in Fig.~\ref{fig:08_kstar-iter_flow-prof}(i)--(l), the magnetization flows $\Gamma_{\rm M,tor}$ and $\Gamma_{\rm M,pol}$ nearly vanish in the plasma interior as we already discussed in Section~\ref{sec:example_dt} above (cf.~Fig.~\ref{fig:06_iter_E0_residual-flow-dt}).

Notice that our KSTAR case in Fig.~\ref{fig:08_kstar-iter_flow-prof}(d) has a net positive flux-surface-averaged toroidal flow $\Gamma_{\rm M,tor}(\psi_{\rm P})$ that is an order of magnitude larger than the poloidal component. Its magnitude corresponds to a toroidal velocity $V_{\rm M,tor} \approx 0.05 v_0 \approx 30\,{\rm km/s}$. The poloidal component $\Gamma_{\rm M.pol}(\psi_{\rm P})$ would however become dominant if one were to include the diamagnetic flow, which amounts to about $V_* \approx 70\,{\rm km/s}$ at mid-radius (cf.~Fig.~\ref{fig:a01_kstar_profs}). The same is true for our JT-60U case in Figs.~\ref{fig:09_jt60u_flow-prof-sign}(d) and \ref{fig:a02_jt60u_profs}.

The lower parts of Figs.~\ref{fig:08_kstar-iter_flow-prof} and \ref{fig:09_jt60u_flow-prof-sign} show how the relative strength of the poloidal flow $\Gamma_{\rm pol}$ increases in the presence of an ambient radial electric field $E_r$. For instance, in the KSTAR case with $E_{r0} = 30\,{\rm kV/m}$, Fig.~\ref{fig:08_kstar-iter_flow-prof}(g) shows a ratio $|\Gamma_{\rm tor}|/|\Gamma_{\rm pol}|$ of $\approx 2$ on the HFS and $\approx 3$ on the LFS.\footnote{If one subtracts ${\bm \Gamma}_{\rm M} \equiv {\bm \Gamma}(E_{r0}=0)$, these ratios reduce to $|\Gamma_{\rm tor} - \Gamma_{\rm M,tor}|/|\Gamma_{\rm pol} - \Gamma_{\rm M,pol}| \approx 1.5$ on the HFS and $\approx 2.5$ on the LFS.}
In contrast, the flux-surface-averaged toroidal electric flows tend to be comparable to (see Figs.~\ref{fig:08_kstar-iter_flow-prof}(p) and \ref{fig:09_jt60u_flow-prof-sign}(h)) or somewhat smaller than the poloidal ones (see Figs.~\ref{fig:08_kstar-iter_flow-prof}(h) and \ref{fig:09_jt60u_flow-prof-sign}(l,p)).

As a consistency check, we verify that the components of the calculated mean electric drift velocity
\begin{equation}
	{\bm V}_{\rm E} \equiv {\bm \Gamma}_{\rm E}/N \approx {\bm \Gamma}(E_{r0}) - {\bm \Gamma}_{\rm M},
	\label{eq:example_vE}
\end{equation}

\noindent substituted into the formula $E_{\rm rad} = V_{\rm E,tor} B_{\rm pol} - V_{\rm E,pol} B_{\rm tor}$ (from \ref{apdx:Erad}) yield the strength of the applied electric field $E_{\rm rad} \equiv {\bm E}\cdot\hat{\bm e}_r = E_r|\nablab r| \approx E_r$. In the JT-60U case ($v_0 \approx 350\,{\rm km/s}$) the flow densities in Fig.~\ref{fig:09_jt60u_flow-prof-sign} at $r/a \approx 0.55$ --- where $N\approx 0.8...0.75$ in Fig.~\ref{fig:07_summary_n_t_mk}(w), and $B_{\rm pol} \approx 0.13\,{\rm T}$ in Fig.~\ref{fig:a02_jt60u_profs}(b) --- give\vspace{-0.05cm}
\begin{align}
	&E_{\rm rad} = (\Gamma_{\rm E,tor} B_{\rm pol} - \Gamma_{\rm E,pol} B_{\rm tor})/N \stackrel{r/a\approx0.55}{\approx} \\
	&\left\{\begin{array}{c@{\hspace{0.05cm}}l}
		\frac{0.0044\, v_0\times 0.13\,{\rm T} - (-0.0054\, v_0)\times 1.17\,{\rm T}}{0.8} & \approx 3.0\,{\rm kV/m}, \\
		\frac{-0.013\, v_0\times 0.13\,{\rm T} - (-0.052\, v_0)\times 1.17\,{\rm T}}{0.75} & \approx 28\,{\rm kV/m},
	\end{array} \right. \nonumber
\end{align}

\noindent The flow densities at $r/a \approx 0.55$ in Fig.\ref{fig:08_kstar-iter_flow-prof} for the KSTAR case ($v_0 = 550\,{\rm km/s}$, $N \approx 0.75$ from Fig.~\ref{fig:07_summary_n_t_mk}(v), and $B_{\rm pol} \approx 0.24\,{\rm T}$ from Fig.~\ref{fig:a01_kstar_profs}(b)) and for the ITER case ($v_0 = 700\,{\rm km/s}$, $N = 1$ from Fig.~\ref{fig:07_summary_n_t_mk}(x), and $B_{\rm pol} \approx 0.55\,{\rm T}$ from Fig.~\ref{fig:a03_iter_profs}(b)) give\vspace{-0.1cm}
\begin{align}
	&E_{\rm rad}  \stackrel{r/a\approx0.55}{\approx} \\
	&\left\{\begin{array}{c@{\hspace{0.05cm}}l}
		\frac{0.007\, v_0\times 0.24\,{\rm T} - (-0.023\, v_0)\times 1.77\,{\rm T}}{0.75} & \approx 31\,{\rm kV/m} \;\text{(KSTAR)}, \\
		\frac{0.014\, v_0\times 0.58\,{\rm T} - (-0.015\, v_0)\times 2.66\,{\rm T}}{1.0} & \approx 34\,{\rm kV/m} \;\text{(ITER)}.
	\end{array} \right. \nonumber
\end{align}

\noindent The contribution of the $V_{\rm E,tor} B_{\rm pol}$ term is $< 10\%$ in JT-60U, $\approx 4\%$ in KSTAR, and $\approx 17\%$ in the ITER case.

The agreement between the above estimates and the actual values of $E_{r0}$ used is satisfactory, considering that deviations up to about $a/R_0 \sim 30\%$ may be tolerable due to flux-surface-averaging, the ignored factor $|\nablab r|$, limited resolution (``noise''), and due to the fact that Eq.~(\ref{eq:example_vE}) assumes for simplicity that the underlying electric and magnetic drift effects can be additively separated, ignoring their interdependence (see the last paragraph of \ref{apdx:gc_dr_du}). If one does not subtract the residual magnetization flow ${\bm V}_{\rm M} = \Gamma_{\rm M}/N$ in Eq.~(\ref{eq:example_vE}) the estimated $E_{\rm rad}$ values change by a few percent in the present cases, so the above-mentioned caveats of Eq.~(\ref{eq:example_vE}) are not substantial here. However, for smaller values of $E_r$, the contribution of ${\bm V}_{\rm M}$ can be significant.

In the following paragraphs we discuss some interesting physical features in more detail. One should bear in mind that the results presented here are ``physical'' only in the sense that they are based on the superposition of Hamiltonian GC orbits. However, our results are unlikely to be realistic because we independently imposed orbit weights based on the isotropic Maxwellian-like model distribution (\ref{eq:fM}) and the electric field $E_r$ in Fig.~\ref{fig:03_kstar_E30_model}, without any concern for self-consistency. Our $E_r$ field and the moments of the resulting GC distribution $f_{\rm gc}$ --- such as the toroidal flow field $\Gamma_{\rm tor}$ that we discuss in the following Section~\ref{sec:example_vtor} --- are thus unlikely to arise in this combination in a real plasma.

\subsection{Discussion of toroidal flow field structure}
\label{sec:example_vtor}

The in-out asymmetry of the toroidal flow density field $\Gamma_{\rm tor}(R,z)$ in Figs.~\ref{fig:08_kstar-iter_flow-prof} and \ref{fig:09_jt60u_flow-prof-sign} has two reasons. First, the toroidal rotation is dominated by the electric precession of mirror-trapped particles (details will follow in Section~\ref{sec:gc}). For positive $E_{r0} > 0$, this contributes $\Gamma_{\rm tor} > 0$ (red contours) on the LFS, which corresponds to out-of-plane flow that is indicated by $\odot$ in Fig.~\ref{fig:08_kstar-iter_flow-prof}(e). Second, positive $E_{r0} > 0$ shifts the trapped-passing boundary towards positive pitch angles, thus, shrinking the co-passing domain and enlarging the counter-passing domain. This gives $\Gamma_{\rm tor} < 0$ (blue contours), which dominates on the HFS and is indicated by $\otimes$ in Fig.~\ref{fig:08_kstar-iter_flow-prof}(e). The signs are reversed for negative $E_{r0}$, as shown in Fig.~\ref{fig:09_jt60u_flow-prof-sign}(m): trapped orbits yield $\Gamma_{\rm tor} < 0$ on the LFS, and a deficit of counter-passing and surplus of co-passing orbits yields $\Gamma_{\rm tor} > 0$ that dominates on the HFS.

The amount by which the trapped-passing boundary is shifted depends on the ratio $v_{\rm E}/v_{\rm T}$ of the electric drift and thermal velocity $v_{\rm T} = \sqrt{2T/M}$. In the case of our Maxwellian-like distribution (\ref{eq:fM}) with nonuniform temperature $T_{\rm ref}(r)$, the resulting toroidal flow field can become more complicated than the mere bipolar structure described in the previous paragraph. One concrete example is the multi-peaked toroidal flow field in Fig.~\ref{fig:09_jt60u_flow-prof-sign}(i)--(k) that we obtained in the JT-60U case with exaggerated electric field $E_{r0} = 30\,{\rm kV/m}$.

\begin{figure}
	[tb]\vspace{-2.4cm}
	\centering
	\includegraphics[width=0.48\textwidth]{\figures/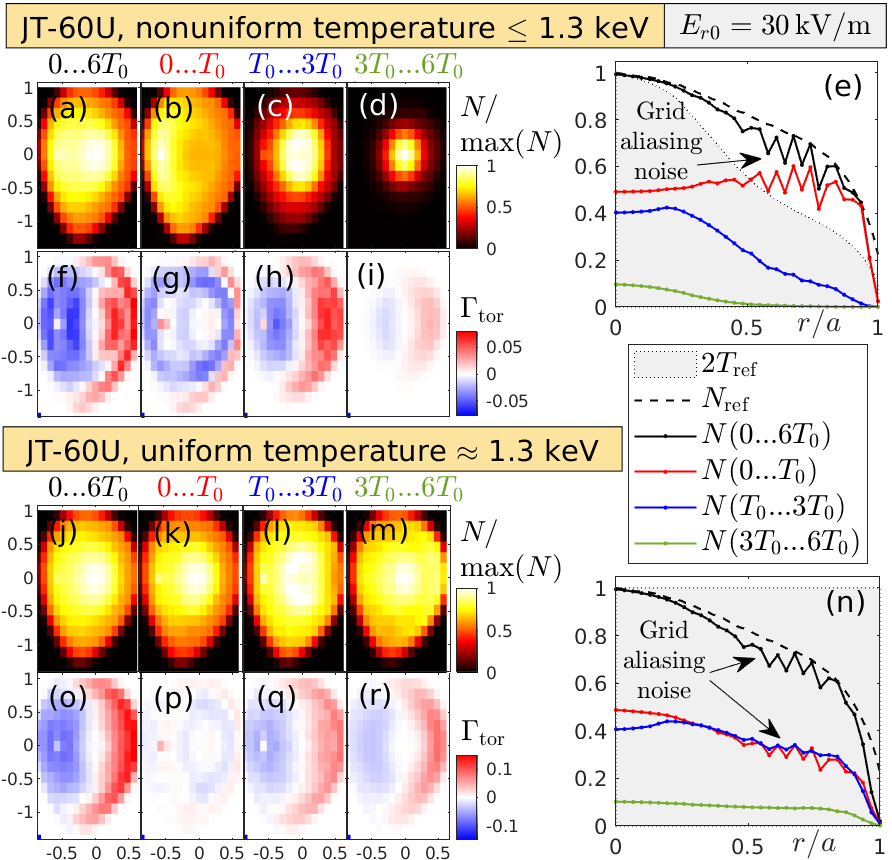}\vspace{-0.15cm}
	\caption{Contributions of particles with different kinetic energies $\ekin$ to the number density $N$ and toroidal flow density $\Gamma_{\rm tor}$ in the JT-60U case with $E_{r0} = 30\,{\rm kV/m}$ with nonuniform (top) and uniform (bottom) reference temperature $T_{\rm ref}(r)$. Panels (a) and (f) reproduce $N$ and $\Gamma_{\rm tor}$ from Figs.~\protect\ref{fig:07_summary_n_t_mk}(m) and \protect\ref{fig:09_jt60u_flow-prof-sign}(i) for the energy range $\ekin \in [0,6T_0]$. In panels (b)--(d) and (g)--(i), the contributions of the energy windows $[0,T_0]$, $[T_0,3T_0]$ and $[3T_0,6T_0]$ are plotted. Panel (e) shows the respective density profiles (solid) in comparison with the reference profile $N_{\rm ref}(r)$ (dashed). The reference temperature profile $T_{\rm ref}(r)$ is plotted as a gray-shaded area. Panels (j)--(r) for the case with uniform temperature $T_{\rm ref} = T_0 = 0.5$ are arranged in the same way. The cause and properties of the ``grid aliasing noise'' that is indicated by arrows in panels (e) and (n) were discussed in Section~\protect\ref{sec:method_grid_alias}.}\vspace{-0.2cm}
	\label{fig:10_jt60u_E30_n-vtor_K-scan}\vspace{-0.2cm}
\end{figure}

This structure can be understood by inspecting the contributions of particles with different kinetic energies as in Fig.~\ref{fig:10_jt60u_E30_n-vtor_K-scan}. The key point is that our sharply peaked reference temperature profile $T_{\rm ref}(r)$ that is shown as a gray-shaded area in Fig.~\ref{fig:10_jt60u_E30_n-vtor_K-scan}(e) implies that particles in the energy range $\ekin \in [0,T_0]$ dominate in the outer half of the plasma, $0.5 \lesssim r/a \lesssim 1$, as one can see by comparing the red and black curves in Fig.~\ref{fig:10_jt60u_E30_n-vtor_K-scan}(e). In that part of the orbit space, the trapped-particle boundary is strongly up-shifted (cf.~Fig.~\ref{fig:b01_jt60_orbtypes}(c) of \ref{apdx:benchmark}), so that negative toroidal flow $\Gamma_{\rm tor} < 0$ dominates in a large portion of the poloidal plane in Fig.~\ref{fig:10_jt60u_E30_n-vtor_K-scan}(g). That negative flow peaks off-axis in the region where the density gradient is steep, and we conjecture that this produces the local minimum on the LFS of Figs.~\ref{fig:10_jt60u_E30_n-vtor_K-scan}(f) and \ref{fig:09_jt60u_flow-prof-sign}(i,k).

\begin{figure}
	[tb]\vspace{-2.4cm}
	\centering
	\includegraphics[width=0.48\textwidth]{\figures/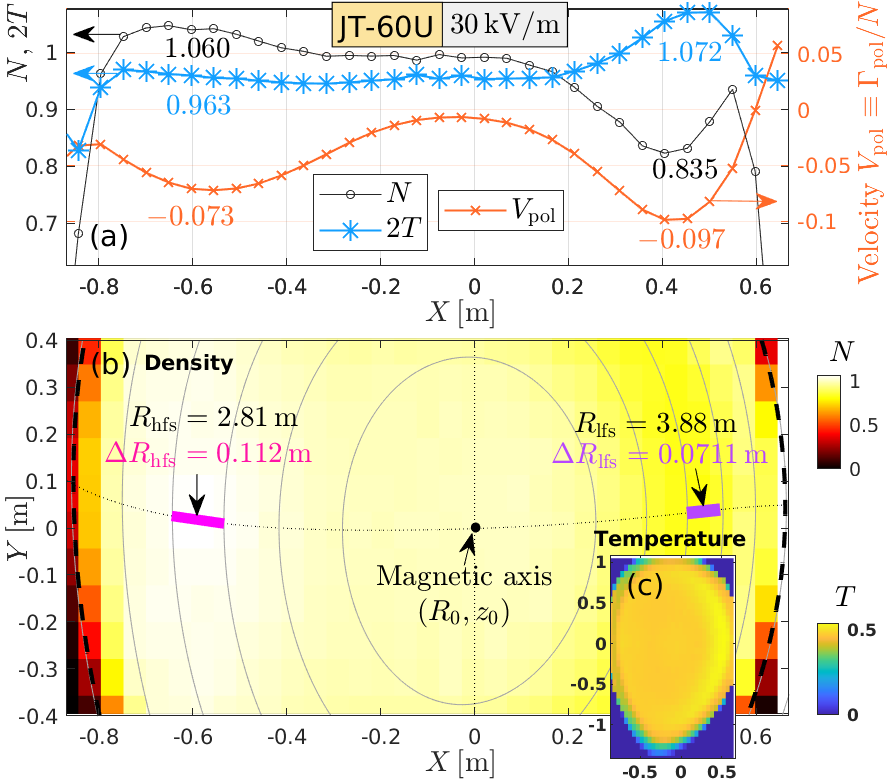}\vspace{-0.15cm}
	\caption{$E_r$-induced modification of the number density $N$ and temperature $T$ in the JT-60U case with exaggerated $E_{r0} = 30\,{\rm kV/m}$. This is a high-resolution run with $N_{\ekin}\times N_\Lambda \times N_X = 32\times 128\times 128$ cells (21M markers). In order to isolate the effect of $E_r$, the analysis was performed using uniform reference profiles $N_{\rm ref} = 1$ and $T_{\rm ref} = 0.5$. Panel (a) shows the profiles of the number density $N$ (black), temperature $T$ (blue) and the poloidal component $V_{\rm pol}$ (orange) of the mean flow velocity ${\bm V} \equiv {\bm \Gamma}/N$ as functions of $X \equiv R-R_0$ at the height of the magnetic axis ($Y \equiv z-z_0$). Panel (b) shows a contour plot of the density field $N(X,Y)$ near the midplane. The magenta bars indicate the radial length elements $\Delta R$ that are used in the analysis in Eq.~(\protect\ref{eq:example_continity}). The data in panel (b) are the same as in Fig.~\protect\ref{fig:b02_jt60_flat30_benchmark}(b) of \protect\ref{apdx:benchmark}, except that a slightly modified color scale was used here to make the small peak on the HFS better visible. The inset panel (c) shows the full poloidal cross-section of the temperature field $T(X,Y)$.}\vspace{-0.2cm}
	\label{fig:11_jt60u_X-prof-tubes}%
\end{figure}

In contrast, when we use a uniform reference temperature $T_{\rm ref} = T_0$, the results in Fig.~\ref{fig:10_jt60u_E30_n-vtor_K-scan}(n) show that particles in the energy windows $\ekin \in [0,T_0]$ and $[T_0,3T_0]$ each contribute nearly half of the particle population. Since the flow density ${\bm \Gamma}$ in Eq.~(\ref{eq:example_mom_gamma}) is weighted towards higher energies, particles in the energy range $\ekin \in [T_0,6T_0]$ dominate the toroidal flow as can be inferred from Fig.~\ref{fig:10_jt60u_E30_n-vtor_K-scan}(p)--(r). The resulting overall $\Gamma_{\rm tor}$ field in panel (o) has a relatively simple bipolar structure. The local maximum of $\Gamma_{\rm tor}$ on the HFS midplane in panels (f)--(h) and (o)--(q) of Fig.~\ref{fig:10_jt60u_E30_n-vtor_K-scan} may be caused by the turning points of barely trapped potato orbits.

\subsection{Discussion of number density field structure}
\label{sec:example_dens}

In the JT-60U case with strong $E_{r0} = 30\,{\rm kV/m}$, the contours of the density $N(R,z)$ in Fig.~\ref{fig:07_summary_n_t_mk}(m) are noticeably shifted inward in $R$. A similar result (not shown) was obtained for $-30\,{\rm kV/m}$. This distortion is energy-dependent, as one can infer from Fig.~\ref{fig:10_jt60u_E30_n-vtor_K-scan}(k)--(m).

A similar distortion occurs also with uniform density and temperature profiles, $N_{\rm ref} = N_0$ and $T_{\rm ref} = T_0$. A detailed view of this case is given in Fig.~\ref{fig:11_jt60u_X-prof-tubes}. One can see that, near the midplane ($Y \equiv z - z_0 \approx 0$), the number density field $N(R,z)$ in Fig.~\ref{fig:11_jt60u_X-prof-tubes}(b) has a minor depression around $X \approx 0.4\,{\rm m}$ on the LFS and a minor elevation around $X \approx -0.6\,{\rm m}$ on the HFS, in the region where the electric field is strong and the mean poloidal velocity $V_{\rm pol} = \Gamma_{\rm pol}/N$ in Fig.~\ref{fig:11_jt60u_X-prof-tubes}(a) has (negative) maxima.

Such a poloidal modulation of the number density $N$ must be accompanied by a corresponding poloidal modulation of the mean flow, which in the case under consideration in Fig.~\ref{fig:11_jt60u_X-prof-tubes} is dominated by electric flow ${\bm V} \approx {\bm V}_{\rm E}$. Its poloidal component ${\bm V}_{\rm E,pol}$, in turn, is dominated by passing particles, so it is meaningful to inspect the poloidal component of the GC drift velocity in Eq.~(\ref{eq:gc_vE_norm}), which has the form $\hat{\bm v}_{\rm E,pol} = -\hat{E}_{\hat{\Psi}_{\rm P}} \hat{I} \hat{\bm B}_{\rm pol}/\hat{B}^2 \approx -\hat{E}_{\hat{\Psi}_{\rm P}} \hat{\bm B}_{\rm pol}R^2/R_0$. The structure of $|B_{\rm pol}|$ in Fig.~\ref{fig:a02_jt60u_profs}(a) multiplied by $|\hat{E}_{\hat{\Psi}_{\rm P}}| R^2$ implies that $v_{\rm E}$ is faster on the LFS and slower on the HFS. By virtue of the continuity equation (particle conservation), the plasma must be diluted on the LFS and compressed on the HFS. This is consistent with the observation in Fig.~\ref{fig:11_jt60u_X-prof-tubes}.

Let us verify that the mean flows really satisfy the continuity equation $-\partial_t N = \nablab\cdot{\bm \Gamma} = 0$ quantitatively. Due to axisymmetry, the contribution of the toroidal flow $\Gamma_{\rm tor}$ is zero. In the midplane, the radial flow must vanish under steady-state conditions (cf.~Section~\ref{sec:method_mid}), so the particle conservation law, with $\smallint{\rm d}^3{\bm x}\,\nablab\cdot{\bm \Gamma}$ rewritten using Gauss' divergence theorem, reduces to\vspace{-0.05cm}
\begin{equation}
	\smallint{\rm d}R\,R\hat{\bm e}_z\cdot{\bm \Gamma} \approx R\Delta R N V_z = {\rm const}. \;\; \text{(on midplane)}.
	\label{eq:example_continuity}\vspace{-0.05cm}
\end{equation}

\noindent The density maximum is located roughly between the third and fourth magnetic surface contours in Fig.~\ref{fig:11_jt60u_X-prof-tubes}, whose radial separations in the midplane, $\Delta R_{\rm hfs}$ and $\Delta R_{\rm lfs}$, are indicated by violet bars. Substituting their values into Eq.~(\ref{eq:example_continuity}), and letting  $|V_z| \approx |V_{\rm pol}|$, we find
\begin{subequations}
	\begin{align}
		|R \Delta R N V_{\rm pol}|_{\rm hfs} \approx&\; 2.81\,{\rm m}\times 0.112\,{\rm m}\times 1.060\times 0.073 \nonumber \\
		\approx&\; 244\,{\rm cm}^2, \\
		|R \Delta R N V_{\rm pol}|_{\rm lfs} \approx&\; 3.88\,{\rm m}\times 0.071\,{\rm m}\times 0.835\times 0.097 \nonumber \\
		\approx&\; 223\,{\rm cm}^2,
	\end{align}
	\label{eq:example_continity}\vspace{-0.5cm}
\end{subequations}

\noindent in units of $N_0 V_0$. At the present level of accuracy a 10\% discrepancy can be ignored, so these values may be regarded as being identical. This agreement can be taken as evidence that the observed density modulation is both physical and correctly computed. Notice that the Shafranov shift contributes to the poloidal modulation of $N$ and ${\bm V}$, since it makes $\Delta R_{\rm hfs} > \Delta R_{\rm lfs}$ in Eq.~(\ref{eq:example_continity}).

One may wonder whether the elevation of the number density $N(R,z)$ on the HFS and its depression on the LFS of Fig.~\ref{fig:11_jt60u_X-prof-tubes} cancel when computing the flux-surface average $N(r)$. However, Figs.~\ref{fig:07_summary_n_t_mk}(w) and \ref{fig:10_jt60u_E30_n-vtor_K-scan}(e,n) all indicate that this is not the case. We always observe a systematic reduction of $N$ that becomes readily visible when the electric field strength $|E_r|$ is sufficiently large. In order to rule out mistakes and numerical problems, we performed various tests, including those in \ref{apdx:benchmark}. Our current understanding is that the radial electric field indeed reduces the net number of physical particles in the system when orbits are weighted using a distribution function that decreases monotonically towards higher energies as in our Eq.~(\ref{eq:fM}). This is because the electric drift increases the mean kinetic energy $\ekin$ of an orbit, whose weight then decreases by virtue of Eq.~(\ref{eq:fM}). To verify this, we reversed the trend by changing the $\ekin$-dependence of $f_{\rm mdl}$ in Eq.~(\ref{eq:fM}) as\vspace{-0.1cm}
\begin{equation}
	\exp\left(-\frac{\left<\ekin\right>}{T_{\rm ref}(\left<r\right>)}\right) \;\; \rightarrow \;\; \exp\left(-\left|\frac{\left<\ekin\right>}{T_{\rm ref}(\left<r\right>)} - 1\right|\right),
	\label{eq:fM_shift}
\end{equation}

\noindent so that the peak of $f_{\rm mdl}$ is shifted from $\ekin = 0$ to $\ekin \approx T_0$. In the absence of an electric field ($E_{r0} = 0$) the shifted distribution (\ref{eq:fM_shift}) produces a density profile $N(r)$ that is somewhat narrower (more sharply peaked) than $N_{\rm ref}(r)$. It was then found that with $E_{r0} = 30\,{\rm kV/m}$ the computed density $N$ increased around the $E_r$ peak, consistently with our prediction that the electric drift now shifts low-energy orbits into regions of larger $f_{\rm mdl}$, thus, increasing their weight according to Eq.~(\ref{eq:fM_shift}).

The reduction in the number density due to $E_r$ shifting orbits into regions of higher energy (and lower $f_{\rm mdl}$) should lead to a net increase in the effective temperature in that region. This is confirmed in Fig.~\ref{fig:11_jt60u_X-prof-tubes}(a) and (c), where we plot the profile and contours of the temperature field $T \equiv P/N$, which can be seen to develop a crescent peak on the LFS. This is presumably the consequence of the efficient energization of trapped orbits whose turning points are located near the peak of $E_r$. The following Section~\ref{sec:gc} contains more details on this and other effects of $E_r$ at the level of individual orbits.

\section{Electric modification of individual GC orbits}
\label{sec:gc}

In the previous Section~\ref{sec:example}, we examined the effect of $E_r$ on the moments of our orbit-based GC distributions. In this section, we examine how $E_r$ affects individual GC orbits from which those distributions were constructed. As a preparation, we begin in Sections~\ref{sec:gc_intro}--\ref{sec:gc_smooth} with some definitions and remarks concerning the distinction between local and orbit-based electric frequency shifts. In Sections~\ref{sec:gc_er0frame}--\ref{sec:gc_orb}, we analyze a few concrete test orbits and make comparisons with analytical estimates whose derivation can be found in \ref{apdx:gc_dr_du}. In Section~\ref{sec:gc_eaccel}, we discuss a problem that we call ``reference point bias'', which arises from the electric modulation of the mirror force in Eqs.~(\ref{eq:mdl_du_dt}) and (\ref{eq:gc_norm_u}) and requires some care when performing $E_r$-scans in the space of relative CoM denoted by ${\bm C}$.\footnote{The conservation laws associated with absolute CoM ($\overline{\bm C}$) are used in \protect\ref{apdx:gc} to derive approximate formulas against which we verify our numerical results here. Meanwhile, we do not presently have a recipe for analyzing the effects of $E_r$ in an intelligible way in terms of absolute CoM, where the reference point bias may be avoided.}
We close by discussing wave-particle resonances in Section~\ref{sec:gc_doppler}.

The scenario considered is based on JT-60U, whose parameters can be found in Table~\ref{tab:tok} and Fig.~\ref{fig:a02_jt60u_profs}. The deuteron density at the magnetic axis is taken to be $N_0 = 1.69\times 10^{19}\,{\rm m}^{-3}$, which gives the Alfv\'{e}n frequency $\omega_{\rm A0} = v_{\rm A0}/R_0 = B_0/(R_0\sqrt{\mu_0 M N_0}) = 2\pi\times 205.11\,{\rm kHz}$ that will be used to normalize time traces.

The initial values of the GC velocity $v_{\rm I}$, pitch angle $\alpha_{\rm I} \equiv \sin^{-1}(\uGC_{\rm I}/v_{\rm I})$ and position $X_{\rm I}$ on the $Y=0$ plane from which we launched the test orbits that are used in the following examples are summarized in Table~\ref{tab:gc_ex_orb}. The corresponding values of $\hat{\uGC}_{\rm I}$ and $\hat{\mu}$ are also shown.

\subsection{Preliminaries: Transit frequencies and $E_r$-induced local \& orbit-based frequency shifts}
\label{sec:gc_intro}

\begin{figure}
	[tb]\vspace{-2.4cm}
	\centering
	\includegraphics[width=0.48\textwidth]{\figures/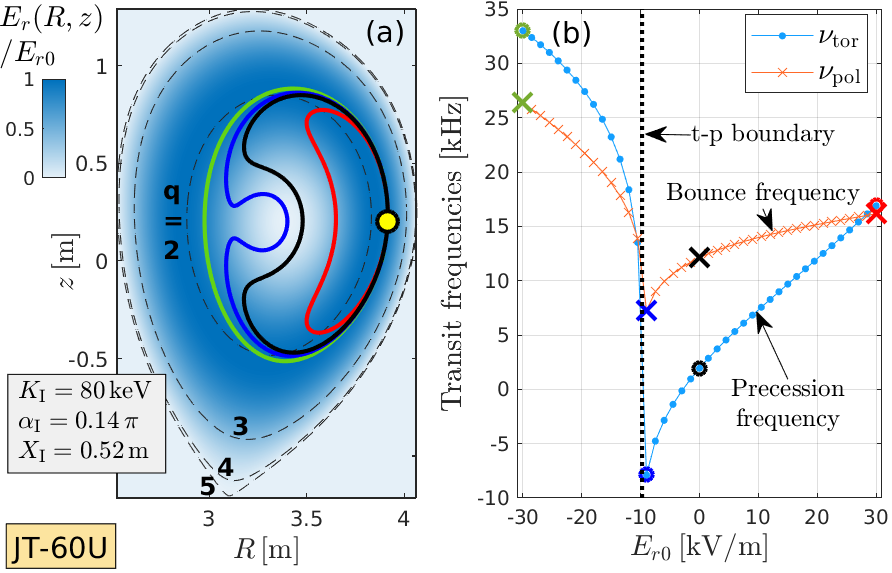}\\
	\includegraphics[width=0.48\textwidth]{\figures/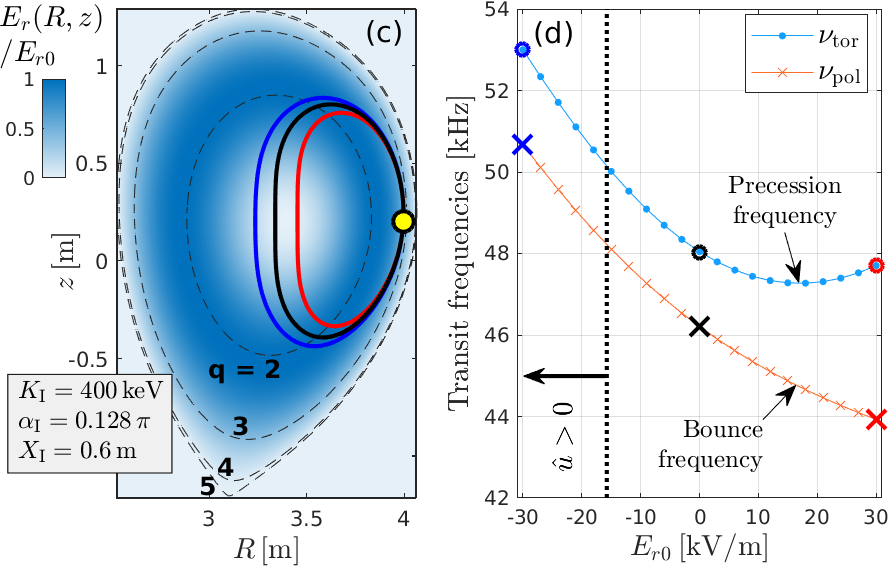}\vspace{-0.2cm}
	\caption{$E_r$-dependence of the orbit contours (left) and transit frequencies (right) at two given points in CoM space that correspond to barely trapped orbits when $E_{r0} = 0$ (black). Results are shown for fast deuterons in JT-60U, with initial kinetic energies $\ekin_{\rm I} = 80\,{\rm keV}$ (top) and $400\,{\rm keV}$ (bottom). Panels (a) and (c) show the poloidal cross-section of the plasma, with dashed black lines indicating the magnetic surfaces where the safety factor equals $q=2,3,4,5$. The blue-shaded contours show the structure of the $E_r$ field model, whose amplitude $E_{r0}$ is varied in our parameter scans. The yellow circle indicates the orbits' launch point near the LFS midplane. Solid red, black, green and blue contours show the form of the GC orbits at the values of $E_{r0}$ that are indicated by the equally-colored symbols in panels (b) and (d), where we plot the $E_r$-dependence of the toroidal (blue) and poloidal (orange) transit frequencies, $\nu_{\rm tor}$ and $\nu_{\rm pol}$, as defined in Eq.~(\protect\ref{eq:gc_transit}).}\vspace{-0.25cm}
	\label{fig:12_jt60u_Er-scan_trapped}%
\end{figure}

In the following analyses, it is important to bear in mind that a point ${\bm C}$ in the space of relative CoM represents different GC orbits with different contours $\{R_{\rm gc}(\tau|{\bm C},E_r),z_{\rm gc}(\tau|{\bm C},E_r)\}$ when $E_r$ is varied. This was already evident from the shift of the trapped-passing boundary in Figs.~\ref{fig:05_kstar_orbtypes} and \ref{fig:b01_jt60_orbtypes}, and is illustrated more clearly in Fig.~\ref{fig:12_jt60u_Er-scan_trapped} for fast deuterons ($80$ and $400\,{\rm keV}$) in a JT-60U plasma. In both examples we have chosen a certain reference point $\{{\bm C}_{\rm I},\tau_{\rm I}\}$ that serves as an initial position (``I'') for the orbit calculation. The orbit time coordinate $\tau \in \tau_{\rm I} + [0,\tau_{\rm pol})$ is periodic relative to the launch point $\tau_{\rm I}$, which is indicated by a yellow circle in Fig.~\ref{fig:12_jt60u_Er-scan_trapped}, here chosen to be the orbits' outer (LFS) midplane crossing.

\begin{table}[tbp]
	\centering\vspace{-2.4cm}
	\begin{tabular}{@{\hspace{0.0cm}}c@{\hspace{0.1cm}}|c|c|c|c|c@{\hspace{0.05cm}}}
		\hline\hline Orbit type & $\hat{v}^2_{\rm I}$ & $\alpha_{\rm I}/\pi$ & $\hat{u}_{\rm I}$ & $\hat{\mu}$ & $X_{\rm I}$ \\
		\hline Trapped & $6$ & $0.02$ & $0.15$ & $3.4$ & $0.52\,{\rm m}$ \\
		\footnotesize{($E_r = 0$ frame)} & \footnotesize{($7.07$)} & \footnotesize{($-0.076$)} & \footnotesize{($-0.63$)} & \footnotesize{($3.8$)} & \footnotesize{($0.52\,{\rm m}$)} \\
		\hline Passing & $6$ & $0.45$ & $2.4$ & $0.08$ & $0.52\,{\rm m}$ \\
		\hline\hline
	\end{tabular}\vspace{-0.15cm}
	\caption{GC orbit parameters for the test orbits analyzed in Figs.~\protect\ref{fig:13_jt60_deep-trapped} and \protect\ref{fig:14_jt60_deep-pass}. The pitch angle is defined as $\alpha \equiv \sin^{-1}(u/v)$. The initial kinetic energy $\ekin_{\rm I} = M v_{\rm I}^2/2 = T_0\hat{v}_{\rm I}^2 = 7.8\,{\rm keV}$ is six times the thermal energy $T_0$. That is also the value we used for $K_{\rm max}$ when sampling the Maxwellian-like distribution (\protect\ref{eq:fM}) in Section~\protect\ref{sec:example} and \protect\ref{apdx:benchmark}. Being not too small and not too large, this value facilitates comparisons of computed orbit widths with analytical estimates. For simplicity and ease of reproducibility, these test orbits were launched here from the geometric midplane ($Y_{\rm I}=0$), not the magnetic midplane.}
	\label{tab:gc_ex_orb}\vspace{-0.3cm}
\end{table}

Panels (a) and (c) show how the GC orbits (that is, the shapes of their contours) change when the equations of motion are solved for electric field strengths in the range $-30\,{\rm kV/m} \leq E_{r0} \leq 30\,{\rm kV/m}$. Panels (b) and (d) show the $E_{r0}$-dependence of the toroidal and poloidal transit frequencies,\vspace{-0.1cm}
\begin{subequations}
	\begin{align}
		\omega_{\rm tor} = 2\pi\nu_{\rm tor} \equiv&\; [\zeta(\tau_{\rm pol} + \tau_{\rm I}) - \zeta( \tau_{\rm I})]/\tau_{\rm pol},
		\label{eq:gc_transit_tor} \\
		\omega_{\rm pol} = 2\pi\nu_{\rm pol} \equiv&\; 2\pi/\tau_{\rm pol},
		\label{eq:gc_transit_pol}
	\end{align}
	\label{eq:gc_transit}\vspace{-0.5cm}
\end{subequations}

\noindent when measured at our chosen point $\{{\bm C}_{\rm I},\tau_{\rm I}\}$ in the lab frame. The orbit period $\tau_{\rm pol}$ is the time that a GC needs to return to its initial point in the poloidal $(R,z)$ plane, which for trapped orbits is known as the bounce time. In general, we refer to $\tau_{\rm pol}$ as the poloidal transit time, and its inverse $\nu_{\rm pol}$ in Eq.~(\ref{eq:gc_transit_pol}) is the poloidal transit frequency. $\zeta(\tau_{\rm pol} + \tau_{\rm I}) - \zeta(\tau_{\rm I})$ is the toroidal angular distance that the GC travels during the time interval $\tau_{\rm pol}$, and their ratio in Eq.~(\ref{eq:gc_transit_tor}) divided by $2\pi$ gives the toroidal transit frequency $\nu_{\rm tor}$, which for trapped orbits is known as the (toroidal) precession frequency.

At $80\,{\rm keV}$, Fig.~\ref{fig:12_jt60u_Er-scan_trapped}(b) shows a transition from co-passing to trapped orbits near $E_{r0} \approx -10\,{\rm kV/m}$ in the form of a discontinuity in both transit frequencies. The orbit contours for four values of $E_{r0}$ are shown in corresponding colors in panel (a).

At $400\,{\rm keV}$, which was the maximal energy of the negative-ion-based neutral beams in JT-60U, our initial point in Fig.~\ref{fig:12_jt60u_Er-scan_trapped}(c) lies in the domain of potato orbits, which encircle the magnetic axis with alternating sign of $\uGC(\tau)$. Increasing $E_{r0}$ to positive values yields fat banana (or bean-shaped) orbits. Reducing $E_{r0}$ below $-16\,{\rm kV/m}$ yields co-passing orbits via a transition that appears perfectly smooth in Fig.~\ref{fig:12_jt60u_Er-scan_trapped}(d), in stark contrast to \ref{fig:12_jt60u_Er-scan_trapped}(b).

Parameter scans like those shown in Fig.~\ref{fig:12_jt60u_Er-scan_trapped} tell us the variation of the transit/bounce/precession frequency of GC orbits found at a certain reference point $\{{\bm C}_{\rm I},\tau_{\rm I}\}$. Let us denote such {\it local} frequency shifts in the lab frame as\vspace{-0.1cm}
\begin{equation}
	\Delta\omega_{\rm E}^{\rm (I)} \equiv \omega(E_r|{\bm C}_{\rm I},\tau_{\rm I}) - \omega(0|{\bm C}_{\rm I},\tau_{\rm I}),
	\label{eq:gc_shift_loc}\vspace{-0.1cm}
\end{equation}

\noindent where we omitted the subscripts ``tor'' and ``pol'' identifying toroidal and poloidal components. The superscript (I) in Eq.~(\ref{eq:gc_shift_loc}) is meant to indicate that both frequencies (with and without electric field) are computed by launching the tracer particle always from the same initial position $\{{\bm C}_{\rm I},\tau_{\rm I}\}$ while varying $E_r$.

The additional orbit time argument $\tau_{\rm I}$ is necessary because we will find that the parallel acceleration term $\propto E_\nabla = \varrho_\parallel {\bm v}_{\bm E}\cdot\nabla B$ in Eq.~(\ref{eq:mdl_du_dt}) introduces a reference point bias in the sense that the electric shifts of the transit frequencies depend on the chosen launch point $\{R_{\rm I},z_{\rm I}\} \equiv \{R_{\rm gc}(\tau_{\rm I}),z_{\rm gc}(\tau_{\rm I})\}$ in the poloidal plane, where the orbit calculation is initialized. In particular, we will show in Section~\ref{sec:gc_eaccel} that this is always the case for the toroidal transit frequency of passing orbits, $\Delta\omega_{\rm tor,E,pass}^{\rm (I)}$.

In some applications, such as resonance analyses, we are interested in the electric frequency shift for a ``specific orbit'' that has a certain spatial contour. Such an {\it orbit-based} frequency shift can be interpreted as an electric ``Doppler shift'' and is denoted here as\vspace{-0.05cm}
\begin{equation}
	\Delta\omega_{\rm E}^{\rm (0)} \equiv \omega(E_r|{\bm C}_{\rm I}) - \omega(0|{\bm C}_{\rm I}^{(0)}).
	\label{eq:gc_shift_orb}\vspace{-0.1cm}
\end{equation}

\noindent The superscript (0) means that a quantity is evaluated in an idealized ``$E_r = 0$ frame of reference''. For instance, if it can be shown that the electric Doppler shift $\Delta\omega_{\rm E}^{\rm (0)}$ of potentially resonant orbits is similar to the rotation frequency of the wave-carrying background plasma, then the resonance analysis --- where we match candidate modes with candidate GC orbits --- could be performed in the absence of $E_r$. This simplified treatment can facilitate quick and inexpensive surveys (perhaps on the `back of an envelope') of potential resonances before deploying heavy computational machinery.

However, the concept of an ``$E_r = 0$ frame'' is an idealization. Its validity is limited by the fact that Eq.~(\ref{eq:gc_shift_orb}) presupposes the existence of identically shaped ``dual'' orbits ${\bm C}_{\rm I}$ and ${\bm C}_{\rm I}^{\rm (0)}$ that can be treated as a unique object whose identity does not depend on the value of $E_r$. However, this is not guaranteed. For instance, in the extreme case where $E_r$ varies significantly along the orbit contour, one obtains distorted orbits \cite{Hinton95} that do not exist when $E_r = 0$. In such cases, the concept of an orbit-based ``Doppler shift'' loses its meaning because one can no longer attribute the original $\omega(0|{\bm C}_{\rm I}^{(0)})$ and shifted frequency $\omega(E_r|{\bm C}_{\rm I})$ to a ``specific orbit''.

Thus, in general, wave-particle resonances must be evaluated in the lab frame, where $E_r$ rotates the wave-carrying background plasma (in addition to diamagnetic rotation), while also modifying the orbits of the resonant particle species. When $E_r$ is so nonuniform that it varies significantly on the resonant particles' gyroradius scale, theses simulations should be done using full Lorentz orbits rather than guiding centers or gyrocenters.

\subsection{Simplification for cases with weakly varying $E_r$}
\label{sec:gc_smooth}

The $E_r$ model field in Fig.~\ref{fig:03_kstar_E30_model} that we use in the present study is sufficiently smooth to be approximately constant along the GC drift orbits of ions in the thermal range of energies, $K \lesssim \O(10\,{\rm keV})$, so we may test the applicability of Eq.~(\ref{eq:gc_shift_orb}). For this purpose, let us turn that formula (\ref{eq:gc_shift_orb}) into more practical forms that are tailored to the specific properties of trapped and passing orbits that are located in regions where $E_r \approx {\rm const}$.

Well-trapped orbits with different coordinates ${\bm C}_{\rm I}$ have similar precessional frequency shifts and (by definition) the bounce frequency for dual orbits does not change significantly either. Eq.~(\ref{eq:gc_shift_orb}) can then be reduced to\vspace{-0.1cm}
\begin{subequations}
	\begin{align}
		\Delta\omega_{\rm trap,E,tor}^{\rm (0)} \approx&\; \Delta\omega_{\rm trap,E,tor}^{\rm (I)} \quad \text{for } E_r \approx {\rm const.},
		\label{eq:gc_shift_precess}
		\\
		\Delta\omega_{\rm trap,E,pol}^{\rm (0)} \approx&\; 0 \quad \text{for dual trapped orbits}.
		\label{eq:gc_shift_bounce}
	\end{align}
	\label{eq:gc_shift_trap}
\end{subequations}\vspace{-0.5cm}

\noindent Although the two quantities in Eq.~(\ref{eq:gc_shift_precess}) are still conceptually different, their quantitative similarity means that one can directly use the local frequency shift $\Delta\omega_{\rm E}^{\rm (I)}$ to find the ``original'' coordinates ${\bm C}_{\rm I}^{(0)} = \{\ekin^{(0)}_{\rm I},\alpha^{(0)}_{\rm I},X^{(0)}_{\rm I}\}$ of a ``specific orbit'' in the ``$E_r=0$ frame'', which then allows to compute the orbit-based frequency shift $\Delta\omega_{\rm E}^{\rm (0)}$. An example is presented in Section~\ref{sec:gc_er0frame} below.

For passing orbits, the poloidal frequency shift is simply the orbit-averaged ${\bm E}\times{\bm B}$ frequency, whereas a little more thought is required for the toroidal component. As will be shown in Section~\ref{sec:gc_eaccel}, $\Delta\omega_{\rm pass,E,tor}^{\rm (I)}(E_r|{\bm C}_{\rm I},\tau_{\rm I})$ in Eq.~(\ref{eq:gc_shift_loc}) is prone to a reference point bias, which required us to specify the launch point $\tau_{\rm I}$ in its argument. This bias should disappear as follows. With the possible exception of local interactions in connection with (enhanced) prompt losses \cite{ZhangRB15}, wave-particle interactions whose analysis is the purpose of Eq.~(\ref{eq:gc_shift_orb}) can usually be thought of as involving all particles that populate the resonant orbits. To be precise, the net energy transfer is determined by the rate at which collective density modulations pass through the wave, so that only the mean orbit transit frequency matters, not the locally varying speed of individual particles. Consequently, we should sum over all possible launch points $\{R_{\rm I},z_{\rm I}\} \equiv \{R_{\rm gc}(\tau_{\rm I}),z_{\rm gc}(\tau_{\rm I})\}$, and this is precisely what is done in Step 4 of our workflow in Fig.~\ref{fig:02_vstart}, where marker particles are loaded around the orbit contour at uniform time intervals $\Delta\tau$. Since the reference point bias originates from the periodic electric acceleration and deceleration via the $E_\nabla$ term in Eq.~(\ref{eq:mdl_du_dt}), whose $\tau_{\rm pol}$-average is zero, the biases end up canceling, so that the orbit-based frequency shift $\Delta\omega_{\rm pass,E,tor}^{\rm (0)}$ vanishes.\footnote{It is worth noting that the result $\omega_{\rm pass,E,tor}^{\rm (0)} \approx 0$ in Eq.~(\protect\ref{eq:gc_shift_tor}) does not necessarily require summation over many launch points $\tau_{\rm I}$. It suffices to add the frequency shifts at the two points where ${\bm B}\cdot\nabla B$ vanishes; that is, at the orbit's midplane crossings defined in Section~\protect\ref{sec:method_mid}. We will find later (Fig.~\protect\ref{fig:15_jt60u_u1pass_uEB0}) that the inner (HFS) and outer (LFS) midplane crossings yield $\Delta\omega_{\rm pass,E,tor}(E_r|{\bm C},\tau_{\rm hfs}) + \Delta\omega_{\rm pass,E,tor}(E_r|{\bm C},\tau_{\rm lfs}) \approx 0$. Here, the ``(I)'' superscript is omitted because $\Delta\omega_{\rm pass,E,tor}$ is evaluated at two {\it different} points of the same orbit.}
Thus, the orbit-based frequency shifts for passing orbits are\vspace{-0.1cm}
\begin{subequations}
	\begin{align}
		\Delta\omega_{\rm pass,E,tor}^{\rm (0)} \approx&\; 0 \quad \text{from sum over launch points},
		\label{eq:gc_shift_tor}
		\\
		\Delta\omega_{\rm pass,E,pol}^{\rm (0)} \approx&\; \Delta\omega_{\rm pass,E,pol}^{\rm (I)} \approx \overline{v_{\rm E,pol}/r} \equiv \frac{1}{\tau_{\rm pol}} \oint{\rm d}\tau \frac{v_{\rm E,pol}}{r}.
		\label{eq:gc_shift_pol}
	\end{align}
	\label{eq:gc_shift_pass}
\end{subequations}\vspace{-0.3cm}

\noindent Let us now proceed to some concrete examples.

\begin{figure*}
	[tb]
	\centering\vspace{-2.6cm}
	\includegraphics[width=0.96\textwidth]{\figures/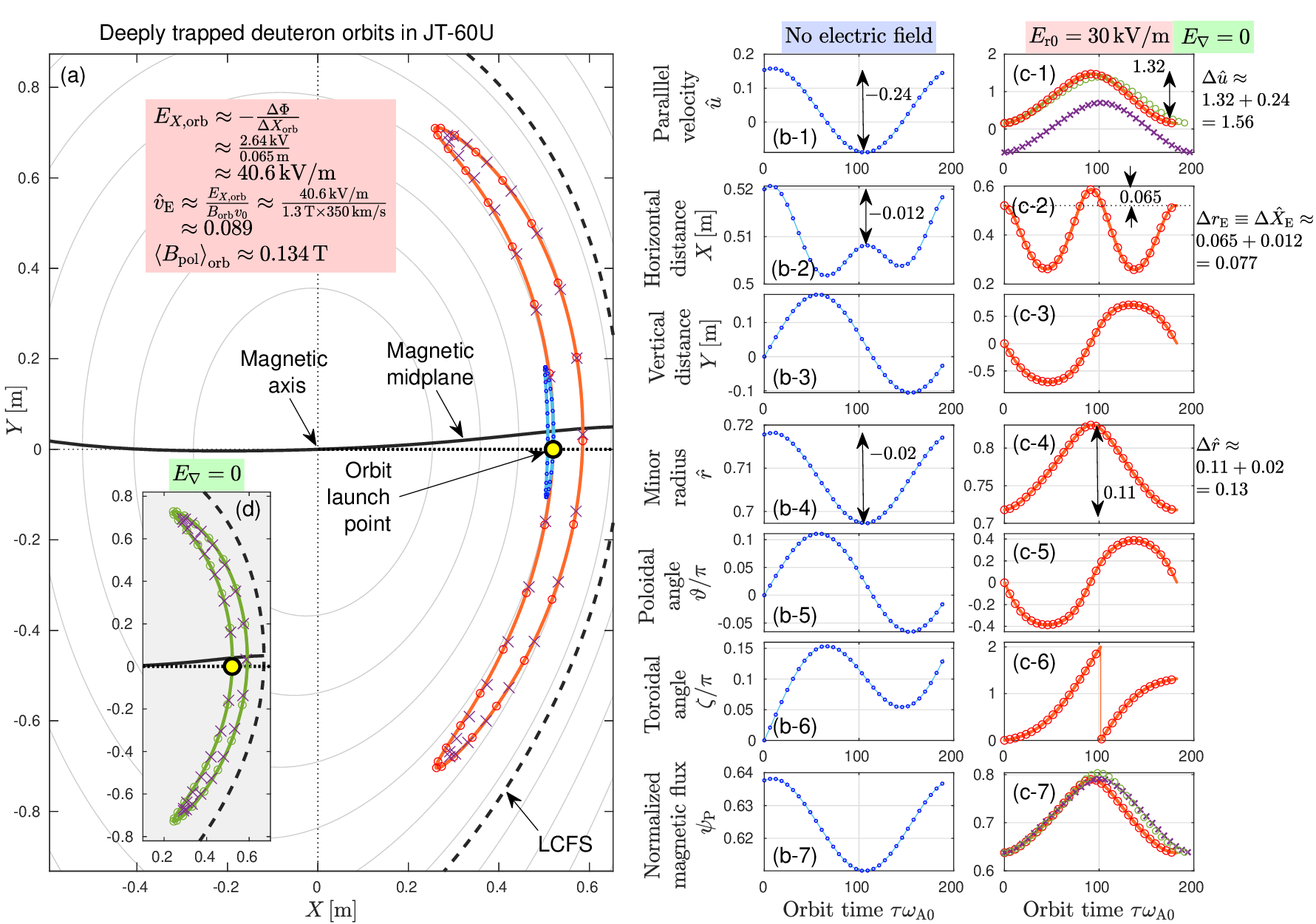}\vspace{-0.25cm}
	\caption{Deeply trapped deuteron orbits in JT-60U. Using the initial conditions in Table~\protect\ref{tab:gc_ex_orb}, we obtained the blue orbit with $E_{r0} = 0$ and the red orbit with $E_{r0} = 30\,{\rm kV/m}$. The violet orbit in panels (c-1), (c-7) and (d) is our estimated rotating-frame counterpart of the red lab-frame orbit. Ignoring the parallel acceleration term $\propto E_\nabla$ in Eq.~(\protect\ref{eq:mdl_du_dt}) yields the green orbit. In contrast to the procedure in Section~\protect\ref{sec:method}, where we constructed databases of orbits launched from the magnetic midplane to model GC distributions, the present test orbits were launched from the yellow circle in panel (a), which lies on the geometric midplane ($Y=0$, dotted black line) for simplicity and easier reproducibility. Panel (a) shows the orbit contours in the poloidal $(X,Y)$ plane in comparison with magnetic flux contours (gray). The symbols are separated by constant time intervals. Rows (b) and (c) show time traces of several GC phase space coordinates. The poloidal transit time of these orbits is about $\tau_{\rm pol} \approx 200\omega_{\rm A0}^{-1} \approx 0.15\,{\rm ms}$.}
	\label{fig:13_jt60_deep-trapped}\vspace{-0.35cm}
\end{figure*}

\subsection{Finding a dual orbit in the ``$E_r=0$ frame''}
\label{sec:gc_er0frame}

The poloidal contour $\{R_{\rm gc}(\tau|{\bm C},E_r),z_{\rm gc}(\tau|{\bm C},E_r)\}$ of our deeply trapped test orbit in JT-60U with $E_{r0} = 0$ is shown in blue color in Fig.~\ref{fig:13_jt60_deep-trapped}. The relatively low kinetic energy $K_{\rm I} = 7.8\,{\rm keV}$ and the small positive value $\hat{\uGC}_{\rm I} \approx 0.15$ of the initial parallel velocity imply that the starting point lies on the outer leg of a narrow, short banana orbit that is localized near the outer midplane.

When we apply a radial electric field with strength $E_{r0} = 30\,{\rm kV/m}$, the same initial condition yields an entirely different orbit, whose trajectory is shown in red color in Fig.~\ref{fig:13_jt60_deep-trapped}. In this case, the $z$-component of ${\bm v}_{\rm E} = {\bm E}\times{\bm B}/B^2$ is negative and larger than the positive $z$-component of $\hat{\bm u}_{\rm I} \equiv \hat{\uGC}_{\rm I}{\bm B}/B$. Thus, the initial GC motion is now downward in $z$, which means that the starting point is now located on the inner leg of a banana orbit. One can also see that the red banana orbit for $E_{r0} = 30\,{\rm kV/m}$ is much longer and much wider than the blue banana orbit that was obtained for $E_{r0} = 0$.

With this, we have an example of a case where the positive radial electric field has shifted the trapped particle domain towards positive pitch $\uGC/v$ to such an extent that our launch point $\{v_{\rm I},\uGC_{\rm I},X_{\rm I}\}$, which lay in the positive-pitch domain for $E_r=0$, is now effectively situated in what used to be the negative-pitch domain.

In the time trace of $\hat{\uGC}(\tau)$ in Fig.~\ref{fig:13_jt60_deep-trapped}(c-1), the mirror-trapped nature of the red orbit is obscured by its rapid electric toroidal precession. Here, the parallel velocity $\hat{\uGC}$ varies between two positive values, $0.15 \lesssim \hat{u} \lesssim 1.68$, whose difference is denoted by $\Delta\hat{\uGC}_{\rm orb} \approx 1.53$ as defined in Eq.~(\ref{eq:gc_du2_du_orb}) of \ref{apdx:gc_dr_du}. From Fig.~\ref{fig:13_jt60_deep-trapped}(b-1), we infer that the magnetic contribution due to the mirror force is $\Delta\hat{\uGC}_{\rm orb,M} \approx -0.24$, so the electric component is\footnote{A caveat of this estimate is discussed at the end of \protect\ref{apdx:gc_dr_du}.}
\begin{equation}
	\Delta\hat{\uGC}_{\rm orb,E} = \Delta\hat{\uGC}_{\rm orb} - \Delta\hat{\uGC}_{\rm orb,M} \approx 1.32 - (-0.24) = 1.56.
\end{equation}

\begin{figure*}
	[tb]
	\centering\vspace{-2.6cm}
	\includegraphics[width=0.96\textwidth]{\figures/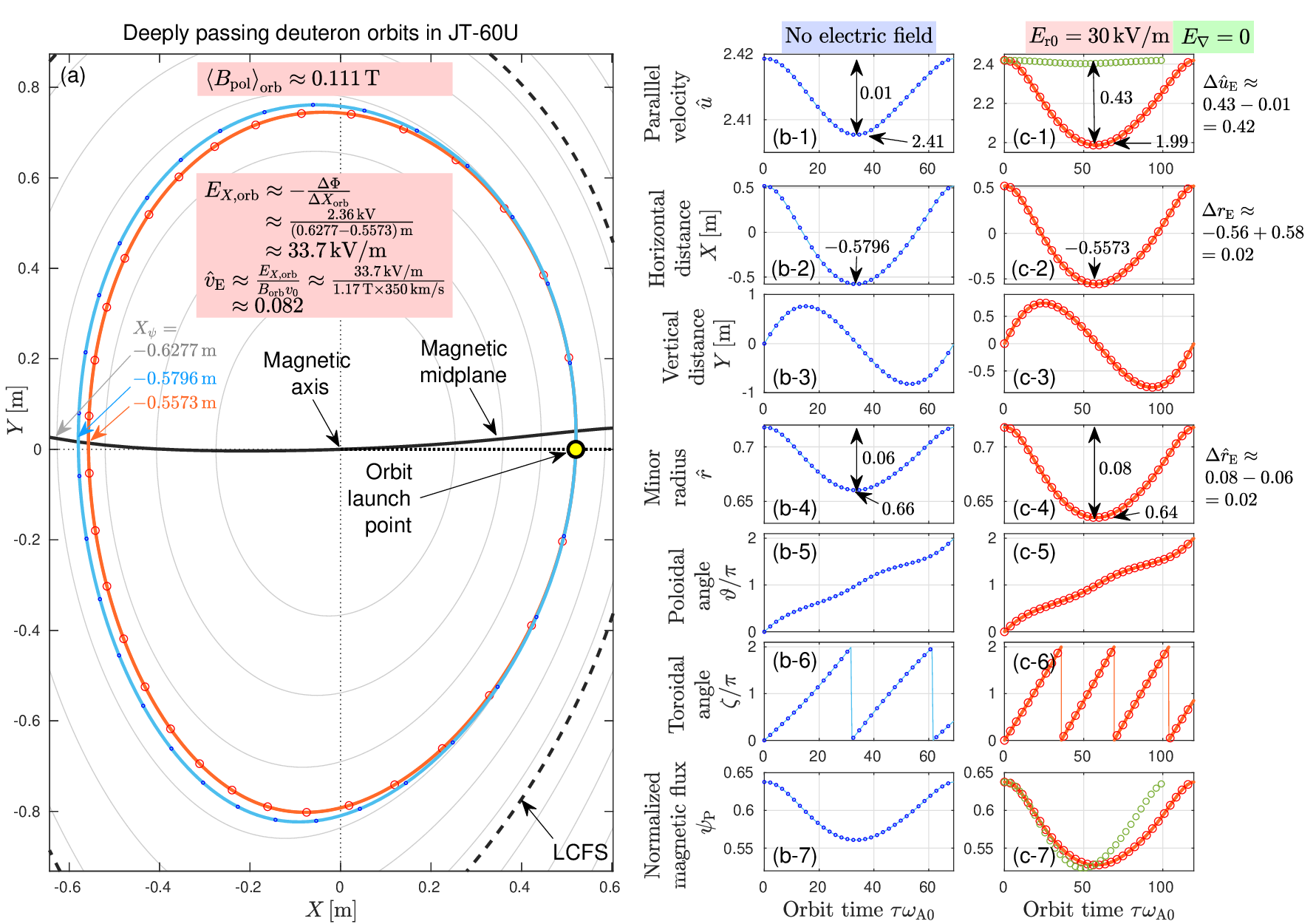}\vspace{-0.25cm}
	\caption{Deeply passing deuteron orbits in JT-60U with $E_{r0} = 0$ (blue) and $30\,{\rm kV/m}$ (red), with launch points from Table~\protect\ref{tab:gc_ex_orb}. Arranged as Fig.~\protect\ref{fig:13_jt60_deep-trapped}.}
	\label{fig:14_jt60_deep-pass}\vspace{-0.35cm}
\end{figure*}

\noindent Since our $E_r$ field is nearly constant in the domain of these orbits, we can take advantage of Eq.~(\ref{eq:gc_shift_precess}) and estimate the parallel precessional Doppler shift using Eq.~(\ref{eq:gc_OmE}) of \ref{apdx:gc_dw}:\vspace{-0.1cm}
\begin{equation}
	\Delta\omega_{u{\rm E}}^{\rm (0)} R_0 \stackrel{(\ref{eq:gc_shift_precess})}{\approx} \Delta\omega_{u{\rm E}}^{\rm (I)} R_0 \stackrel{(\ref{eq:gc_OmE})}{\approx} \frac{\Delta\hat{\uGC}_{\rm orb,E}}{2} \sim \frac{1.56}{2} = 0.78.
	\label{eq:gc_doppler0}\vspace{-0.1cm}
\end{equation}

\noindent While this estimate of the parallel velocity offset,
\begin{equation}
	\Delta \uGC_{\rm I}^{\rm (0)} \equiv \hat{\uGC}_{\rm I} - \hat{\uGC}_{\rm I}^{\rm (0)} \approx \Delta\hat{\uGC}_{\rm orb,E}/2,
\end{equation}

\noindent is based on the direct measurement of the modulation of the numerically computed parallel velocity in Fig.~\ref{fig:13_jt60_deep-trapped}, the same result can be obtained from the analytically derived Eq.~(\ref{eq:gc_du0_trap}), which can be written as\vspace{-0.cm}
\begin{align}
	\Delta\hat{\uGC}_{\rm I}^{\rm (0)} \approx&\; \hat{v}_{\rm E}\frac{B_0}{B_{\rm pol}} \left(1 + \frac{\Delta\hat{\mu}^{\rm (0)}}{\hat{v}_{\rm E}}\frac{\varrho_0}{R_0}\right)
	\label{eq:gc_doppler0_bench}
	\\
	\approx&\; 0.089 \times \frac{1.17}{0.134} \left(1 +  \frac{\Delta\hat{\mu}^{\rm (0)}}{0.089}\frac{0.006}{3.4}\right) \nonumber \\
	\approx&\; 0.78 \times (1 + 0.02\times\Delta\hat{\mu}^{\rm (0)}) \approx 0.78. \nonumber
\end{align}\vspace{-0.5cm}

\noindent where the $\Delta\hat{\mu}^{\rm (0)}$ term is ignorably small. Here, the values $\hat{v}_{\rm E} \approx 0.089$ and $B_{\rm pol} \approx 0.134\,{\rm T}$ were computed by orbit-averaging the radial electric field in Fig.~\ref{fig:03_kstar_E30_model}(a) and the poloidal magnetic field in Fig.~\ref{fig:a02_jt60u_profs}(a). The agreement between the results in Eqs.~(\ref{eq:gc_doppler0}) and (\ref{eq:gc_doppler0_bench}) confirms the assumption underlying Eq.~(\ref{eq:gc_shift_precess}) that the local frequency shift at a certain point ${\bm C}$ in the lab frame can be used to estimate the Doppler shift of a certain orbit.

With this, we are now able to determine the coordinates ${\bm C}^{\rm (0)}$ of our trapped test orbit in the rotating frame.

The corresponding total velocity $\hat{v}^{(0)}_{\rm I}$ and pitch angle $\alpha^{(0)}_{\rm I} = {\rm asin}(\hat{\uGC}^{(0)}_{\rm I}/v^{(0)}_{\rm I})$ in the ``$E_r = 0$ frame'' can be reconstructed as follows. Using Eqs.~(\ref{eq:u_transform}) and (\ref{eq:gc_OmE}) of \ref{apdx:gc_dr_du}, we find the parallel velocity by subtracting the above Doppler shift:
\begin{equation}
	\hat{\uGC}^{(0)}_{\rm I} \approx \hat{\uGC}_{\rm I} - \Delta\hat{\uGC}_{\rm orb,E}/2 \sim 0.15 - 0.78 = -0.63.
\end{equation}

\noindent In order to estimate the total velocity $\hat{v}^{(0)}_{\rm I}$, we recall that $\ekin = M v^2/2$ is conserved on the blue orbit while $\etot = \ekin + Ze\Phi$ is conserved on the red orbit. We estimate $v^{(0)}_{\rm I}$ from the mean value of the former, that is $\overline{\ekin} \approx \etot - Ze\Delta\Phi_{\rm orb}/2$. The potential difference in Fig.~\ref{fig:03_kstar_E30_model}(b) across the orbit width $\psi_{\rm out} - \psi_{\rm in} \approx 0.814-0.638 \approx 0.18$ in Fig.~\ref{fig:13_jt60_deep-trapped}(c-7) is approximately $\Delta\Phi_{\rm orb} \approx -2.64\,{\rm kV}$, so that
\begin{align}
	|\hat{v}^{(0)}_{\rm I}|^2 \approx&\; \hat{v}_{\rm I}^2 - \Delta\Phi/(\varrho_0 B_0 v_0) \nonumber \\
	=&\; 2\times 3 + \frac{2.64\,{\rm kV}}{0.006\,{\rm m}\times 1.17\,{\rm T}\times 350\,{\rm km/s}} \nonumber \\
	\approx&\; 6 + 1.07 = 7.07.
\end{align}
	
\noindent The ``$E_r=0$ frame'' pitch angle is then
\begin{equation}
	\alpha^{(0)}_{\rm I} \approx {\rm asin}(-0.63/\sqrt{7.07}) \approx -0.076\pi.
	\label{eq:gc_alpha0}
\end{equation}

\noindent The GC orbit obtained with these parameters in the rotating frame is indicated by violet crosses in Fig.~\ref{fig:13_jt60_deep-trapped}(a), and the corresponding time traces of $\hat{\uGC}(\tau)$ and $\psi_{\rm P}(\tau)$ are shown in panels (c-1) and (c-7). The offset in $\hat{\uGC}(\tau)$ corresponds to the Doppler shift $0.78$ in Eqs.~(\ref{eq:gc_doppler0}) and (\ref{eq:gc_doppler0_bench}).

One can see that our estimated violet rotating-frame orbit is similar but not quite identical to the red lab-frame orbit. The violet ``banana'' in Fig.~\ref{fig:13_jt60_deep-trapped}(a) is slightly shorter and the transit time in column (c) is about $11\%$ longer: $\tau_{\rm pol}^{(0)}\omega_{\rm A0} = 202.1$ instead of $182.0$. The orbit contour in panel (a) is very sensitive to the pitch angle: using $\alpha^{(0)}_{\rm I} \approx -0.08\pi$ instead of $-0.076\pi$ in Eq.~(\ref{eq:gc_alpha0}) would give an almost perfectly matching orbit contour.

The discrepancy in the poloidal transit time is more robust with respect to the orbit coordinates. However, it turns out that this difference can be largely eliminated if the lab-frame orbit is computed without the parallel electric acceleration term $\propto E_\nabla = \varrho_\parallel {\bm v}_{\bm E}\cdot\nabla B$ in Eq.~(\ref{eq:mdl_du_dt}). The resulting time traces $\hat{\uGC}(\tau)$ and $\psi_{\rm P}(\tau)$ are plotted in panels (c-1) and (c-7) of Fig.~\ref{fig:13_jt60_deep-trapped} using green circles. We find that the neglect of $E_\nabla$ increases the poloidal transit time to $\tau_{\rm pol}\omega_{\rm A0} = 196.4$, which is only 3\% (instead of 11\%) shorter than our estimate $\tau_{\rm pol}^{(0)}\omega_{\rm A0} = 202.1$ for the rotating-frame orbit. The orbit contours are compared in panel (d), noting again that a closer match could be obtained with a slightly different pitch angle.

This indicates that an exact one-to-one correspondence (duality) between GC orbits in the presence and absence of $E_r$ is prevented not only by the nonuniformity of $E_r$ that we discussed n Section~\ref{sec:gc_intro} but also by the periodic parallel acceleration and deceleration via $E_\nablab$ (whose omission would violate the model's Hamiltonian character). Thus, an {\it inertial} ``$E_r = 0$ frame'' does not strictly exist at the level of individual GC orbits.

\subsection{Further verification against analytical estimates}
\label{sec:gc_orb}

For thermal trapped orbits ($\hat{u}\varrho_0/R_0 \ll B_{\rm pol}/B_0$) and sufficiently deeply passing orbits ($\hat{\mu} \ll \hat{u}_{\rm I}^2$), our formulas (\ref{eq:gc_dr_trap}) and (\ref{eq:gc_dr_pass}) for estimating the drift orbit width $\Delta r_{\rm orb}$ and the parallel velocity modulation $\Delta\hat{\uGC}_{\rm orb,E}$ can be summarized as
\begin{subequations}
	\begin{align}
		\Delta r_{\rm orb} \approx&\; \left\{
		\begin{array}{l@{\hspace{0.1cm}}l}
			2\varrho_0 \frac{B_0^2}{B_{\rm pol}^2} \left(\hat{v}_{\rm E} + \hat{\mu} \frac{\varrho_0}{R_0} - \hat{\uGC}_{\rm I}\frac{B_{\rm pol}}{B_0}\right) & : \text{trapped}, \\
			2\varrho_0 \frac{\overline{r}}{R_0} \frac{B_0^2}{B_{\rm pol}^2} \left(\hat{v}_{\rm E} + \hat{\overline{\uGC}}\frac{B_{\rm pol}}{B_0}\right) & : \text{passing}
		\end{array}\right. \nonumber \\
		\approx&\, \Delta r_{\rm orb,E} + \Delta r_{\rm orb,M},
		\label{eq:gc_examples_dr}
		\\
		\Delta\hat{\uGC}_{\rm orb,E} \approx&\; 2R_0\Delta\omega_{u,{\rm E}}^{\rm (I)} \approx 2\hat{v}_{\rm E} \times \left\{
		\begin{array}{l@{\hspace{0.1cm}}l}
			\frac{B_{\rm pol}}{B_0} & : \text{trapped}, \\
			\frac{\overline{r} B_{\rm pol}}{R_0 B_0} \approx q & : \text{passing},
		\end{array}\right.
		\label{eq:gc_examples_duE}
	\end{align}
	\label{eq:gc_examples}
\end{subequations}\vspace{-0.3cm}

\noindent where the subscripts ``E'' and ``M'' identify electric and magnetic drift contributions. We emphasize that ``$r$'' is measured along the $X$ axis (approximately the midplane) as defined in Eq.~(\ref{eq:gc_Rr}), so its metric is ${\rm d}r \approx {\rm d}X$ and differs from that of the volume-averaged minor radius $\hat{r} \equiv r/a$ that is plotted in panels (b-4) and (c-4) of Figs.~\ref{fig:13_jt60_deep-trapped} and \ref{fig:14_jt60_deep-pass}. $\overline{r}$ and $\overline{\uGC}$ denote the mean radius and mean parallel velocity as defined in Eqs.~(\ref{eq:gc_rorb}) and (\ref{eq:gc_umean}). See \ref{apdx:gc_dr_du} for details.

The consistency between Eq.~(\ref{eq:gc_examples}) and our estimated ``$E_r=0$ frame'' coordinates from Section~\ref{sec:gc_er0frame} has already been verified in Eq.~(\ref{eq:gc_doppler0_bench}), because that equation follows from Eq.~(\ref{eq:gc_examples}) under the assumption that the dual lab-frame and rotating-frame orbits that are being compared have ``identical'' contours, so they must have the same $\Delta r_{\rm orb}$.

Next, let us compare the computed orbit width and velocity shifts with analytical estimates. Substituting the parameters from Table~\ref{tab:gc_ex_orb} for the trapped orbit into Eq.~(\ref{eq:gc_examples}), we obtain the estimates
\begin{subequations}
	\begin{align}
		\Delta r_{\rm trap} \approx&\; \frac{2\varrho_0}{B_{\rm pol}^2/B_0^2} \left(\hat{v}_{\rm E} + \hat{\mu} \frac{\varrho_0}{R_0} - \hat{\uGC}_{\rm I} \frac{B_{\rm pol}}{B_0}\right) \nonumber \\
		\approx&\; \frac{2\times 0.006\,{\rm m}}{0.12^2} \times\left(0.089 + 3.4\times\frac{0.006\,{\rm m}}{3.4\,{\rm m}}\right. \nonumber \\
		&\qquad\qquad\qquad\qquad\;\, - 0.15\times 0.12) \nonumber \\
		\approx&\; 0.83\,{\rm m}\times(0.089 + 0.006 - 0.018) \nonumber \\
		\approx&\; 0.074\,{\rm m}\;\text{(electric}) - 0.010\,{\rm m}\;\text{(magnetic}) \nonumber \\
		\approx&\;0.064\,{\rm m}\; \text{(total)},
		\label{eq:gc_examples_trap_dr} \\
		\Delta \hat{\uGC}_{\rm trap,E} \approx&\;\frac{B_{\rm pol}}{B_0} \frac{\Delta r_{\rm trap,E}}{\varrho_0} \approx 0.12\times\frac{0.074\,{\rm m}}{0.006\,{\rm m}} \nonumber \\
		\approx&\; 1.48.
		\label{eq:gc_examples_trap_du}
	\end{align}
	\label{eq:gc_examples_trap}\vspace{-0.5cm}
\end{subequations}

\noindent The above analytical estimates are in reasonable agreement with the computed values in Fig.~\ref{fig:13_jt60_deep-trapped}(c-1), \ref{fig:13_jt60_deep-trapped}(b-2) and \ref{fig:13_jt60_deep-trapped}(c-2). The magnetic banana orbit width $\Delta r_{\rm trap,M} \approx -0.012\,{\rm m}$ is underestimated in Eq.~(\ref{eq:gc_examples_trap_dr}) by $17\%$. The total banana orbit width $\Delta r_{\rm trap} \approx 0.065\,{\rm m}$ is underestimated in Eq.~(\ref{eq:gc_examples_trap_dr}) by $2\%$. The maximal electric precession offset $|\Delta \hat{\uGC}_{\rm trap,E}| \approx 1.56$ is underestimated in Eq.~(\ref{eq:gc_examples_trap_du}) by $5\%$.

For the passing orbit, Eq.~(\ref{eq:gc_examples}) yields the estimates
\begin{subequations}
	\begin{align}
		\Delta r_{\rm pass} \approx&\; \frac{2\varrho_0}{B_{\rm pol}/B_0} \left(\frac{\hat{v}_{\rm E}}{B_{\rm pol}/B_0} + \hat{\overline{\uGC}}\right)\frac{\overline{r}}{R_0} \nonumber \\
		\approx&\; \frac{2\times 0.006\,{\rm m}}{0.12} \times \left(\frac{0.082}{0.12} + 2.4\right)\times \frac{0.54\,{\rm m}}{3.4\,{\rm m}} \nonumber \\
		\approx&\; 0.016\,{\rm m}\times(0.68 + 2.4) \nonumber \\
		\approx&\; 0.011\,{\rm m}\;\text{(electric}) + 0.038\,{\rm m}\;\text{(magnetic}) \nonumber \\
		=&\; 0.049\,{\rm m}\; \text{(total)},
		\label{eq:gc_examples_pass_dr} \\
		\Delta\hat{\uGC}_{\rm pass,E} \approx&\; \frac{B_{\rm pol}}{B_0} \frac{\Delta r_{\rm pass,E}}{\varrho_0} \approx 0.12\times\frac{0.011\,{\rm m}}{0.006\,{\rm m}} \nonumber \\
		\approx&\; 0.22.
		\label{eq:gc_examples_pass_du}
	\end{align}
	\label{eq:gc_examples_pass}\vspace{-0.5cm}
\end{subequations}

\noindent Due to the large Shafranov shift, Eq.~(\ref{eq:gc_rorb}) cannot be used to measure the radial shift in the computed data. Instead, we measure the radial shift relative to the magnetic flux surface $\psi_{\rm I}$ where the orbit was launched. On the HFS of Fig.~\ref{fig:14_jt60_deep-pass}(a), that flux surface crosses the midplane at $X_\psi \approx -0.6277\,{\rm m}$ as indicated by the gray arrow. The blue orbit ($E_r=0$) and red orbit ($30\,{\rm kV/m}$) cross the midplane at $-0.5796\,{\rm m}$ and $-0.5573\,{\rm m}$, which gives $\Delta r_{\rm pass,M} = 0.048\,{\rm m}$ and $\Delta r_{\rm pass} = 0.070\,{\rm m}$, respectively. These values are underestimated by $20...30\%$ in Eq.~(\ref{eq:gc_examples_pass_dr}), which is reasonable given the approximations made and the complications associated with the relatively large Shafranov shift. If we had used the safety factor $q \approx 2$ from Fig.~\ref{fig:a02_jt60u_profs}(c) instead of $\overline{r} B_0/(R_0B_{\rm pol}) \approx 1.32$, then we would have overestimated the computed values by $20...10\%$. The maximal electric precession offset inferred from Fig.~\ref{fig:14_jt60_deep-pass}(c) is $\Delta\hat{\uGC}_{\rm pass,E} \approx 0.42$, which is underestimated in Eq.~(\ref{eq:gc_examples_pass_du}) by nearly $50\%$ if we use $\overline{r} B_0/(R_0B_{\rm pol}) \approx 1.32$, or by $25\%$ if we use $q\approx 2$.

\subsection{Reference point bias and resulting ambiguity of electric frequency shifts in the passing domain}
\label{sec:gc_eaccel}

It must be emphasized that Eq.~(\ref{eq:gc_examples_pass_du}) for passing orbits tells us only the overall magnitude of the $E_r$-induced modulation of the parallel GC velocity, which in the case of Fig.~\ref{fig:14_jt60_deep-pass}(c-1) is $|\Delta\hat{\uGC}_{\rm orb}| \approx 0.42$. Its sign and its orbit-averaged value depend on the chosen launch point. In the present example, the mean parallel electric frequency shift can be anywhere in the range $-0.21 \lesssim \Delta\omega_{\uGC,{\rm E}}^{\rm (I)}R \lesssim 0.21 = |\Delta\hat{\uGC}_{\rm orb}|/2$. We call this ambiguity ``reference point bias''. The physical reason for this ambiguity is the term $\propto E_\nabla = \varrho_\parallel {\bm v}_{\bm E}\cdot\nabla B$ in Eq.~(\ref{eq:mdl_du_dt}), which --- just like the mirror force $\propto \mu\nablab B\cdot{\bm B}^*/B_\parallel^*$ in the same equation--- decelerates a GC above the midplane and accelerates it below the midplane when $E_r > 0$ (and {\it vice versa} for $E_r < 0$).

\begin{figure}
	[tb]
	\centering
	\includegraphics[width=0.48\textwidth]{\figures/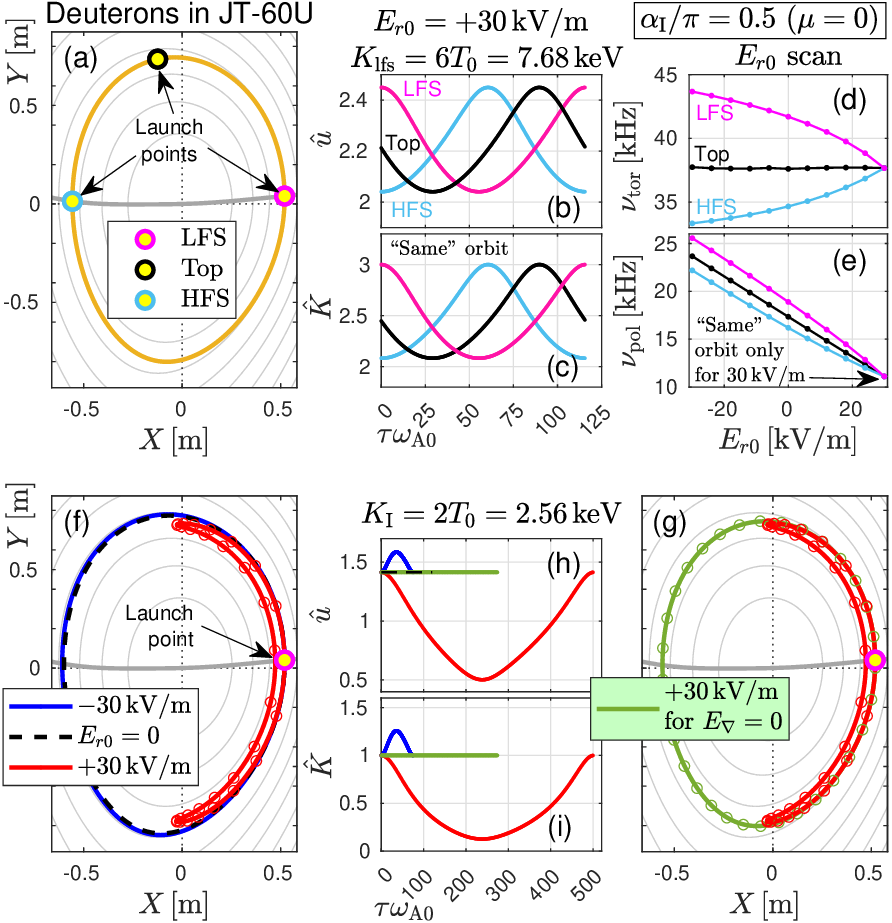}
	\caption{Demonstration of the reference point bias and effect of $E_\nabla$ for deuterons with zero magnetic moment ($\mu=0$) in the JT-60U case. While fixing the kinetic energy on the LFS midplane crossing ($K_{\rm lfs} = 6T_0 = 7.68\,{\rm keV}$) and setting $E_{r0} = 30\,{\rm kV/m}$, the top row demonstrates the influence of the launch point location, which is chosen to be on the orbit's LFS, top, or HFS, as shown in (a). Panels (b) and (c) show the corresponding time traces of the parallel GC velocity $\hat{\uGC}(\tau)$ and kinetic energy $\hat{K}(\tau)$. The three curves in (b) and (c) represent the same orbit, just starting from a different point on its $(X,Y)$ contour. Panels (d) and (e) show the $E_{r0}$-dependence of the respective toroidal and poloidal transit frequencies, $\nu_{\rm tor}$ and $\nu_{\rm pol}$, while fixing $\{{\bm C}_{\rm I},X_{\rm I},Y_{\rm I}\}$ at the values of panel (a). Consequently, the results of this $E_{r0}$-scan are reference-point-biased, and each launch point yields a different orbit for $E_{r0} < 30\,{\rm kV/m}$ in (d) and (e). For a lower initial kinetic energy ($2.56\,{\rm keV}$), the bottom row demonstrates how the sign of $E_{r0} = \pm30\,{\rm kV/m}$ (blue \& red) and the omission of $E_\nabla$ (green) influences the orbit contour in (f,g), $\hat{\uGC}(\tau)$ in (h), and $\hat{K}(\tau)$ in (i).}
	\label{fig:15_jt60u_u1pass_uEB0}%
\end{figure}

Illustrative examples for this phenomenon are presented in Fig.~\ref{fig:15_jt60u_u1pass_uEB0}, where have set $\mu=0$ in order to disable the mirror force and isolate the effect of $E_\nabla$.

If we initialize the orbit calculation near the LFS midplane (as we did in Fig.~\ref{fig:14_jt60_deep-pass}), a positive $E_r$ field decelerates the GC during the first half of the orbit (above the midplane) and accelerates it during the second half (below the midplane). This situation is shown in the upper row of Fig.~\ref{fig:15_jt60u_u1pass_uEB0} in magenta color, where one can see that the entire $\hat{\uGC}(\tau)$ curve in panel (b) lies below the initial value $\hat{\uGC}_{\rm I} = 2.4$. From this, one may na\"{i}vely infer that the mean parallel electric frequency shift is negative: $\Delta\omega_{\uGC,{\rm E}}^{\rm (I)}R \approx -|\Delta\hat{\uGC}_{\rm orb}|/2 \approx -0.2$. However, launching the GC near the HFS midplane yields the opposite result: as the cyan $\hat{\uGC}(\tau)$ curve in panel (b) shows, the GC is first accelerated (below the midplane) and then decelerated (above the midplane), giving a net positive parallel electric frequency shift: $\Delta\omega_{\uGC,{\rm E}}^{\rm (I)}R \approx |\Delta\hat{\uGC}_{\rm orb}|/2 \approx +0.2$. If we launch the GC in the upper (or lower) half-plane at one of the points where $\hat{\uGC}(\tau) \approx \hat{\uGC}_{\rm I} \pm |\Delta\hat{\uGC}_{\rm orb}|/2$, the acceleration and deceleration cancel on average, giving $\Delta\omega_{\uGC,{\rm E}}^{\rm (I)}R \approx 0$ as for the black $\hat{\uGC}(\tau)$ curve in panel (b).

The orientation of the $\hat{\uGC}(\tau)$ curves is the same for co- and counter-passing orbits because the $E_\nabla$ terms is proportional to $\hat{\uGC}$. For instance, when $E_\nabla$ makes a co-passing $\hat{\uGC}(\tau)$ less positive on average --- as on the magenta curve in Fig.~\ref{fig:15_jt60u_u1pass_uEB0}(b) --- then a counter-passing $\hat{\uGC}(\tau)$ becomes more negative. The kinetic energy $\hat{K}(\tau)$ is then reduced on average for co-passing orbits in this example --- as on the magenta curve in Fig.~\ref{fig:15_jt60u_u1pass_uEB0}(c) ---  but increased for counter-passing orbits.

For completeness, the bottom row of Fig.~\ref{fig:15_jt60u_u1pass_uEB0} shows how the sign of $E_{r0}$ and the artificial omission of $E_\nabla$ affects the orbit. This is done for a lower initial kinetic energy ($2.56\,{\rm keV}$ instead of $7.68\,{\rm keV}$), so that the LFS launch point already lies deep in the trapped domain when $E_{r0} = 30\,{\rm kV/m}$. Since we have set $\hat{\mu} = 0$, there is no magnetic mirror effect, so the GC on the red banana orbit in Fig.~\ref{fig:15_jt60u_u1pass_uEB0}(f,g) is electrically trapped via the $E_\nabla$ term. This is confirmed by the green curves, which show that omitting the electric mirror effect by artificially setting $E_\nabla = 0$ yields constant $\hat{\uGC}$ and $\hat{K}$ in panels (h) and (i), and a circulating orbit contour in panel (g).

The reference point bias caused by $E_\Delta$ impacts the results of $E_{r0}$ scans, which includes the estimation of electric frequency shifts, where one compares frequencies in the presence and absence of $E_r$. As an example, panels (d) and (e) of Fig.~\ref{fig:15_jt60u_u1pass_uEB0} show the $E_{r0}$-dependence of the toroidal and poloidal transit frequencies. In panel (d), one can see that the launch point determines whether $\nu_{\rm tor}$ decreases (HFS), increases (LFS) or remains unchanged when $E_{r0}$ is varied. Meanwhile, the $E_{r0}$-dependence of $\nu_{\rm pol}$ in panel (e) is less sensitive to the launch point; at least as long as the orbits remain deeply passing.\footnote{Scanning $E_{r0}$ to higher values ($>30\,{\rm kV/m}$) would move the trapped-passing boundary towards the LFS launch point (magenta in Fig.~\protect\ref{fig:15_jt60u_u1pass_uEB0}) and eventually lead to the situation shown in Fig.~\protect\ref{fig:12_jt60u_Er-scan_trapped}(a,b).}

Obviously, one cannot simply average the magenta and cyan curves in Fig.~\ref{fig:15_jt60u_u1pass_uEB0}(d,e), because they represent different orbits: fixing the values ${\bm C}_{\rm I}$ of our relative CoM at the LFS midplane crossing leaves them free to vary at the HFS crossing when $E_{r0}$ changes, and {\it vice versa}.

Nevertheless, the results in Fig.~\ref{fig:15_jt60u_u1pass_uEB0} allow us to verify Eq.~(\ref{eq:gc_shift_pass}) of Section~\ref{sec:gc_smooth}. First, the black curve in Fig.~\ref{fig:15_jt60u_u1pass_uEB0}(d) shows that $E_r$ does not affect the toroidal transit frequency $\nu_{\rm tor}$ of deeply passing orbits when the reference point bias is eliminated, which is here achieved by placing the launch point at one of the locations where $E_\nabla$ has an extremum, which may be called the orbit's magnetic apex (top) or nadir (bottom). This confirms that $\Delta\nu_{\rm pass,E,tor}^{\rm (0)} \approx 0$ as we predicted in Eq.~(\ref{eq:gc_shift_tor}). Second, the black curve in Fig.~\ref{fig:15_jt60u_u1pass_uEB0}(e) shows that the poloidal transit frequency $\nu_{\rm pol}$ varies by about $6.2\,{\rm kHz}$ when $E_{r0}$ is reduced by $30\,{\rm kV/m}$. This value is within $5\%$ of the prediction\vspace{-0.1cm}
\begin{equation}
	\left|\Delta\nu_{\rm pass,E,pol}^{\rm (0))}\right| \approx \left|\frac{v_0 \overline{\hat{v}_{\rm E}}}{2\pi \overline{r}}\right| \approx \frac{350\,\tfrac{\rm km}{\rm s} \times0.071}{2\pi \times 0.67\,{\rm m}} \approx 5.9\,{\rm kHz}.
	\label{eq:gc_wpol_ve_r}
\end{equation}

\noindent that is based on Eq.~(\ref{eq:gc_shift_pol}) with approximate $v_{\rm E} \approx E_r / B_0$.

\subsection{Comparison of electric precession frequencies with mean rotation velocities, and implications}
\label{sec:gc_doppler}

In Sections~\ref{sec:gc_er0frame} and \ref{sec:gc_orb}, we found that our trapped test orbit in Fig.~\ref{fig:13_jt60_deep-trapped} has a precessional Doppler shift of
\begin{equation}
	\Delta\hat{\omega}_{\rm tor,E}^{(0)} \approx \Delta\hat{\omega}_{u,{\rm E}}^{(0)} \approx \frac{1}{R}\times\left\{
	\begin{array}{l@{\hspace{0.2cm}}c@{\hspace{0.2cm}}l}
		1.56/2 = 0.78 & : & \text{computed}, \\
		1.48/2 = 0.74 & : & \text{estimated}, \\
	\end{array}\right.
	\label{eq:gc_doppler_tor}
\end{equation}

\noindent when $E_{r0} = 30\,{\rm kV/m}$. For the same JT-60U scenario, we found in Fig.~\ref{fig:09_jt60u_flow-prof-sign}(i)--(l) of Section~\ref{sec:example_results} the mean toroidal and poloidal rotation velocities of our Maxwellian-like population of thermal deuterons,
\begin{subequations}
	\begin{align}
		\hat{V}_{\rm tor} =&\; \hat{\Gamma}_{\rm tor}/\hat{N} \approx 0.06/0.7 \approx 0.086 \approx \hat{V}_{\rm E,tor},
		\label{eq:gc_vrot_tor} \\
		\hat{V}_{\rm pol} =&\; \hat{\Gamma}_{\rm pol}/\hat{N} \approx -0.05/0.7 \approx -0.071 \approx \hat{V}_{\rm E,pol},
		\label{eq:gc_vrot_pol}
	\end{align}
	\label{eq:gc_vrot}\vspace{-0.45cm}
\end{subequations}

\noindent in the domain of the red orbit in Fig.~\ref{fig:13_jt60_deep-trapped} ($X \approx 0.5\,{\rm m}$ at $Y=0$). The fact that $V_{\rm E,tor}$ in Eq.~(\ref{eq:gc_vrot_tor}) is one order of magnitude smaller than $R\Delta\omega_{\rm tor,E}^{(0)}$ in Eq.~(\ref{eq:gc_doppler_tor}) means that our trapped test orbit performs rapid toroidal precession while the toroidal rotation of the bulk plasma is slow. This illustrates that the electric precession of individual GC orbits and the electric toroidal rotation of the bulk plasma in the same ambient $E_r$ field can be entirely different in the absence of collisions. This is related to the remark we made at the end of Section~\ref{sec:example_results}, where we noted that our results are physical (within the framework of the model used), but not necessarily realistic.

It is generally assumed that, in the core of a real plasma (away from transport barriers and other anomalies), the poloidal component of the electric rotation is rapidly converted into toroidal flow. Various theoretical models exist for different collisionality regimes, but here we shall be content with assuming that this flow conversion is an empirical fact. In an axisymmetric system like a tokamak, the conversion of poloidal to toroidal flow is thought to occur in a conservative manner, which means that the electric poloidal rotation is ultimately canceled by the poloidal component of a parallel flow ${\bm U} = U\hat{\bm b}$, which satisfies $\hat{\bm e}_r\cdot({\bm U}\times{\bm B}) = U_{\rm tor} B_{\rm pol} - U_{\rm pol}B_{\rm tor} = 0$ by definition. In our case, the cancellation condition $V_{\rm E,pol} + U_{\rm pol} \approx 0$ yields
\begin{align}
	\hat{V}_{\rm tor,E}^{\rm relaxed} \approx&\; \hat{V}_{\rm E,tor} + \hat{U}_{\rm tor} \approx \hat{V}_{\rm E,tor} - \hat{V}_{\rm E,pol} \frac{B_{\rm tor}}{B_{\rm pol}} \nonumber \\
	\approx&\; 0.086 - (-0.071) / 0.12 = 0.68,
\end{align}

\noindent which now agrees with $\Delta\omega_{\rm tor,E}^{(0)}R \approx 0.74...0.78$ in Eq.~(\ref{eq:gc_doppler_tor}) with an accuracy of about $10\%$. Thus, one may indeed assume that sufficiently thin trapped orbits in a sufficiently strong $E_r$ field precess at approximately the same speed as the local bulk plasma rotates toroidally.

This assumption is often used to explain low-frequency modes in terms of precessional resonances, where the condition of similar precessional and rotational Doppler shifts,
\begin{equation}
	\underbrace{\Delta\omega_{\rm tor,E}}\limits_{\text{orbit precession}} \sim \;\;\; \underbrace{\Omega_{\rm tor,E} \equiv V_{\rm tor,E}^{\rm relaxed} / R}\limits_{\text{plasma rotation}},
	\label{eq:gc_similar_doppler}
\end{equation}

\noindent allows to compare precession and mode frequencies in the (rotating) plasma frame; that is, without explicitly including the presence of $E_r$ in the calculations. A recent example is Ref.~\cite{Lee23}, where this assumption (\ref{eq:gc_similar_doppler}) has been applied to the study of peculiar low-frequency ``fishbone-like'' Alfv\'{e}nic modes in KSTAR (which, in fact, motivated the present work).

While this convenient assumption is widely used in the literature,\footnote{For instance, see the related statement on p.~1021 of Ref.~\protect\cite{Heidbrink90}. One exception that we found is the study of energetic-particle-driven wall modes (or ``off-axis fishbone modes'') in Refs.~\protect\cite{Matsunaga09, Matsunaga10}, where the {\it lab-frame} mode frequency was compared to the precession frequency in the ``$E_r = 0$ frame''. However, this choice was abandoned in favor of Eq.~(\protect\ref{eq:gc_similar_doppler}) in subsequent work \protect\cite{Okabayashi11, Heidbrink11}. In practice, neither of the two limits may be entirely correct, especially for wide orbits that average over the nonuniform $E_r$ field. Therefore, one should perhaps treat such estimates as upper and lower bounds.}
it hinges on the above-mentioned underlying assumption that {\it all} electric poloidal rotation has been conservatively converted into toroidal rotation via the formation of a parallel ``return flow'' ${\bm U}$. This should be verified case-by-case, since it affects the mode frequency. Meanwhile the trapped orbits do not rely on this mechanism, because they effectively perform this conversion independently of the bulk plasma.

The assumptions underlying Eq.~(\ref{eq:gc_similar_doppler}) should also be questioned in cases where the potentially resonant orbits drift across a large portion of the plasma radius, where they encounter different electric field strengths as in Fig.~\ref{fig:12_jt60u_Er-scan_trapped}, so that the effective (orbit-averaged) electric field $\left<E_r\right>_{\rm orb}$ acting on them may differ significantly from the effective (mode-width-averaged) electric field $\left<E_r\right>_{\rm mode}$ in the domain of the mode. In the case of so-called energetic particle modes (EPM) \cite{Chen94}, Eq.~(\ref{eq:gc_similar_doppler}) may still hold because the EPM width is thought to be determined by the resonant orbit width. In the case of KSTAR's double-peaked fishbone-like modes \cite{Lee23}, where both the mode and potentially resonant orbits span the entire plasma radius, the situation is currently unclear and will be subject to further investigation. For modes whose width is largely independent of the orbit width (such as internal kinks stabilized by trapped fast ions), assumption (\ref{eq:gc_similar_doppler}) should also receive scrutiny.

When analyzing and interpreting experimental results, the applicability of assumption (\ref{eq:gc_similar_doppler}) and the assumed profile of $E_r$ should be verified by measurements of the toroidal and poloidal plasma rotation velocities. Unfortunately, the poloidal component $V_{\rm pol}$ is often plagued by high uncertainties. Neoclassical transport models are often used as a substitute, but the reliability of their predictions is uncertain in the presence of fluctuations like the above-mentioned Alfv\'{e}nic modes.

The situation becomes particularly complicated when an $E_r$ field is produced by and interacts with the mode activity itself. For instance, the resulting transient flows may temporarily defy the neoclassical decay of poloidal flow. The consequences for the resonances driving the underlying mode activity is currently being studied via numerical simulations of fishbone-like modes \cite{Brochard24a,Brochard24b}.

\section{Summary and conclusion}
\label{sec:summary}

In this paper we have described our method for constructing GC distributions in axisymmetric tokamak plasmas with a time-independent ambient radial electric field $E_r$ (Sections~\ref{sec:intro}--\ref{sec:method}). The method was then applied to test cases with realistic geometries based on the tokamaks KSTAR, JT-60U and ITER (Section~\ref{sec:example}), and it underwent thorough tests with respect to physical validity and numerical accuracy (Section~\ref{sec:gc} \& \ref{apdx:benchmark}).

The method has been implemented in the research code {\tt VisualStart} \cite{Bierwage22a}, which may be viewed as a working prototype. This code was routinely used in the past to initialize GC distributions for simulations of fast-ion-driven Alfv\'{e}nic instabilities and to analyze their resonances (e.g., Refs.~\cite{Bierwage17a,Bierwage18,Bierwage22b}). The inclusion of $E_r$ effects should allow us to apply this code also to the study of low-frequency modes in rotating plasmas. Our discussion of Doppler-shifted resonances in the final Section~\ref{sec:gc_doppler} of this paper is relevant for such efforts.

Being written in {\tt MATLAB} \cite{MATLAB, MATLAB_Parallel}, {\tt VisualStart} is easy to maintain and extend, and its graphical user interface makes it suitable for the manual design and analysis of simulation scenarios on a PC, workstation or small cluster. In order to apply our methods in an automated workflow that constructs or processes large numbers of charged particle distributions, as is the plan for ITER IMAS, a faster code with similar functionality should be written. {\tt VisualStart} can serve as a template and reference for benchmarks. Such work is currently in progress at ITER for stationary plasmas \cite{BrochardEPS24}, and it may later be extended to include $E_r$ using the present method.

The techniques developed and the experience gathered here may find use in ambitious efforts dedicated to the construction of self-consistent models of burning fusion plasmas, where the electric and magnetic fields, the distributions of all relevant particle species with their respective electric currents and mass flows, as well as the action of collisions, sources and sinks form a mutually consistent set. Some parts of this challenging problem are already being tackled. For instance, the problem of a gyrokinetic equilibrium was discussed in Ref.~\cite{Qin00} and gyrokinetic theory with (potentially strong) flows was developed \cite{Brizard07}, including also collisions \cite{Sugama17}. Work related to the consistency between external torque, plasma rotation and the ambient electric field can be found in Refs.~\cite{Honda09, Honda13, Honda16} using the fluid-type moment approach. Transport analysis workflows based on the concept of phase space zonal structures are developed in Refs.~\cite{Zonca15b, Falessi19a, Falessi23, Lauber24, Meng24}, which incorporates the effects of wave-particle interactions and employs GC constants of motion (CoM). Another example is the inclusion of the fast ion contribution into the construction of magnetohydrodynamic equilibria, which has been addressed in Ref.~\cite{Belova03}. As noted in Section~\ref{sec:example_results}, such multi-physics self-consistency constraints are currently not enforced in {\tt VisualStart} and would have to be established beforehand for the inputs, or through integration with other codes in an iterative workflow. When applying the present framework to electrons, the parallel component $E_\parallel$ of the ambient electric field should also be considered.\footnote{A distribution of ions that depends only on their CoM has contours that deviate from magnetic flux surfaces, which implies that the electron distribution must be modified by an intrinsic parallel electric field $E_\parallel$ that modulates the electron density by precisely the amount needed to neutralize the ion's charge beyond the Debye scale to the degree permitted by electron inertia. In axisymmetric systems, ${\bm E} = -\nablab\Phi = -\partial_r\Phi\nablab r - \partial_\vartheta\Phi\nablab\vartheta$ has radial ($\nablab r$), parallel ($\hat{\bm b}$) and binormal ($\hat{\bm b}\times\nablab r \approx r\nablab\vartheta$) components. The necessary presence of a binormal component whenever $E_\parallel \neq 0$ and $E_\zeta = 0$ implies that the electric drift ${\bm v}_{\rm E} = {\bm E}\times{\bm B}/B^2$ must have a radial component $v_{{\rm E}r}$. While its flux surface average must vanish in a steady state ($\left<v_{{\rm E}r}\right>_\psi = 0$), we expect this radial electric drift to contribute to the shift of all species' density contours with respect to magnetic flux contours.}

\section*{Acknowledgments}

A.B.\ is grateful for helpful discussions with Zhixin Lu and Thomas Hayward-Schneider (MPG IPP Garching), Mitsuru Honda (Kyoto University), Yasutaro Nishimura (NCKU, Tainan), Kimin Kim, Won-ha Ko and Tongnyeol Rhee (KFE, Daejeon). This work was partly supported by MEXT as ``Program for Promoting Researches on the Supercomputer Fugaku'' (Exploration of burning plasma confinement physics, JPMXP1020200103). This work has been partially carried out within the framework of the EUROfusion Consortium, funded by the European Union via the Euratom Research and Training Programme (Grant Agreement No.\ 101052200-EUROfusion). Views and opinions expressed are however those of the author(s) only and do not necessarily reflect those of the European Union or the European Commission. Neither the European Union nor the European Commission can be held responsible for them. ITER is the Nuclear Facility INB No.\ 174. The views and opinions expressed herein do not necessarily reflect those of the ITER Organisation. Y.C.G.\ and W.L.\ are supported by the National R\&D Program through the National Research Foundation of Korea (NRF) funded by the Ministry of Science and ICT (Grant Nos.\ RS-2022-00155917 and NRF-2021R1A2C2005654).

\addcontentsline{toc}{section}{Appendices with supplementary information}
\appendix
\addtocontents{toc}{\setcounter{tocdepth}{-1}}

\setcounter{figure}{0}
\section{Setup of working examples}
\label{apdx:examples}

\subsection{Geometry and profiles}

\begin{figure}
	[tb]
	\centering
	\includegraphics[width=0.48\textwidth]{\figures/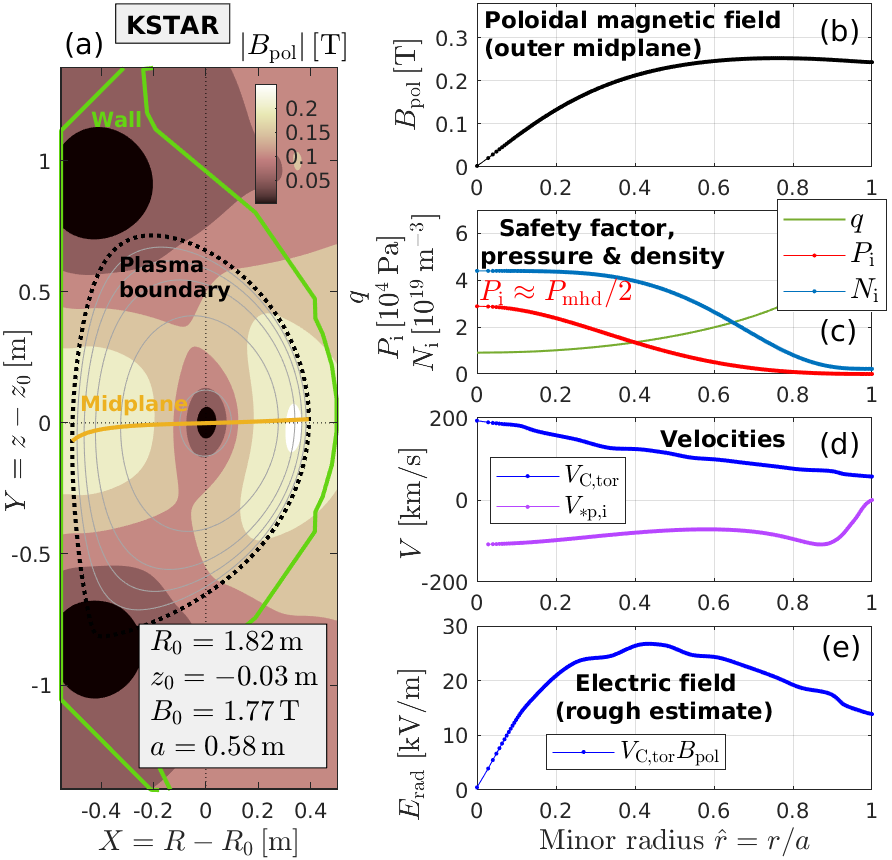}\vspace{-0.2cm}
	\caption{Model of KSTAR shot \#18567 \protect\cite{Lee23}. Panels (a) and (b) show the structure of the poloidal magnetic field $|B_{\rm pol}|(R,z)$ and its radial profile $B_{\rm pol}(\hat{r})$ near the outer midplane. Panel (c) shows the safety factor, ion pressure and density profiles. The latter two yield the diamagnetic velocity $V_{\rm i*}$ that is plotted together with the measured toroidal rotation profile $V_{\rm C,tor}$ of the carbon VI impurity in panel (d). Panel (e) shows the resulting estimate of the radial electric field based on purely toroidal flow, $E_{\rm rad} \leq V_{\rm C,tor} B_{\rm pol}$, using Eq.~(\protect\ref{eq:er_ms_approx_l}).}
	\label{fig:a01_kstar_profs}\vspace{-0.2cm}
\end{figure}

Fig.~\ref{fig:a01_kstar_profs} shows our plasma model based on KSTAR shot \#18567. This and similar plasmas were found to produce peculiar double-peaked low-frequency modes in the core and peripheral regions \cite{Lee23}, which appear to be driven by fast ions but still await a complete explanation. These deuterium plasmas were driven by positive-ion-based neutral beams (P-NB) with net co-current injection that served as an effective source of toroidal momentum. There was no reliable measurement of the poloidal rotation profile $V_{\rm pol}$, so we estimated the electric field using Eq.~(\ref{eq:er_ms_approx_l}) below, which yields $E_{\rm rad} \lesssim +25\,{\rm kV/m}$ in the limit of vanishing poloidal flow.

\begin{figure}
	[tb]
	\centering\vspace{-0.8cm}
	\includegraphics[width=0.48\textwidth]{\figures/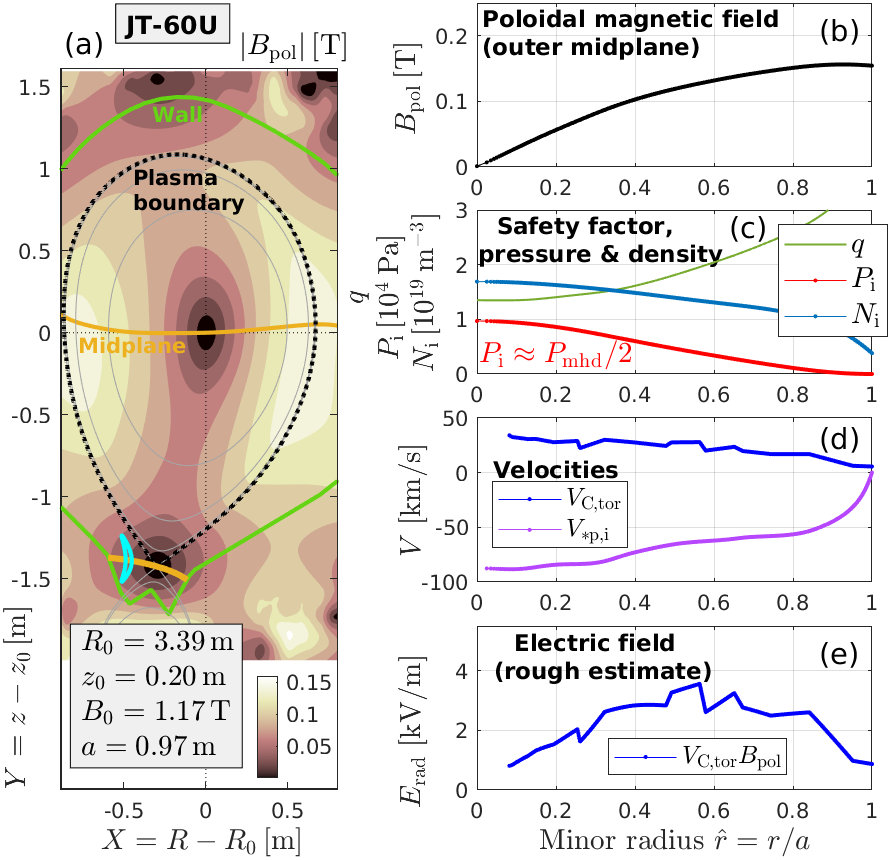}\vspace{-0.15cm}
	\caption{Model of JT-60U shot \#E039672 \protect\cite{Shinohara02, Bierwage18}, with ion profiles taken from Fig.~3 of Ref.~\cite{Bierwage17a}. Arranged as Fig.~\protect\ref{fig:a01_kstar_profs}. This configuration has three magnetic midplanes satisfying Eq.~(\protect\ref{eq:method_mid}), two of which can be seen as orange lines in panel (a). (The third one is located at the outer wall near $X \approx 1\,{\rm m}$.) In the present work, we consider only GC orbits that cross the internal midplane near $Y \approx 0$. However, it is interesting to note that some mirror-trapped GC orbits can also be found on the external midplane in the divertor region, an example of which is plotted in cyan color in panel (a). This orbit was launched from $(X_{\rm I},Y_{\rm I}) = (-0.49\,{\rm m},-1.4\,{\rm m})$ with kinetic energy $\hat{\ekin}_{\rm I} = 0.35$ and pitch angle $\alpha_{\rm I} = 0.03\pi$.}
	\label{fig:a02_jt60u_profs}%
\end{figure}

\begin{figure}
	[tbp]
	\centering\vspace{-0.8cm}
	\includegraphics[width=0.48\textwidth]{\figures/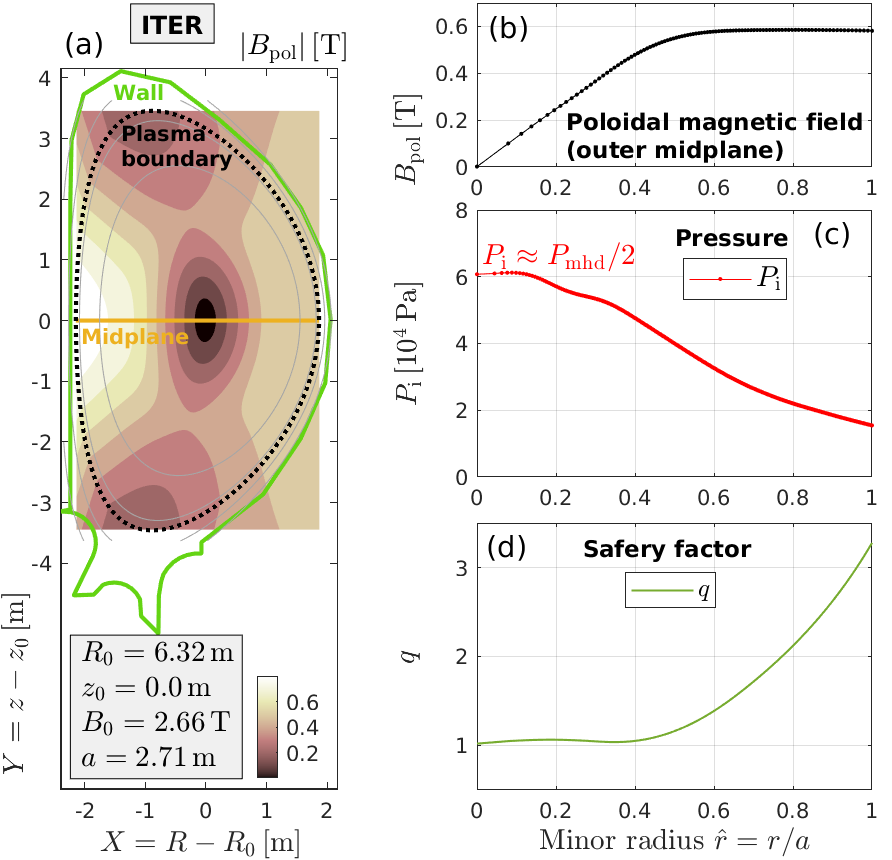}\vspace{-0.15cm}
	\caption{Simplified up-down symmetric version of ITER model shot 101006r50. Panels (a) and (b) show the structure of the poloidal magnetic field $|B_{\rm pol}|(R,z)$ and its radial profile $B_{\rm pol}(\hat{r})$ near the outer midplane. Panel (c) shows the ion pressure profile $P_{\rm i} \approx P_{\rm mhd}/2$, and panel (c) the safety factor profile $q$.}
	\label{fig:a03_iter_profs}%
\end{figure}

In JT-60U plasmas without significant external toroidal torque injection, toroidal velocities on the order of $\lesssim 30\,{\rm km/s}$ were measured and we expect $E_{\rm rad} \sim {\rm few}\,{\rm kV/m}$ in such cases. As a concrete example, we present in Fig.~\ref{fig:a02_jt60u_profs} data from JT-60U shot E039672, which was studied in Refs.~\cite{Shinohara02, Bierwage17a, Bierwage18, Shinohara04, Ishikawa05} in the context of Alfv\'{e}nic instabilities driven by negative-ion-based neutral beams (N-NB). Note that the energy $300...400\,{\rm keV}$ of the two N-NBs in JT-60U significantly exceeded the critical energy for ion-ion collisions, so they did not constitute an efficient source of torque for the bulk ions. The poloidal rotation profile $V_{\rm pol}$ could not be measured reliably for this deuterium plasma. In the limit of vanishing poloidal flow, Eq.~(\ref{eq:er_ms_approx_l}) yields $E_{\rm rad} \lesssim +3\,{\rm kV/m}$. For the analysis of guiding center motion in this paper, we often use $E_{\rm rad} = 30\,{\rm kV/m}$, which may not be representative for this JT-60U case, but helped us to test the methods that we introduced in the present paper.

The diamagnetic velocity profiles in Figs.~\ref{fig:a01_kstar_profs}(d) and \ref{fig:a02_jt60u_profs}(d) were estimated from $V_{\rm *p,i} \approx R_0 P_{\rm i}'/(Z_{\rm i} e N_{\rm i} \Psi_{\rm P}')$ based on Eq.~(\ref{eq:vdiamag}) below.
 
Finally, the numerical model for our ITER test case is shown in Fig.~\ref{fig:a03_iter_profs}. This is a simplified up-down symmetric version of ITER model shot 101006r50.

\subsection{Estimation of the radial electric field}
\label{apdx:Erad}

The ambient electric field inside a hot tokamak plasma cannot be measured directly. Instead, it must be inferred from the balance between the mean Lorentz force acting on the fluid model of a charged particle species $s$ and its pressure gradient across magnetic surfaces ($\nablab r$ direction):
\begin{equation}
	\nablab r\cdot[Z_s e (N_s{\bm E} + {\bm \Gamma}_s\times{\bm B}) - \nablab P_s] \stackrel{\text{radial force balance}}{\longrightarrow} 0,
\end{equation}

\noindent where the convective nonlinearity ${\bm V}_s\cdot\nablab{\bm V}_s$ has been ignored and a scalar pressure is assumed for simplicity. The condition $E_\parallel \rightarrow 0$ together with the assumption of axisymmetry ($\partial_\zeta\Phi = 0$) implies that the ambient electric field only has a radial component:
\begin{equation}
	{\bm E} \rightarrow E_{\rm rad}\hat{\bm e}_r = E_r(r)\nablab r = -\Phi'(r)\nablab r,
\end{equation}

\noindent where $\Phi' \equiv {\rm d}\Phi/{\rm d}r$. The sign and magnitude of $E_{\rm rad}$ can then be computed from the mean flow velocity ${\bm V}_s$, number density $N_s(r)$ and thermal pressure $P_s(r)$ via
\begin{align}
	E_{\rm rad} =&\; \hat{\bm e}_r\cdot({\bm B}\times{\bm V}_s) + |\nablab r|P_s'/(Z_s e N_s) \nonumber \\
	=&\; V_{{\rm tor},s} B_{\rm pol} - V_{{\rm pol},s} B_{\rm tor} + V_{*{\rm p},s} B_{\rm pol}.
	\label{eq:forcebal_Erad}
\end{align}

\noindent Here, the diamagnetic flow velocity is defined as
\begin{equation}
	V_{*{\rm p},s} \equiv \frac{|\nablab r| P_s'}{Z_s e N_s B_{\rm pol}} = \frac{R P_s'}{Z_s e N_s \Psi_{\rm P}'},
	\label{eq:vdiamag}
\end{equation}

\noindent where we used $B_{\rm pol} = |\nablab r|\Psi_{\rm P}'/R$. Strictly speaking, only the perpendicular pressure is associated with diamagnetic flow, but Eq.~(\ref{eq:vdiamag}) holds in the isotropic limit.

Charge exchange recombination spectroscopy (CHERS, CER, CES or CXRS) is one of the main techniques to measure the plasma rotation velocities that are needed to evaluate Eq.~(\ref{eq:forcebal_Erad}). In plasmas where neutral beams inject particles of the same species as the bulk ions, this diagnostic allows to measure the overall mean velocity (which can be useful for fast ion velocity space tomography \cite{Saleweski17, Madsen20}), but it does not directly yield the rotation speed of the bulk ions (which carry most of the inertia and dominate the Doppler shift of Alfv\'{e}n waves that we wish to estimate). Instead, one measures the rotation of some impurity ion species such as carbon, boron, helium or neon.

According to neoclassical theory, the rotation speeds of different species are generally expected to differ \cite{Kim91, Valanju92}, but they can be related to each other by noting that Eq.~(\ref{eq:forcebal_Erad}) holds independently for each species $s$ as long as $E_{\rm rad}$ is approximately constant on the scale of the gyroradius and radial magnetic drifts. Equating $E_{\rm rad}$ obtained from Eq.~(\ref{eq:forcebal_Erad}) of two different species $s=1,2$, the difference between their toroidal velocity components can then be estimated from
\begin{align}
	\Delta V_{\rm tor,12} \equiv&\; V_{\rm tor,1} - V_{\rm tor,2}
	\label{eq:dV12}
	\\
	=&\; \frac{1}{B_{\rm pol}} \left(\Delta V_{\rm pol,12} B_{\rm tor} - \frac{P'_1}{Z_1 e n_1} + \frac{P'_2}{Z_2 e n_2}\right). \nonumber
\end{align}

\noindent For plasmas that appear to have negligible anomalous transport, Eq.~(\ref{eq:dV12}) has been reported to yield agreement between neoclassical estimates and experimental measurements \cite{Kim94, Baylor04} (in part, with surprisingly high accuracy). In the limit where the poloidal flows $V_{\rm pol,1}$ and $V_{\rm pol,2}$ both vanish or cancel, we have
\begin{equation}
	\Delta V_{\rm tor,12} B_{\rm pol} \approx - \left(\frac{P'_1}{Z_1 e n_1} - \frac{P'_2}{Z_2 e n_2}\right),
	\label{eq:dV12_lim_l}
\end{equation}

\noindent The diamagnetic drift of high-$Z$ impurities may be ignored. In our KSTAR and JT-60U cases, CES measures the motion of carbon (C) in a deuteron (D) plasma, and we let $Z_{\rm D}/Z_{\rm C} = 1/6 \ll 1$. Substituting Eq.~(\ref{eq:dV12_lim_l}) into Eq.~(\ref{eq:forcebal_Erad}) for $E_{\rm rad}$ yields
\begin{equation}
	E_{\rm rad} \approx V_{\rm tor,C} B_{\rm pol} \quad (\text{for small } \Delta V_{\rm pol,CD}B_{\rm tor}).
	\label{eq:er_ms_approx_l}
\end{equation}

\noindent This approximation is used to obtain the $E_{\rm rad}$ profiles shown in Figs.~\ref{fig:a01_kstar_profs}(e) and \ref{fig:a02_jt60u_profs}(e).
	
It is however questionable that the assumption of small $\Delta V_{\rm pol,CD}B_{\rm tor}$ generally holds. First, cancellation via $V_{\rm pol,C} \approx V_{\rm pol,D}$ is unlikely. Second, it is evident that the contribution of poloidal flow is only negligible if $V_{{\rm pol},s}/V_{{\rm tor},s} \sim B_{\rm pol}/B_{\rm tor}\times\O(0.1) \sim \O(0.01)$. In cases where $V_{{\rm pol},s}/V_{{\rm tor},s} \sim \O(0.1)$ --- which is commonly assumed to be the case in experiments --- the contribution of $V_{{\rm pol},s}$ in the radial force balance can be significant. For instance, in our KSTAR case in Fig.~\ref{fig:a01_kstar_profs}, the plasma rotation is fairly rapid and probably largely parallel, so that the neglect of $V_{{\rm pol},s}$ in Eqs.~(\ref{eq:er_ms_approx_l}) is likely to cause an overestimation of the radial electric field, because the contribution of the parallel flow component,
\begin{equation}
	\hat{\bm e}_r\cdot({\bm U} \times {\bm B}) = U_{\rm tor}B_{\rm pol} - U_{\rm pol} B_{\rm tor} = 0,
	\label{eq:vparxB}
\end{equation}

\noindent with ${\bm U} \equiv U\hat{\bm b}$, no longer vanishes when $V_{\rm pol}$ (which contains $U_{\rm pol}$) is ignored.

In future practical applications of our methods, one should thus include $V_{{\rm pol},s}$ based on experimental measurements (if available) or numerical models as in Refs.~\cite{Honda13,Aiba17}. In the present study, this is not essential because $E_{\rm rad}$ is used as a free parameter.

As a side note, we point out that, with $B_{\rm tor} = I/R$ and $E_{\rm rad} = -\Phi'|\nablab r|$, Eq.~(\ref{eq:forcebal_Erad}) can be written as
\begin{equation}
	-\Phi' = \Omega_{{\rm tor},s} \Psi_{\rm P}' - \frac{V_{{\rm pol},s}}{B_{\rm pol}R^2}I + \frac{P_s'}{Z_s e N_s}.
	\label{eq:forcebal_dPhidr}
\end{equation}

\noindent Thus, the toroidal angular velocity $\Omega_{{\rm tor},s} \equiv V_{{\rm tor},s}/R$ becomes a flux function when the poloidal flow $V_{{\rm pol},s}$ vanishes or its angular dependence cancels that of $B_{\rm pol}R^2$, whose form can be seen in panel (a) of Figs.~\ref{fig:a01_kstar_profs}--\ref{fig:a03_iter_profs}.

\subsection{Typical values of $E_{\rm rad}$ in the literature}

In the present study, we scanned the peak value of the global ambient $E_r$ field through the range $(-30...30)\,{\rm kV/m}$. This choice is based on a cursory survey of the literature, which indicated to us that values for the ambient radial electric field in typical tokamak plasmas can be expected to lie in the range\vspace{-0.2cm}
\begin{equation}
	\underbrace{-50\,\tfrac{\rm kV}{\rm m}}\limits_{\rm ETB} ... \underbrace{-20\,\tfrac{\rm kV}{\rm m} \lesssim E_{\rm rad} \lesssim 20\,\tfrac{\rm kV}{\rm m}}\limits_{\text{typical}} ... \underbrace{55\,\tfrac{\rm kV}{\rm m}}\limits_{\text{outliers?}}.
	\label{eq:er_exp}\vspace{-0.25cm}
\end{equation}

\noindent The lowest negative values were found in Refs.~\cite{Plank23, Viezzer14}, both of which examined edge transport barriers (ETB) in H-mode plasmas on ASDEX-Upgrade. Near such transport barriers, it is often estimated that ${\bm V}_s\times{\bm B}$ is small, which means that the electric drift happens to be largely balanced by the diamagnetic drift: $E_{\rm rad} \approx V_{\rm *i} B_{\rm pol}$ (see Ref.~\cite{Plank23} and references therein). In the core of Ohmic KSTAR plasmas, values around $-10\,{\rm kV/m}$ were estimated in Ref.~\cite{Lee22}. In TEXTOR L-mode plasmas, values in the range $-25\,{\rm kV/m} \lesssim E_{\rm rad} \lesssim 10\,{\rm kV/m}$ were reported in the region $0.5 \lesssim \hat{r} \lesssim 1$ \cite{Coenen10}. Modeling of beam-driven plasmas in EAST gave peak values around $E_{\rm rad} \sim \pm 15\,{\rm kV/m}$ \cite{Xu21}.

We found two papers showing values that significantly exceed $20\,{\rm kV/m}$: A recent analysis of rotating KSTAR plasmas \cite{Kim23} and a classical paper from DIII-D \cite{Kim91} both show values up to $E_{\rm rad} \lesssim 55\,{\rm kV/m}$ in the outer core, with toroidal velocities on the order of $(100...300)\,{\rm km/s}$. The 2022 KSTAR experiments \cite{Kim23} did not have reliable measurements of poloidal rotation, and $E_r$ in Fig.~8 of that paper was computed using the code {\tt TRANSP}\footnote{\href{https://transp.pppl.gov/}{https://transp.pppl.gov/}}
while neglecting $V_{\rm pol}$, so it is possible that $E_r$ was overestimated by missing the cancellation in Eq.~(\ref{eq:vparxB}). The DIII-D study in Ref.~\cite{Kim91} of 1991 reports poloidal rotation measurements, but many technical challenges and sources of inaccuracies associated with such measurements have been discussed in the intervening decades. Therefore, it is possible that the value of $55\,{\rm kV/m}$ on the right-hand side of Eq.~(\ref{eq:er_exp}) is due to inaccurate outliers and the upper limit may be lower by a factor 2 or so, perhaps around $30\,{\rm kV/m}$.

\setcounter{figure}{0}
\setcounter{table}{0}
\section{Benchmark of orbit space slicing in relative \& absolute CoM coordinates}
\label{apdx:benchmark}

\subsection{Rationale and setup of the benchmark}
\label{apdx:benchmark_setup}

The presence of a radial electric field $E_r$ alters the Jacobian factors in Eq.~(\ref{eq:model_dVol_transf}) or the structure of the GC phase space mesh in Fig.~\ref{fig:04_kstar_grid} or both, and we anticipated in Section~\ref{sec:method_slice} that different choices of coordinates have different numerical accuracy and convergence properties. These differences are analyzed here in detail. The following benchmarking exercises also serve the purpose of convincing ourselves that no mistakes have been made in the derivations and implementations.

The application examples in Section~\ref{sec:example} of the main paper were computed by slicing the GC orbit space along lines of constant midplane-based relative CoM $\{\sigma_\uGC\ekin,\Lambda,X\}$, with volume elements given by Eq.~(\ref{eq:model_dVol_transf_X}). The pitch coordinate $\Lambda \equiv \mu B_0/\ekin$ or, equivalently, the pitch angle $\alpha \equiv \sigma_\uGC{\rm asin}(\sqrt{1 - \Lambda\hat{B}})$ took the place of the magnetic moment $\mu$. The mixed (relative + absolute) CoM set $\{\sigma_\uGC\ekin,\Lambda,P_\zeta\}$ in Eq.~(\ref{eq:model_dVol_transf_P}) gave identical results.

\begin{table}[tbp]
	\centering
	\begin{tabular}{@{\hspace{0.1cm}}c|c|c@{\hspace{0.1cm}}}
		\hline\hline Filter & $\Delta P_{\zeta,i} = P_{\zeta,i+1} - P_{\zeta,i}$ invalid & Figures \\
		\hline {\tt AND} logic: & {\tt if} $P_{\zeta,i}$ {\tt AND} $P_{\zeta,i+1}$ {\tt invalid} & \protect\ref{fig:b02_jt60_flat30_benchmark}(o)--(u) \\
		\hline {\tt OR} logic: & {\tt if} $P_{\zeta,i}$ {\tt OR} $P_{\zeta,i+1}$ {\tt invalid} & \protect\ref{fig:b03_jt60_flat30_can-dP_convergence} \& \protect\ref{fig:b04_jt60_flat3_can-dP_convergence} \\
		\hline\hline
	\end{tabular}\vspace{-0.1cm}
	\caption{Two possible filters were used to discard invalid cells $\Delta P_\zeta$ when slicing the space of absolute CoM space along lines of constant $\{\sigma_\uGC\etot,\overline{\Lambda},P_\zeta\}$. The less restrictive {\tt AND}-filter requires that Eq.~(\protect\ref{eq:bench_u20}) is satisfied only at one cell boundary. The cell is discarded only if one boundary {\it and} the cell center lie in the invalid domain. At the invalid boundary, $P_\zeta$ is then evaluated by artificially letting $\uGC = 0$, which introduces inaccuracies. The most restrictive {\tt OR}-filter discards cells as soon as one boundary violates Eq.~(\protect\ref{eq:bench_u20}). The resulting losses cause a systematic underestimation of moments like $N$ and ${\bm \Gamma}$ and a lack of samples near $\uGC = 0$, as can be seen in the right column of Fig.~\protect\ref{fig:b02_jt60_flat30_benchmark}.}
	\label{tab:can_dP_filter}
\end{table}

In this section, we demonstrate that the above coordinates sets that utilize the relative CoM $\ekin$ \& $\Lambda$ yield significantly higher accuracy and better numerical convergence than Eqs.~(\ref{eq:model_dVol_transf_can}) and (\ref{eq:model_dVol_transf_canX}) with $\etot$ \& $\overline{\Lambda} \equiv \mu B_0/\etot$, at least for our relatively simple implementation. The set $\{\sigma_\uGC\etot,\overline{\Lambda},P_\zeta\}$ that consist exclusively of absolute CoM performed the worst. As we mentioned in Section~\ref{sec:method_slice}, the main culprit is the boundary of the physically valid GC phase space domain, in particular the constraint\vspace{-0.1cm}
\begin{equation}
	\hat{\uGC}^2/2 = \hat{\ekin} - \hat{\mu}\hat{B} = \hat{\etot} - \hat{\Phi}/\varrho_0 - \hat{\mu}\hat{B} \geq 0.
	\label{eq:bench_u20}\vspace{-0.1cm}
\end{equation}

\noindent For relative CoM the red line in Fig.~\ref{fig:04_kstar_grid}(b) representing the limit (\ref{eq:bench_u20}) is straight, so this line can be directly used as a boundary of $\Delta\Lambda_j\Delta X_k$ cells, and the cell sizes can be computed accurately with ease. In contrast, the red line in Fig.~\ref{fig:04_kstar_grid}(e) intersects the absolute CoM space slices in a curved fashion, so the boundary cells can have more complicated shapes that depend on the form of the electrostatic potential $\Phi$ and would require more sophistication in order to be sampled accurately. In our simple implementation --- which is only meant to serve demonstration purposes --- the cells are defined independently of $\Phi$, so the red line in Fig.~\ref{fig:04_kstar_grid}(e) divides most of our boundary cells into valid and invalid portions. When using the coordinate set $\{\sigma_\uGC\etot,\overline{\Lambda},X\}$, the Jacobian factor $|{\rm d}P_\zeta/{\rm d}X|$ that converts $\Delta X$ into $\Delta P_\zeta$ is evaluated at the cell center, so the condition (\ref{eq:bench_u20}) needs to be satisfied only at that location. If this is not the case, the cell is labeled as invalid and discarded. This introduces some errors via missing boundary cells or inaccurate cell sizes. These problems become more severe when using exclusively absolute CoM $\{\sigma_\uGC\etot,\overline{\Lambda},P_\zeta\}$. In order to compute the increment $\Delta P_\zeta$ directly, the constraint (\ref{eq:bench_u20}) should ideally be satisfied at both cell boundaries. The resulting need to discard or truncate cells will be shown to cause systematic underestimations of moments like $N$ and ${\bm \Gamma}$ and a severe lack of samples near $\uGC = 0$. The two criteria described in Table~\ref{tab:can_dP_filter} were implemented for discarding invalid cells: a highly restrictive {\tt OR}-filter and a less restrictive {\tt AND}-filter. Again, in principle, all this could be done accurately if one invests more effort into the meshing algorithm, which we did not do because we already have a simpler, more elegant way to obtain accurate results by slicing the GC orbit space in relative CoM coordinates on the magnetic midplane.

For the following benchmark and convergence tests, we use the JT-60U case with low ion temperature $T_{\rm i0} = 1.27\,{\rm keV}$ and an exaggerated electric field strength up to $E_{r0} = 30\,{\rm kV/m}$, because that was found to be a fairly challenging example that may be regarded as a worst-case scenario. This choice magnifies numerical inaccuracies, thus, making them clearly visible.

Our Maxwellian-like distribution in Eq.~(\ref{eq:fM}) was evaluated here using uniform profiles for the reference density and temperature, $N_{\rm ref} = N_0$ and $T_{\rm ref} = T_0$. This minimizes the contribution of non-electric magnetization flows and will also help to make deviations more easily visible. As before, number densities, temperatures and flow densities are normalized as $\hat{N} = N/N_0$, $\hat{T} = T/(Mv_0^2) = T/(2T_0)$, and $\hat{\bm \Gamma} = {\bm \Gamma}/(N_0 v_0)$. The hats will usually be omitted.

\begin{figure}
	[tb]\vspace{-0.3cm}
	\centering
	\includegraphics[width=0.48\textwidth]{\figures/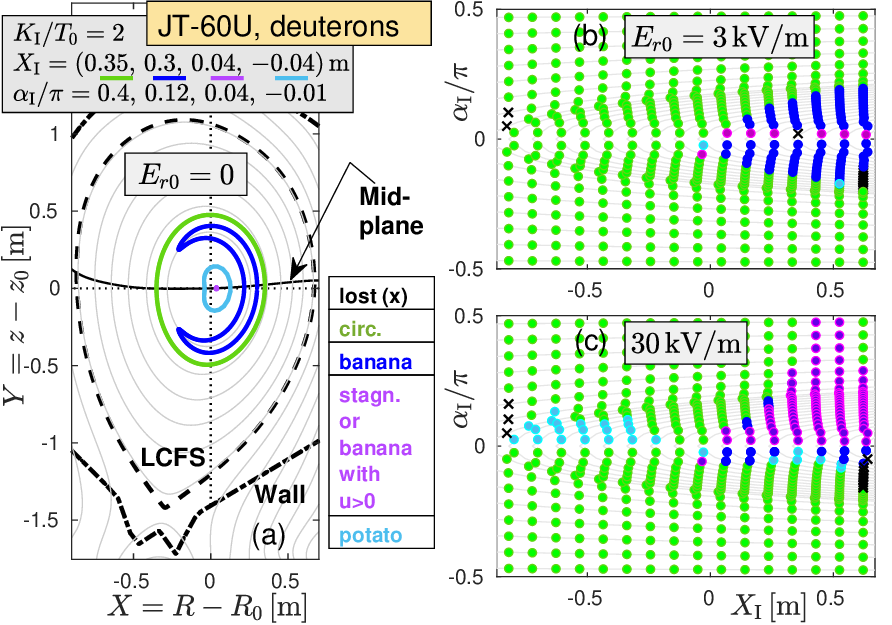}\vspace{-0.3cm}
	\caption{CoM space grid and distribution of orbit classes in the JT-60U case with major radius $R_0 \approx 3.4\,{\rm m}$. Other parameters are given in Table~\protect\ref{tab:tok} and Fig.~\protect\ref{fig:a02_jt60u_profs}. The deuteron orbits shown were initialized with $\ekin_{\rm I} = 2T_0 \approx 2.5\,{\rm keV}$. This figure is arranged as Fig.~\protect\ref{fig:05_kstar_orbtypes}, except that panel (b) shows results for a weak electric field $E_{r0} = 3\,{\rm kV/m}$ which is considered to be realistic for this JT-60U plasma. $E_{r0} = 30\,{\rm kV/m}$ in panel (c) is an exaggerated ``worst-case'' scenario that we use for most of our benchmark and convergence tests in the present \protect\ref{apdx:benchmark}. The present $X_{\rm I}$-mesh covers the midplane only within the LCFS.}\vspace{-0.3cm}
	\label{fig:b01_jt60_orbtypes}%
\end{figure}

\begin{figure*}
	[tb]\vspace{-2.5cm}
	\centering
	\includegraphics[width=0.96\textwidth]{\figures/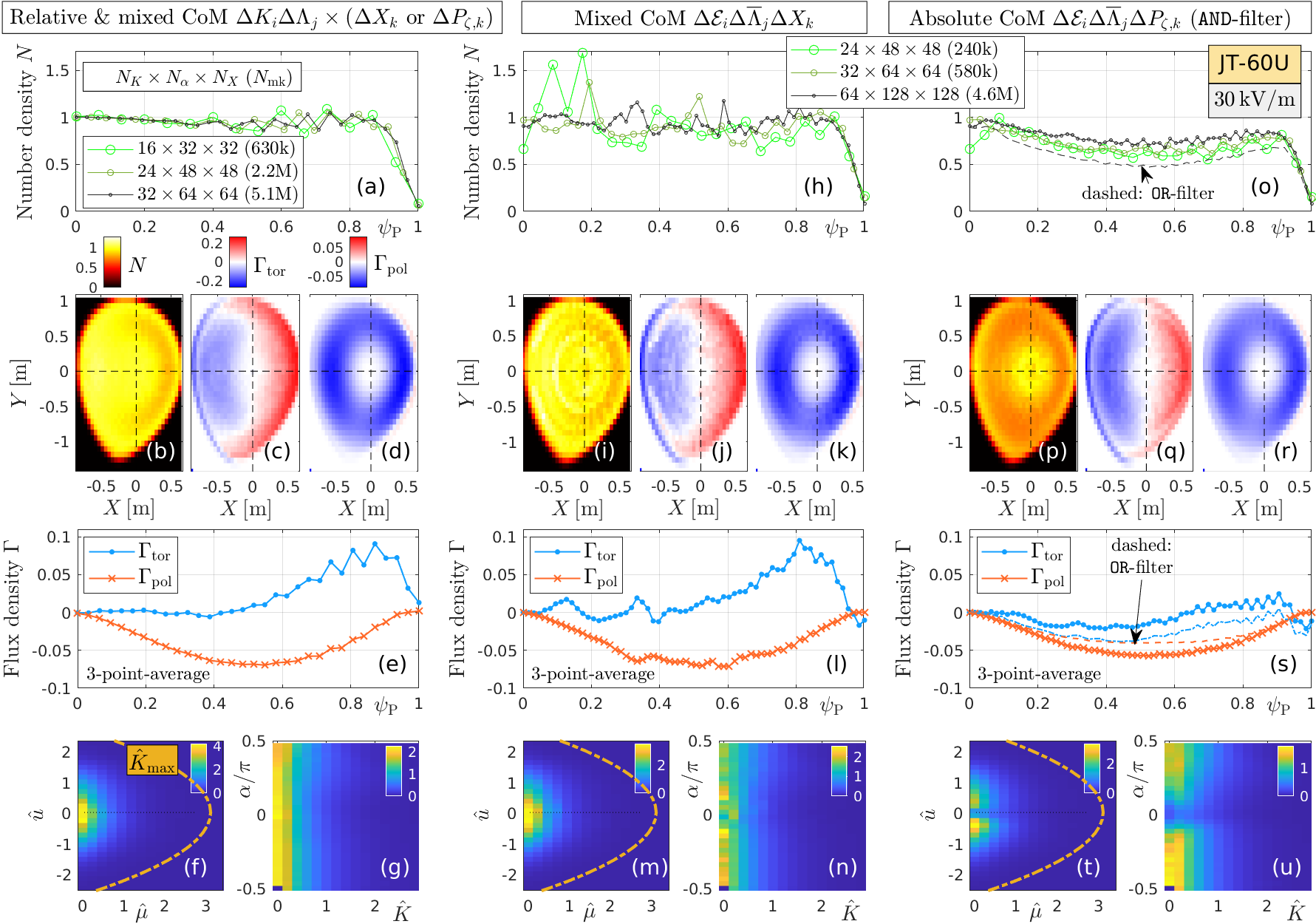}\vspace{-0.25cm}
	\caption{Benchmark in our worst-case scenario: JT-60U with exaggerated $E_{r0} = 30\,{\rm kV/m}$. To simplify the comparison with expected values, we used uniform reference profiles, $N_{\rm ref}=1$ and $T_{\rm ref}=0.5$, for the density and temperature in Eq.~(\protect\ref{eq:fM}). Results obtained after slicing the GC orbit space along lines of constant $\{\sigma_\uGC\ekin,\Lambda,X\}$ or $\{\sigma_\uGC\ekin,\Lambda,P_\zeta\}$, as described in Fig.~\protect\ref{fig:04_kstar_grid}(a)--(c), are summarized in the left column. Panel (a) shows the number density profile $N(\psi_{\rm P})$ for three different resolutions as indicated in the legend. For the case with the highest resolution (black line), panels (b)--(d) show contour plots of the number density $N$, and toroidal and poloidal flow densities $\Gamma_{\rm tor}$ and $\Gamma_{\rm pol}$ in the poloidal plane. Panel (e) shows the flux-surface-averaged radial profiles of $\Gamma_{\rm tor}$ and $\Gamma_{\rm pol}$, smoothed via 3-point-averaging. Panels (f) and (g) show the spatially integrated velocity distributions $f_{\rm gc}(\mu,\uGC)$ and $f_{\rm gc}(\ekin,\alpha)$. The central and right column are arranged in the same manner, showing results obtained with the mixed CoM coordinates $\{\sigma_\uGC\etot,\overline{\Lambda},X\}$ and absolute CoM coordinates $\{\sigma_\uGC\etot,\overline{\Lambda},P_\zeta\}$, slicing the GC orbit space as shown in Fig.~\protect\ref{fig:04_kstar_grid}(d)--(f). The results for the absolute CoM case in (o)--(u) were obtained with {\tt AND}-filter logic, which discards cells only if the cell center and one cell boundary is invalid. For comparison, the dashed lines in panels (o) and (s) show the profiles one obtains with a more restrictive {\tt OR}-filter logic, where a CoM cell is discarded when at least one boundary is invalid. The full results of convergence tests obtained with {\tt OR}-filter logic can be found in Fig.~\protect\ref{fig:b03_jt60_flat30_can-dP_convergence}.}
	\label{fig:b02_jt60_flat30_benchmark}\vspace{-0.2cm}
\end{figure*}

Fig.~\ref{fig:b01_jt60_orbtypes}(a) shows the plasma cross-section of our JT-60U test case and a few examples of GC orbit contours for deuterons at twice the thermal energy $\ekin = 2T_0$ ($2.54\,{\rm keV}$). The distribution of orbit types on a low-resolution mesh in midplane-based relative CoM coordinates $(\alpha,X)$ is shown in Fig.~\ref{fig:b01_jt60_orbtypes}(b) for the realistic value $E_{r0} = 3\,{\rm kV/m}$, and in Fig.~\ref{fig:b01_jt60_orbtypes}(c) for the exaggerated value $E_{r0} = 30\,{\rm kV/m}$. The 2D slice of CoM space in panels (b) and (c) was also taken at twice the value of the central thermal energy, $\ekin = 2T_0$, and one can see that the $30\,{\rm kV/m}$ field in (c) is strong enough to shift the trapped-passing boundary all the way to the maximal pitch angle $\alpha = \pi/2$, thus, significantly diminishing the domain of co-passing orbits and enlarging the counter-passing domain.

\subsection{Comparison of results}
\label{apdx:benchmark_methods}

The benchmark results for the $E_{r0} = 30\,{\rm kV/m}$ case are summarized in Fig.~\ref{fig:b02_jt60_flat30_benchmark}. The sets of relative and mixed CoM coordinates $\{\sigma_\uGC\ekin,\Lambda,X\}$ and $\{\sigma_\uGC\ekin,\Lambda,P_\zeta\}$ gave identical results, which are shown in the left column of Fig.~\ref{fig:b02_jt60_flat30_benchmark}. Panels (a) and (b) show that the computed number density field $N$ is close to the uniform reference $N_{\rm ref} = 1$ that was used in Eq.~(\ref{eq:fM}). Deviations near the plasma boundary ($\psi_{\rm P} = 1$) are due to losses caused by magnetic drifts as expected. The small reduction of $N(\psi_{\rm P})$ in the region $0.3 \lesssim \psi_{\rm P} \lesssim 0.7$ was explained at the end of Section~\ref{sec:example_dens}.

The set of $N(\psi_{\rm P})$ profiles in panel (a) shows good convergence even at fairly low resolution. In contrast, the results in panels (h) and (o) show that significantly higher resolution is needed to make $N(\psi_{\rm P})$ approach the expected value $1$ when using $\{\sigma_\uGC\etot,\overline{\Lambda},X\}$ and $\{\sigma_\uGC\etot,\overline{\Lambda},P_\zeta\}$.

The remaining panels in the second, third and fourth row of Fig.~\ref{fig:b02_jt60_flat30_benchmark} show results for the case with the respective highest resolution; namely, for the black curves in panels (a), (h) and (o).

The number density field $N(R,z)$ in Fig.~\ref{fig:b02_jt60_flat30_benchmark}(b) exhibits a minor depression on the low-field-side (LFS) and a minor elevation on the high-field side (HFS), in the region $0.3\lesssim \psi_{\rm P} \lesssim 0.7$, where the electric field is strong. As discussed in Section~\ref{sec:example_dens} of the main text, we attribute this to the poloidal nonuniformity of the ${\bm E}\times{\bm B}$ velocity and the Shafranov shift.

The toroidal and poloidal flow densities $\Gamma_{\rm tor}$ and $\Gamma_{\rm pol}$ in panels (c)--(e) of Fig.~\ref{fig:b02_jt60_flat30_benchmark} may be compared to those in Fig.~\ref{fig:09_jt60u_flow-prof-sign}(i)--(l) of the main text. The differences are due to the use of different $N_{\rm ref}$ and $T_{\rm ref}$ profiles, which were nonuniform in Fig.~\ref{fig:09_jt60u_flow-prof-sign} and are uniform in Fig.~\ref{fig:b02_jt60_flat30_benchmark}.

Figs.~\ref{fig:b02_jt60_flat30_benchmark}(f) and \ref{fig:b02_jt60_flat30_benchmark}(g) show the spatially integrated velocity distributions $f_{\rm gc}(\mu,\uGC)$ and $f_{\rm gc}(\ekin,\alpha)$. The overall form of the isotropic reference distribution (\ref{eq:fM}) is reproduced. A close look at $f_{\rm gc}(\ekin,\alpha)$ in panel (g) reveals a weak asymmetry between co- and counter-passing domains, which is presumably related to the shift of the trapped-passing boundary in Fig.~\ref{fig:b01_jt60_orbtypes}(c). The trend is reversed when the sign of $E_{r0}$ is flipped (not shown).

The results obtained using the coordinate set $\{\sigma_\uGC\etot,\overline{\Lambda},X\}$ are shown in the central column of Fig.~\ref{fig:b02_jt60_flat30_benchmark}. Overall, one can see qualitative and quantitative agreement with the results in the left column, which suggests that the formulas (\ref{eq:model_dVol_transf_P})--(\ref{eq:model_dVol_transf_canX}) and the methods described in Section~\ref{sec:method_slice} have been correctly implemented. However, one can also see that the results obtained with $\{\sigma_\uGC\etot,\overline{\Lambda},X\}$ converge much more slowly and tend to be ``noisy'' due to the inaccuracies of our simple implementation as mentioned in \ref{apdx:benchmark_setup} above.

\begin{figure}
	[tb]\vspace{-1.5cm}
	\centering
	\includegraphics[width=0.48\textwidth]{\figures/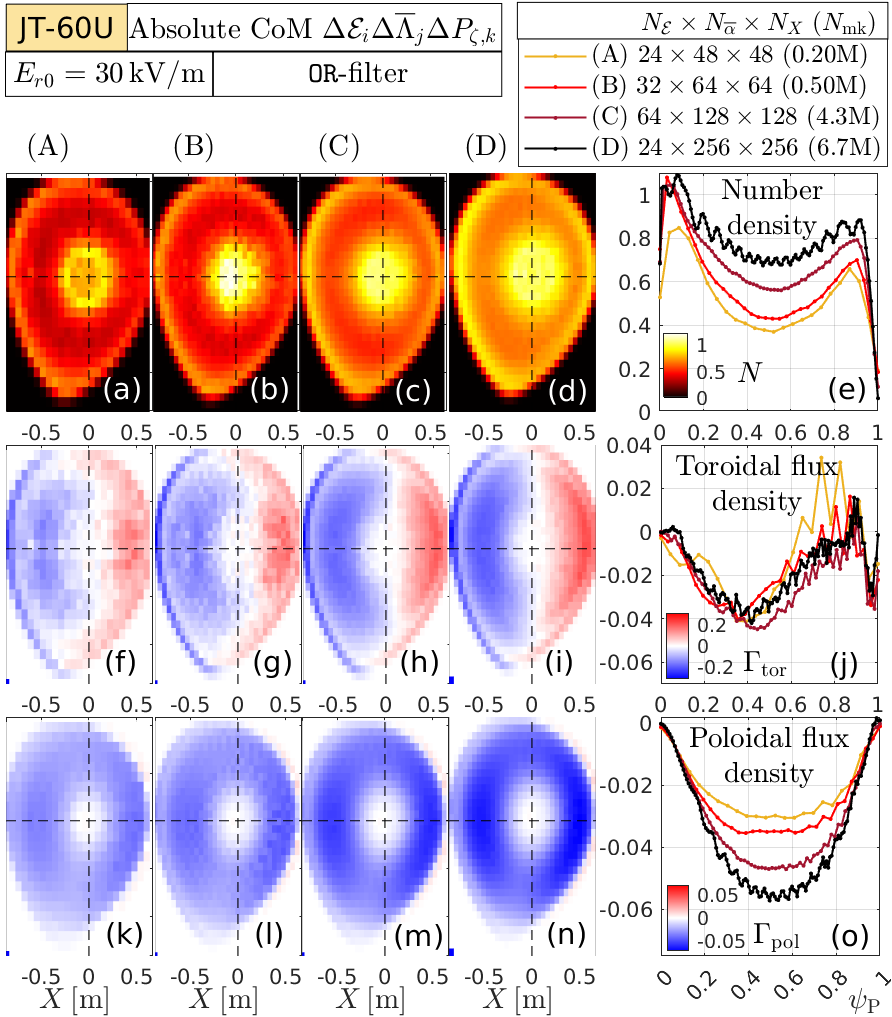}\\
	\includegraphics[width=0.48\textwidth]{\figures/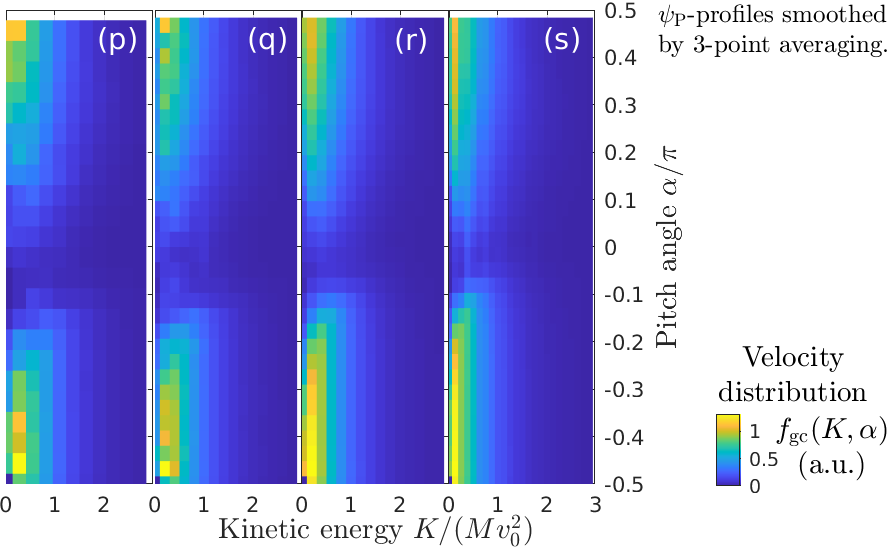}\vspace{-0.25cm}
	\caption{Convergence test for the case with absolute CoM coordinates $\{\sigma_\uGC\etot,\overline{\Lambda},P_\zeta\}$ in Fig.~\protect\ref{fig:b02_jt60_flat30_benchmark}(o)--(u) with exaggerated $E_{r0} = 30\,{\rm kV/m}$. Here, invalid cells are filtered out with the most restrictive {\tt OR} logic. Results are shown for 4 different resolutions, labeled (A)--(D), as indicated in the legend at the top. Panels (a)--(e) show the number density $N$, panels (f)--(o) show the toroidal and poloidal flow densities $\Gamma_{\rm tor}$ and $\Gamma_{\rm pol}$, and panels (p)--(s) show the velocity distribution $f_{\rm gc}(\ekin,\alpha)$.}\vspace{-0.35cm}
	\label{fig:b03_jt60_flat30_can-dP_convergence}%
\end{figure}

\begin{figure}
	[tb]\vspace{-1.5cm}
	\centering
	\includegraphics[width=0.48\textwidth]{\figures/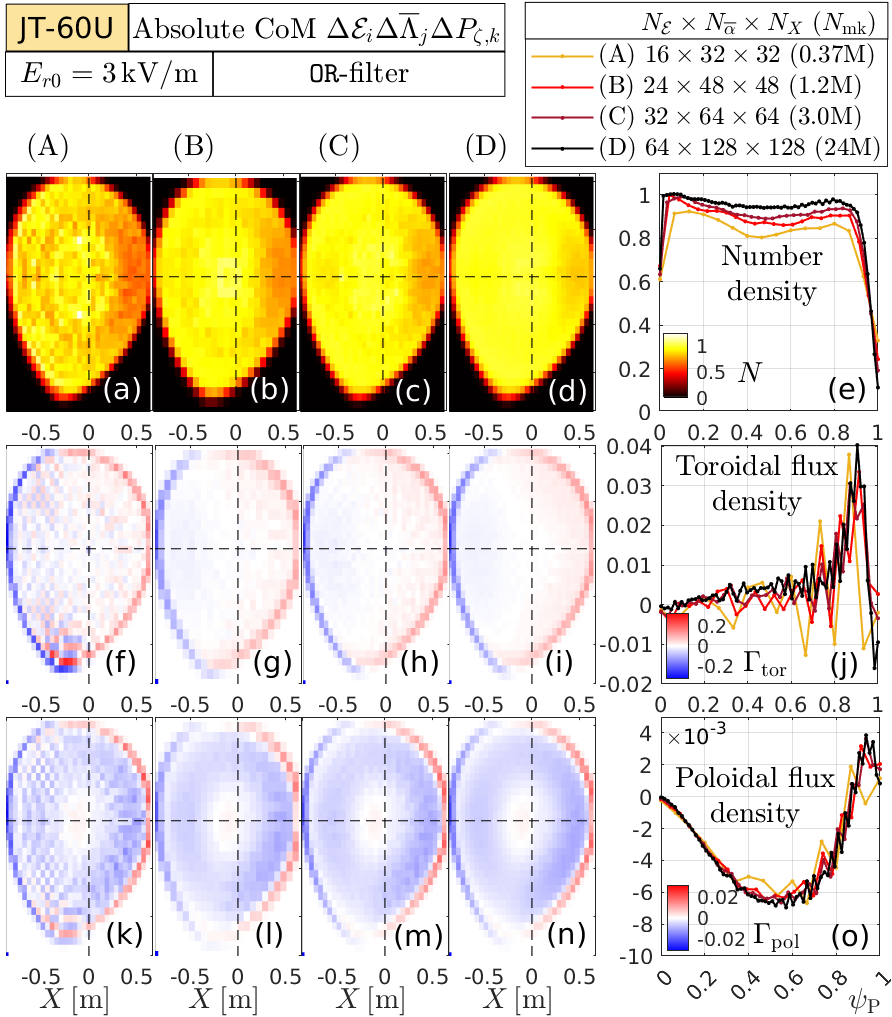}\\
	\includegraphics[width=0.48\textwidth]{\figures/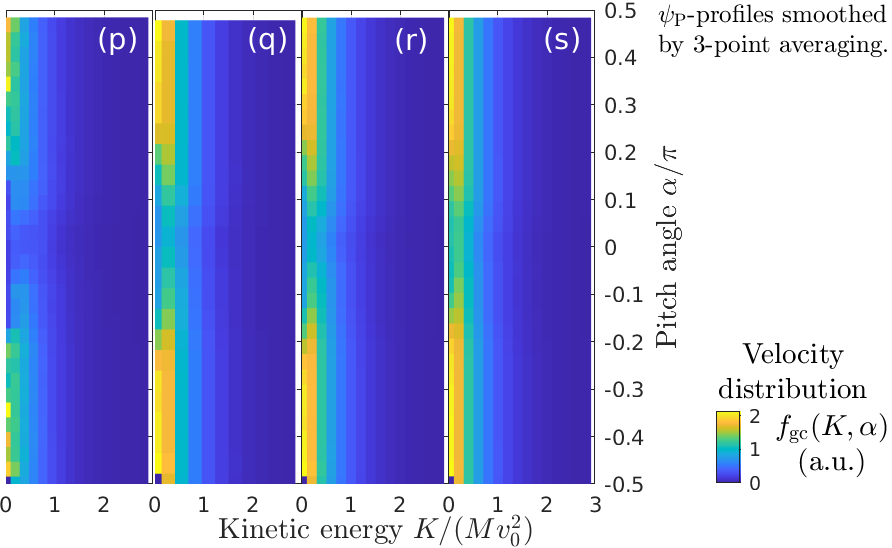}\vspace{-0.25cm}
	\caption{Convergence test for the case with absolute CoM coordinates $\{\sigma_\uGC\etot,\overline{\Lambda},P_\zeta\}$ as in Fig.~\protect\ref{fig:b03_jt60_flat30_can-dP_convergence}, but with more realistic $E_{r0} = 3\,{\rm kV/m}$.}\vspace{-0.3cm}
	\label{fig:b04_jt60_flat3_can-dP_convergence}%
\end{figure}

The situation worsens significantly for the coordinate set $\{\sigma_\uGC\etot,\overline{\Lambda},P_\zeta\}$ as the results in the right column of Fig.~\ref{fig:b02_jt60_flat30_benchmark} show. On the quantitative side, we observe that the magnitudes of all moments (here $N$, $\Gamma_{\rm tor}$, $\Gamma_{\rm pol}$) are severely underestimated. For instance, in Fig.~\ref{fig:b02_jt60_flat30_benchmark}(o) the number density at mid-radius lies about 40\% below the expected value when low resolution $N_{\etot}\times N_{\overline{\Lambda}}\times N_{P_\zeta} = 24\times 48\times 48$ is used. Raising the number of cells to $64\times 128\times 128$ reduces the density shortfall to about 20\%, which is still significant. On the qualitative side, three features stand out:
\begin{enumerate}
	\item[(i)] While the number density field $N(R,z)$ in Fig.~\ref{fig:b02_jt60_flat30_benchmark}(b) is slightly elevated on the HFS and reduced on the LFS, the absolute CoM case in panel (p) shows a broad reduction also on the HFS.
	
	\item[(ii)] The region of positive toroidal flow $\Gamma_{\rm tor} > 0$ (red) on the LFS of Fig.~\ref{fig:b02_jt60_flat30_benchmark}(c) has a crescent shape and dominates over negative flow $\Gamma_{\rm tor} < 0$ (blue) on the HFS, so the flux-surface-averaged profile $\Gamma_{\rm tor}(\psi_{\rm P})$ in panel (e) is positive overall. In contrast, the boundary between the blue and red regions is approximately a straight vertical line in the absolute CoM case in Fig.~\ref{fig:b02_jt60_flat30_benchmark}(q), and the negative component dominates both in panels (q) and (s).
	
	\item[(iii)] The velocity histograms in Fig.~\ref{fig:b02_jt60_flat30_benchmark}(t) and \ref{fig:b02_jt60_flat30_benchmark}(u) exhibit a large deficit of particles around $\uGC \approx 0$.
\end{enumerate}

\noindent If the algorithms are implemented correctly, these discrepancies must gradually decrease with increasingly finer CoM space meshes. The results of some of the convergence tests that we performed are shown in Fig.~\ref{fig:b03_jt60_flat30_can-dP_convergence}--\ref{fig:b05_jt60u_flat30_benchmark_K-win} and discussed in the following section.

\subsection{Convergence with absolute CoM}
\label{apdx:benchmark_convergence}

The Jacobian factor in Eq.~(\ref{eq:model_dVol_transf_can}) is a constant, so it cannot cause qualitative discrepancies. Furthermore, the quadruple benchmark in Fig.~\ref{fig:b02_jt60_flat30_benchmark} rules out some possible mistakes. First, Eqs.~(\ref{eq:model_dVol_transf_P}) and (\ref{eq:model_dVol_transf_X}) have different Jacobians but give identical results (left column). Second, the same code is used to compute $P_\zeta$ for the left and right column, and their Jacobians in Eqs.~(\ref{eq:model_dVol_transf_X}) and (\ref{eq:model_dVol_transf_canX}) are identical. Third, the same code is used to slice the CoM space into $\Delta\etot\Delta\overline{\Lambda}$ cells for the calculations in the middle and right column. A mistake made in one of these items would affect two of the sets of data and not the others. In this way, many possible mistakes in the code or derivations can be ruled out. It thus remains to explain why the results in the right column of Fig.~\ref{fig:b02_jt60_flat30_benchmark} differ so much, and one should demonstrate that these differences tend to vanish with increasing resolution.

First, we confirmed that the magnitude of the discrepancies is sensitive to the filter used to discard invalid cells in the absolute CoM case (see the description in Table~\ref{tab:can_dP_filter} above). The main results in Fig.~\ref{fig:b02_jt60_flat30_benchmark}(o)--(u) were obtained with the {\tt AND}-filter, which retains some marginal cells, while making some error in estimating their size in the $(\overline{\Lambda},P_\zeta)$-plane. For comparison, panels (o) and (s) also show as dashed lines results obtained with the most restrictive {\tt OR}-filter. Comparing these dashed curves with the equally-colored solid curves illustrates how much the filter used to enforce Eq.~(\ref{eq:bench_u20}) affects the systematic underestimation of $N$, $\Gamma_{\rm tor}$ \& $\Gamma_{\rm pol}$.

Additional results obtained with the most restrictive {\tt OR}-filter are summarized in Fig.~\ref{fig:b03_jt60_flat30_can-dP_convergence} for the exaggerated electric field strength $E_{r0} = 30\,{\rm kV/m}$, and in Fig.~\ref{fig:b04_jt60_flat3_can-dP_convergence} for the more realistic value $E_{r0} = 3\,{\rm kV/m}$. One can see that the deficit of particles around $\uGC = 0$ is gradually alleviated with increasing resolution, albeit at a very slow rate.

\begin{figure}
	[tb]\vspace{-0.2cm}
	\centering
	\includegraphics[width=0.48\textwidth]{\figures/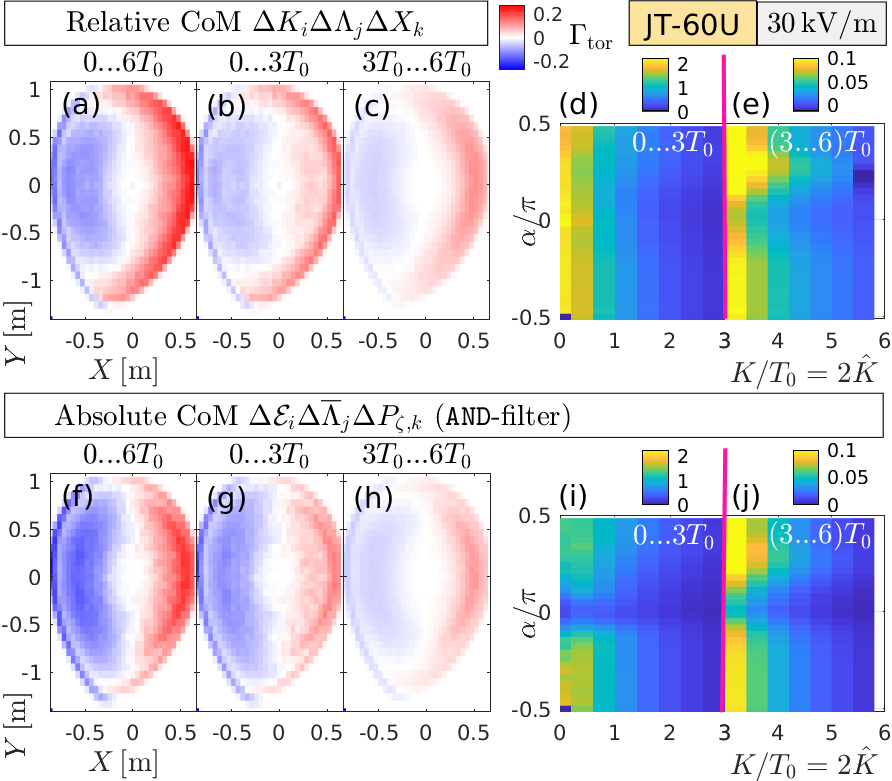}\vspace{-0.2cm}
	\caption{Contributions of thermal and suprathermal particles to the toroidal flow density field $\Gamma_{\rm tor}(X,Y)$ in the JT-60U case with exaggerated $E_{r0} = 30\,{\rm kV/m}$. Results are compared for the orbit space sliced in midplane-based relative CoM (top) and absolute CoM (bottom). For the latter, invalid cells are discarded using the less restrictive {\tt AND}-filter (cf.~Table~\protect\ref{tab:can_dP_filter}). Panels (a) and (f) show the respective $\Gamma_{\rm tor}$ fields integrated over the entire loaded energy range, $\ekin \in [0,6T_0]$. Panels (b) and (g) show the contribution of particles in the thermal range $\ekin \in [0,3T_0]$, and their spatially integrated energy-pitch distributions6 $f_{\rm gc}(\ekin,\alpha)$ are shown in panels (d) and (i). The corresponding data for the suprathermal range $\ekin \in [3T_0,6T_0]$ are shown in panels (c), (e), (h), (j). Note that the color scale of (e) and (j) differs from (d) and (i).}\vspace{-0.2cm}
	\label{fig:b05_jt60u_flat30_benchmark_K-win}%
\end{figure}

Second, we confirmed that the discrepancies are dominated by particles near the thermal energy and below ($\ekin \lesssim T$). Results of this test are summarized in Fig.~\ref{fig:b05_jt60u_flat30_benchmark_K-win}. In the case of relative CoM in the upper row of Fig.~\ref{fig:b05_jt60u_flat30_benchmark_K-win}, the toroidal flow field $\Gamma_{\rm tor}(R,z)$ has a similar form in the thermal energy band (b) $\ekin \in [0,3T_0]$ and in the supra-thermal energy band (c) $\ekin \in [3T_0,6T_0]$. The corresponding results for absolute CoM in the lower row of Fig.~\ref{fig:b05_jt60u_flat30_benchmark_K-win} tend to agree better at higher energy (h) but deviate more at lower energy (g). This discrepancy coincides with a lack of deeply trapped and barely passing particles in the velocity distribution (i) obtained with absolute CoM, while the relative CoM gave a relatively isotropic distribution (d) at low $\ekin$. The velocity distributions (e) and (j) for higher energies are more similar.

\subsection{Summary and technical remarks}
\label{apdx:benchmark_summary}

Overall, one can see that the convergence of the radial profiles in Fig.~\ref{fig:b02_jt60_flat30_benchmark}(o) and (s) goes in the right direction. The same is true for the 2D fields and velocity space histograms in Fig.~\ref{fig:b03_jt60_flat30_can-dP_convergence}. The results converge also towards higher particle energies as shown in Fig.~\ref{fig:b05_jt60u_flat30_benchmark_K-win}, which is of course to be expected because faster particles are less sensitive to the electric field. By the same token, the discrepancies are reduced for weaker $E_{r0}$ at given temperature as can be seen in Fig.~\ref{fig:b04_jt60_flat3_can-dP_convergence}.

Convergence tests are always limited by the law of diminishing returns: the computational cost for our method of slicing in absolute CoM space increases exponentially, while its numerical convergence rate slows down. Given the very slow convergence in Fig.~\ref{fig:b03_jt60_flat30_can-dP_convergence}, one may ask which method --- absolute or relative CoM space slicing --- yields a more accurate qualitative form of the results. In particular, one may ask:
\begin{itemize}
	\item  Could it be that, in the presence of $E_r$, there should remain a certain deficit of particles around $\uGC \approx 0$ at low energies? In other words, could it be that the true GC distribution lies somewhere between the results in Fig.~\ref{fig:b05_jt60u_flat30_benchmark_K-win}(d) and \ref{fig:b05_jt60u_flat30_benchmark_K-win}(i)?
	\item  Could it be that the $E_r$ field really distorts the toroidal flow pattern $\Gamma_{\rm tor}(R,z)$ as in Fig.~\ref{fig:b05_jt60u_flat30_benchmark_K-win}(g), or should it remain largely independent of $E_r$ (and $\ekin$) as in Fig.~\ref{fig:b05_jt60u_flat30_benchmark_K-win}(b) and \ref{fig:b05_jt60u_flat30_benchmark_K-win}(c)?
\end{itemize}

We are presently convinced that the answer to both questions is ``no'', and that the results of the relative CoM space slicing method in the left column of Fig.~\ref{fig:b02_jt60_flat30_benchmark} and upper row of Fig.~\ref{fig:b05_jt60u_flat30_benchmark_K-win} are accurate. This is supported by the following arguments (paraphrasing and elaborating the first paragraph of \ref{apdx:benchmark_convergence}):
\begin{itemize}
	\item  The fact that $\uGC = 0$ is a straight line in Fig.~\ref{fig:04_kstar_grid}(b)--(c) allows us to compute volume elements accurately in relative CoM space. That meshing algorithm is not affected by $E_r$ and has already been verified in Ref.~\cite{Bierwage22a} by transforming the constructed distribution functions repeatedly back and forth between 3D CoM and 4D non-CoM spaces.
	\item  The quantitative and qualitative agreement between the results in the left and central columns of Fig.~\ref{fig:b02_jt60_flat30_benchmark} proves that the modified meshing procedure in Fig.~\ref{fig:04_kstar_grid}(d)--(f) is equivalent to the previously verified procedure in Fig.~\ref{fig:04_kstar_grid}(a)--(c). The only difference between the central and right columns of Fig.~\ref{fig:b02_jt60_flat30_benchmark} is that the former uses $\Delta X$ (with algebraic Jacobian factor $|{\rm d}P_\zeta/{\rm d}X|$) instead of $\Delta P_\zeta$.
	
	\item  The correct form and implementation of the Jacobian factor $\F \equiv {\rm d}P_\zeta/{\rm d}X$ in Eq.~(\ref{eq:method_jac_f}) is verified by the fact that identical results were obtained with the coordinate sets $\{\sigma_\uGC\ekin,\Lambda,X\}$ and $\{\sigma_\uGC\ekin,\Lambda,P_\zeta\}$.
	
	\item  What remains as a source of discrepancy is the loss or corruption of cells around the $\uGC = 0$ boundary of the absolute CoM space, and we have demonstrated in Figs.~\ref{fig:b02_jt60_flat30_benchmark}(o,s) and \ref{fig:b03_jt60_flat30_can-dP_convergence} that the results are indeed sensitive to the filters we use (cf.~Table~\ref{tab:can_dP_filter}).
\end{itemize}

\noindent  Thus, based on the results of the tests performed, we assert that the discrepancies are due to a deficit of deeply trapped, barely passing and potato orbits when slicing the GC orbit space in absolute CoM, where volume elements around $\uGC = 0$ --- namely, near the red line in Fig.~\ref{fig:04_kstar_grid}(e,f) --- are computed inaccurately or lost at an intolerably large rate in our simple implementation.

In conclusion, the use of midplane-based relative CoM for orbit space slicing as in Fig.~\ref{fig:04_kstar_grid}(a)--(c) is attractive in that it is both simple and performs well. That is, it has a relatively low computational cost due to rapid numerical convergence while requiring minimal effort in mesh design. One only needs to ensure that the limits $\ekin \in [\ekin_{\rm min}, \ekin_{\rm max}]$ of the kinetic energy window being sampled are wide enough to capture all relevant orbits. This is important here because our factor $1/2$ in Eq.~(\ref{eq:model_dVol_transf}) assumes that each GC orbit is counted twice --- once at each midplane crossing --- while its kinetic energy at these locations differs in the presence of $E_r$.

\setcounter{figure}{0}
\section{GC motion in the presence of $E_r$}
\label{apdx:gc}

\subsection{Analytical estimates of orbit width and parallel velocity modulation}
\label{apdx:gc_dr_du}

First some preliminaries. The simplified midplane is taken to be located at $z\approx 0$. Each orbit crosses the midplane at two locations along the major radius, which we denote as $R_{\rm in}$ (inner) and $R_{\rm out}$ (outer). We assume that the major radius $R$ is related to the minor radius $r$ and poloidal angle $\vartheta$ as
\begin{equation}
	R \approx R_0 + r\cos\vartheta.
	\label{eq:gc_Rr}
\end{equation}

\noindent The orbit's midplane crossings in minor radius are $r_{\rm in}$ and $r_{\rm out}$. The orbit width $\Delta r_{\rm orb}$ and mean radius $\overline{r}$ are then defined as
\begin{subequations}
	\begin{align}
		\Delta r_{\rm orb} \equiv&\; r_{\rm out} - r_{\rm in} \\
		=&\; \left\{
		\begin{array}{ll}
			R_{\rm out} - R_{\rm in} & : {\rm trapped}, \\
			R_{\rm out} + R_{\rm in} - 2R_0 & : {\rm passing},
		\end{array}\right. \nonumber \\
		\overline{r} \equiv&\; \frac{r_{\rm out} + r_{\rm min}}{2} \\
		=&\; \left\{
		\begin{array}{ll}
			(R_{\rm out} + R_{\rm in} - 2R_0)/2 & : {\rm trapped}, \\
			(R_{\rm out} - R_{\rm in})/2 & : {\rm passing}.
		\end{array}\right. \nonumber
	\end{align}
	\label{eq:gc_rorb}
\end{subequations}

\noindent The reverse relations are
\begin{subequations}
	\begin{align}
		R_{\rm out} \equiv&\; R_0 + \overline{r} + \Delta r_{\rm orb}/2, \\
		R_{\rm in} \equiv&\; \left\{
		\begin{array}{ll}
			R_0 + \overline{r} - \Delta r_{\rm orb}/2 & : {\rm trapped}, \\
			R_0 - \overline{r} - \Delta r_{\rm orb}/2 & : {\rm passing}.
		\end{array}\right.
	\end{align}
	\label{eq:gc_r}
\end{subequations}

\noindent The difference in the magnetic field strength $B \approx B_0 R_0/R$ and the electric potential difference across the orbit width are taken to be
\begin{subequations}
	\begin{align}
		\Delta \hat{B}_{\rm orb} \approx&\; \left\{
		\begin{array}{ll}
			-\Delta r_{\rm orb}/R_0 & : {\rm trapped}, \\
			-2\overline{r}/R_0 & : {\rm passing},
		\end{array}\right. \\
		\Delta\Phi_{\rm orb} \approx&\; -E_r\Delta r_{\rm orb} \quad \text{with} \;\, E_r \equiv v_E B_0 \approx {\rm const}.
	\end{align}
	\label{eq:gc_dB_dE}\vspace{-0.3cm}
\end{subequations}

\noindent Terms of higher order in $\O(r/R_0)$ and $\O(\Delta r_{\rm orb}/r)$ were omitted. For the poloidal magnetic flux function, we let
\begin{equation}
	\Delta\Psi_{\rm P,orb}  \approx B_{\rm pol} R \Delta r_{\rm orb},
\end{equation}

\noindent since $B_{\rm pol}(z\approx 0) \approx B_z = R^{-1}{\rm d}\Psi_{\rm P}/{\rm d}R$ in the midplane. We are now ready for the derivation.

Conservation of total energy $\etot$ in Eq.~(\ref{eq:model_com_etot}) and conservation of toroidal momentum $P_\zeta$ in Eq.~(\ref{eq:model_com_pzeta}) imply that, at an arbitrary point along the GC orbit, the parallel energy gain $\Delta\hat{u}^2 \equiv \hat{u}^2 - \hat{u}_{\rm I}^2$ and the parallel momentum gain $\Delta\hat{u} \equiv \hat{u} - \hat{u}_{\rm I}$ relative to the initial value $\hat{u}_{\rm I}$ are
\begin{subequations}
	\begin{align}
		\frac{\Delta\hat{u}^2}{2} =&\; \hat{v}_{\rm E} \frac{\Delta r}{\varrho_0} - \hat{\mu}\Delta \hat{B} \\
		\Delta\hat{u} \approx&\; \frac{\Delta\Psi_{\rm P}}{\varrho_0 I}\frac{B}{B_0} + \hat{u} \frac{\Delta B}{B},
	\end{align}
	\label{eq:gc_du2_du}
\end{subequations}

\noindent where we ignored $\Delta I/I$ since it is small in usual tokamaks. Applying Eq.~(\ref{eq:gc_du2_du}) to the orbit's midplane crossings,  assuming $B \approx B_{\rm tor}$, and ignoring corrections of higher order in $\O(r/R_0)$, we obtain
\begin{subequations}
	\begin{align}
		\hspace{-0.3cm} \frac{\Delta\hat{u}^2_{\rm orb}}{2} \approx&\; \hat{v}_{\rm E} \frac{\Delta r_{\rm orb}}{\varrho_0} + \hat{\mu}\left\{
		\begin{array}{ll}
			\frac{\Delta r_{\rm orb}}{R_0} & : {\rm trapped}, \\
			2\frac{\overline{r}}{R_0} & : {\rm passing},
		\end{array}\right. \\
		\hspace{-0.3cm} \Delta\hat{u}_{\rm orb} \approx&\; \frac{B_{\rm pol}}{B_0}\frac{\Delta r_{\rm orb}}{\varrho_0} - \hat{u}\left\{
		\begin{array}{ll}
			\frac{\Delta r_{\rm orb}}{R_0} & : {\rm trapped}, \\
			2\frac{\overline{r}}{R_0} & : {\rm passing}.
		\end{array}\right.
	\end{align}
	\label{eq:gc_du2_du_orb}
\end{subequations}

\noindent where $\Delta\hat{u}_{\rm orb}^2 \equiv \hat{u}_{\rm out}^2 - \hat{u}_{\rm in}^2$ and $\Delta\hat{u}_{\rm orb} \equiv \hat{u}_{\rm out} - \hat{u}_{\rm in}$. Introducing the mean velocity,
\begin{equation}
	\hat{u} \rightarrow \overline{\hat{u}} \equiv \hat{u}_{\rm in} + \Delta\hat{u}/2 \equiv \hat{u}_{\rm out} - \Delta\hat{u}/2,
	\label{eq:gc_umean}
\end{equation}

\noindent we obtain the identities\footnote{Equations~(\protect\ref{eq:gc_du2_du_orb}) and (\protect\ref{eq:gc_du2}) yield quadratic equations for $\Delta r$, which have two solutions each. The physical meaning of these solution can be inferred by inspecting their behavior in the limit of vanishing inverse aspect ratio $r/R_0$. One of these solutions yields $\Delta r_{\rm orb} \rightarrow 0$ and corresponds to the cylindrical limit, where $r/R_0$ becomes small as $R_0 \rightarrow \infty$. The other solution yields nonzero $\Delta r_{\rm orb}$ and corresponds to the case where $r \rightarrow 0$ while $R_0$ remains finite. Here we are interested in the latter solution, with nonzero toroidal curvature.}
\begin{subequations}
	\begin{align}
		\hspace{-0.3cm} (\Delta u_{\rm orb})^2 =&\; \Delta u_{\rm orb}^2 - 2 u_{\rm in}\Delta u_{\rm orb}
		\label{eq:gc_du2_trap1} \\
		=&\; 2 u_{\rm out}\Delta u_{\rm orb} - \Delta u_{\rm orb}^2,
		\label{eq:gc_du2_trap2} \\
		\hspace{-0.3cm} \Delta u_{\rm orb}^2 =&\; (u_{\rm out} + u_{\rm in})(u_{\rm out} - u_{\rm in}) = 2\overline{u}\Delta u_{\rm orb}.
		\label{eq:gc_du2_pass}
	\end{align}
	\label{eq:gc_du2}\vspace{-0.3cm}
\end{subequations}

\noindent Eqs.~(\ref{eq:gc_du2_trap1}) or (\ref{eq:gc_du2_trap2}) are useful for trapped orbits, whereas Eq.~(\ref{eq:gc_du2_pass}) is useful for passing orbits.

For trapped orbits, we assume that $\hat{u}\varrho_0/R_0 \ll B_{\rm pol}/B_0$, and obtain
\begin{subequations}
	\begin{align}
		\Delta\hat{u}_{\rm trap} \approx&\; \frac{B_{\rm pol}}{B_0}\frac{\Delta r_{\rm trap}}{\varrho_0},
		\label{eq:gc_du_trap} \\
		\frac{\Delta r_{\rm trap}}{\varrho_0} \approx&\; 2\frac{B_0^2}{B_{\rm pol}^2}\left(\hat{v}_{\rm E} + \hat{\mu}\frac{\varrho_0}{R_0} - \frac{B_{\rm pol}}{B_0} \hat{u}_{\rm in} \right)
		\label{eq:gc_dr_trap}
	\end{align}
	\label{eq:gc_du_dr_trap}\vspace{-0.3cm}
\end{subequations}

\noindent When $\hat{v}_{\rm E}$ is positive or has a negligibly small negative value, negative $\hat{u}_{\rm in} < 0$ yields positive $\Delta r_{\rm trap}$. Only in this case, $\hat{u}_{\rm in}$ retains its original meaning as the trapped orbit's inner midplane crossing. In contrast, if one chooses $\hat{u}_{\rm in} > 0$, or if $\hat{v}_{\rm E}$ has a sufficiently large negative value, one may obtain $\Delta r_{\rm trap} < 0$, which means that $\hat{u}_{\rm in}$ acquires the meaning of the orbit's outer midplane crossing. An example was presented in Fig.~\ref{fig:13_jt60_deep-trapped}.

For passing orbits, we assume that $\hat{u}^2 \gg \hat{\mu}$, and obtain
\begin{subequations}\vspace{-0.5cm}
	\begin{align}
		\hspace{-0.3cm} \Delta\hat{u}_{\rm pass} \approx&\; \frac{B_{\rm pol}}{B_0}\frac{\Delta r_{\rm pass}}{\varrho_0} - 2\overline{\hat{u}}\frac{\overline{r}}{R_0} \approx 2\frac{\overline{r}}{R_0} \frac{B_0}{B_{\rm pol}} \hat{v}_{\rm E},
		\label{eq:gc_du_pass} \\
		\hspace{-0.3cm} \frac{\Delta r_{\rm pass}}{\varrho_0} \approx&\; 2\frac{\overline{r}}{R_0}\frac{B_0}{B_{\rm pol}} \left(\overline{\hat{u}} + \frac{B_0}{B_{\rm pol}}\hat{v}_{\rm E}\right) \rightarrow \frac{2\Delta R_{\rm stag}}{\varrho_0},
		\label{eq:gc_dr_pass}
	\end{align}
	\label{eq:gc_du_dr_pass}\vspace{-0.3cm}
\end{subequations}

\noindent where the large term $-2\overline{\hat{u}}\overline{r}/R_0$ in Eq.~(\ref{eq:gc_du_pass}) was canceled by an equivalent term from Eq.~(\ref{eq:gc_dr_pass}). It is worth noting that the quantity $\Delta r_{\rm pass}/2 \equiv (X_{\rm out} + X_{\rm in})/2$ in Eq.~(\ref{eq:gc_dr_pass}) measures the distance $\Delta R_{\rm stag} = R_{\rm stag} - R_0$ of the stagnation point from the magnetic axis, which is positive/negative for co/counter-current passing orbits ($\overline{\hat{u}} \gtrless 0$). Positive and negative values of $\Delta r_{\rm pass}$ should be interpreted in this way.

\subsection{Frequency shifts}
\label{apdx:gc_dw}

Notice that $\hat{P}_\zeta = -\hat{\Psi}_{\rm P} + \varrho_0\hat{u}\hat{I}/\hat{B}$ is still a constant (though with a different value) when the parallel GC velocity is transformed as
\begin{equation}
	u' = u + 2\Delta\omega_{\rm tor} \frac{R_0^2 B(R,z)}{I(r)} = u + 2\Delta\omega_\uGC(R,z) \frac{R_0^2}{R},
	\label{eq:u_transform}
\end{equation}

\noindent with an arbitrary constant $2\Delta\omega_{\rm tor}$ that has the units of an angular frequency and may be interpreted as a toroidal (precessional) frequency shift. For simplicity, we assume that $\Delta\hat{u}_{\rm orb}$ is related to the mean parallel frequency shift $\Delta\omega_\uGC \equiv \Delta\omega_{\rm tor} B/B_{\rm tor} \approx {\rm const}$.\ in Eq.~(\ref{eq:u_transform}) as
\begin{equation}
	\Delta\hat{\omega}_{\uGC} \approx \frac{1}{R_0} \frac{\Delta\hat{\uGC}_{\rm orb}}{2}.
\end{equation}

\noindent The factor $1/2$ accounts for the fact that $\Delta\hat{u}_{\rm orb}$ represents the full magnitude of the parallel velocity modulation between an orbit's LFS and HFS midplane crossings, so the overall (orbit-averaged) offset is approximately half its value (except perhaps close to the trapped-passing boundary).

For trapped particles, $\Delta\hat{\uGC}_{\rm orb}$ in Eq.~(\ref{eq:gc_du_trap}) contributes the magnetic component (due to the mirror force)
\begin{equation}
	\Delta\hat{\omega}_{u,{\rm M}} \approx \frac{1}{R_0} \frac{\Delta\hat{u}_{\rm orb,M}}{2} \approx \frac{B_0}{B_{\rm pol}} \frac{\varrho_0}{R_0^2}\hat{\mu} - \frac{\hat{u}_{\rm in}}{R_0}.
	\label{eq:gc_OmM}
\end{equation}

\noindent The contribution of the electric component from Eqs.~(\ref{eq:gc_du_trap}) and (\ref{eq:gc_du_pass}) is
\begin{align}
	\Delta\hat{\omega}_{u,{\rm E}}^{\rm (I)} \approx&\; \frac{1}{R_0} \frac{\Delta\hat{u}_{\rm orb,E}}{2} \approx \frac{\hat{v}_{\rm E}}{R_0} \times \left\{
	\begin{array}{ll}
		\frac{B_0}{B_{\rm pol}} & : {\rm trapped}, \\
		\frac{\overline{r}B_0}{R_0 B_{\rm pol}} \approx q & : {\rm passing}.
	\end{array}\right.
	\label{eq:gc_OmE}
\end{align}

\noindent As indicated by the superscript ``(I)'' that we introduced in Eq.~(\ref{eq:gc_shift_loc}) of Section~\ref{sec:gc_intro} of the main text, the quantity $\Delta\hat{\omega}_{u,{\rm E}}^{\rm (I)}$ is interpreted as the electric frequency shift at a specific point $\{{\bm C}_{\rm I},\tau_{\rm I}\}$ in the lab frame. This is because Eq.~(\ref{eq:gc_OmE}) results from taking the difference
\begin{equation}
	\Delta\hat{u}_{\rm orb,E} = \Delta\hat{u}_{\rm orb}(E_r|{\bm C}_{\rm I},\tau_{\rm I}) - \Delta\hat{u}_{\rm orb}(0|{\bm C}_{\rm I},\tau_{\rm I}),
	\label{eq:gc_OmE_I}
\end{equation}

\noindent while fixing $\hat{\mu}$ and $\hat{u}_{\rm in}$ in Eq.~(\ref{eq:gc_du_dr_trap}) for trapped orbits, and $\overline{r}$ and $\overline{\hat{u}}$ in Eq.~(\ref{eq:gc_du_dr_pass}) for passing orbits. It is evident that Eq.~(\ref{eq:gc_OmE_I}) corresponds to Eq.~(\ref{eq:gc_shift_loc}) of Section~\ref{sec:gc_intro}.

The ``Doppler shift'' of a specific orbit, which is identified in terms of its poloidal contour $\{R_{\rm gc}(\tau|{\bm C},E_r),z_{\rm gc}(\tau|{\bm C},E_r)\}$, and which is denoted by $\Delta\omega_{\rm E}^{\rm (0)}$ in Eq.~(\ref{eq:gc_shift_orb}) may be estimated as follows. By definition, the ``original'' and the ``Doppler shifted'' orbit must have the same radial width $\Delta r_{\rm orb}$ and initial position, and differ only in their velocity and pitch.

For trapped orbits, Eq.~(\ref{eq:gc_du_dr_trap}) yields the conditions
\begin{equation}
	\Delta\hat{u}_{\rm in}^{\rm (0)} = \frac{B_0}{B_{\rm pol}}\left(\hat{v}_{\rm E} + \Delta\hat{\mu}^{\rm (0)}\frac{\varrho_0}{R_0}\right);
	\label{eq:gc_du0_trap}
\end{equation}

\noindent where $\Delta\hat{u}_{\rm in}^{\rm (0)} \equiv \hat{u}_{\rm in} - \hat{u}_{\rm in}^{\rm (0)}$ and $\Delta\hat{\mu}^{\rm (0)} \equiv \hat{\mu} - \hat{\mu}^{\rm (0)}$. The superscript ``(0)'' was introduced in Eq.~(\ref{eq:gc_shift_orb}) of Section~\ref{sec:gc_intro} to identify quantities evaluated in the idealized ``$E_r=0$ frame''. Equation~(\ref{eq:gc_du0_trap}) was successfully applied in Eq.~(\ref{eq:gc_doppler0_bench}) of Section~\ref{sec:gc_er0frame}.

For passing orbits, Eq.~(\ref{eq:gc_du_dr_pass}) gives
\begin{equation}
	\Delta\hat{u}^{\rm (0)}_{\rm pass} \approx 0,
	\label{eq:gc_du0_pass}
\end{equation}

\noindent at leading order, which is consistent with the vanishing toroidal frequency shift predicted in Eq.~(\ref{eq:gc_shift_tor}) based on our consideration on cancellation of the reference point bias in Section~\ref{sec:gc_smooth}. The poloidal frequency shift would in principle appear at higher order and was estimated in Eq.~(\ref{eq:gc_shift_pass}) to be approximately the ${\bm E}\times{\bm B}$ rotation velocity. This was confirmed in Eq.~(\ref{eq:gc_wpol_ve_r}) of Section~\ref{sec:gc_eaccel}.

One may wonder whether the additive decomposition into electric and magnetic components in Eqs.~(\ref{eq:gc_OmM}) and (\ref{eq:gc_OmE}) is justified. After all, their contributions depend on each other, as is evident from the underlying equations of motion (\ref{eq:gc_norm}), where the mirror force and magnetic drift determine how much time the GC spends in a region with a certain $E_r$, while the electric acceleration and electric drift determine how much time the GC spends in a region with a certain $\nablab B$. For instance, an orbit that starts in a region where $E_r = 0$ may never see an electric field when magnetic drifts are ignored. Nevertheless, we believe that the additive decomposition in the above Eqs.~(\ref{eq:gc_OmM}) and (\ref{eq:gc_OmE}), and in Eq.~(\ref{eq:gc_examples_dr}) of the main paper, is justified because the underlying Eqs.~(\ref{eq:gc_du_trap}) and (\ref{eq:gc_du_pass}) were derived on the basis of momentum and energy conservation, which hold independently at each point of a GC orbit. One should however keep in mind that this decomposition is artificial in the sense that it is not based on instantaneous drifts but incorporates prior knowledge about the global orbit contour and its overlap with the nonuniform ambient electric field. More care may be required when dealing with the resulting mean flows as in Eq.~(\ref{eq:example_vE}) of Section~\ref{sec:example_results} of the main paper.

\subsection{Alternative equations of motion with $\varrho_\parallel$}
\label{apdx:gc_alt}

The standard GC pushing routine of {\tt VisualStart} was benchmarked against another version that solves for $\rhoPar \equiv u/\omega_{\rm B}$ instead of $\uGC$. The results were identical within the level of accuracy required for this work.\footnote{More detailed analyses indicate that the formulation in terms of $\uGC$ seems to be more accurate than $\rhoPar$, at least in stellarators. The difference may be less pronounced in up-down symmetric tokamaks. (Private communication: Akinobu Matsuyama, Kyoto University/Japan.)}
The equations for $\uGC$ and $\rhoPar$ differ in terms of the explicit or implicit appearance of parallel electric acceleration, whose influence has been of some interest in this work, so we outline and discuss the derivation here. Colors are used to track three relevant terms from their first appearance until they vanish by cancellation or orthogonality.

The GC Lagrangian $\L$ and Hamiltonian $\H$ written in terms of $\rhoPar \equiv u/\omega_{\rm B}$ instead of $u$, are
\begin{subequations}
	\begin{align}
		\L =&\; Ze\overbrace{({\bm A} + \rhoPar{\bm B})}\limits^{{\bm A}^*_{\rhoPar}}\cdot\dot{\bm X} + J_1\dot{\xi} - \H,
		\\
		\H =&\; \underbrace{\frac{(Ze)^2 B^2}{M}}\limits_{M\omega_{\rm B}^2} \frac{\rhoPar^2}{2} + \underbrace{\mu B + Ze\Phi}\limits_{Ze\Phi^*}.
	\end{align}
\end{subequations}

\noindent For stationary fields ($\partial_t\Phi = \partial_t{\bm A} = 0$), variation ${\rm d}_t(\partial_{\dot{\eta}}\L) = \partial_\eta\L$ of the Lagrangian with respect to $\rhoPar$ and $\dot{\bm X}$ gives, respectively,
\begin{subequations}
	\begin{align}
		0 = Ze{\bm B}\cdot\dot{\bm X} - M\omega_{\rm B}^2\rhoPar \quad &\Rightarrow \rhoPar = \hat{\bm b}\cdot\dot{\bm X}/\omega_{\rm B},
		\label{eq:alt_rhoPar}
		\\
		\overbrace{{\bm B}\dot{\varrho}_\parallel + \dot{\bm X}\cdot\nablab{\bm A} + \rhoPar\dot{\bm X}\cdot\nablab{\bm B}}\limits^{{\rm d}_t{\bm A}^*_{\rhoPar}} =&\; \nablab{\bm A}^*_{\rhoPar}\cdot\dot{\bm X} - \overbrace{\nablab\Phi^*}\limits^{-{\bm E}^*} \nonumber \\
		&\; {\cb -\; \omega_{\rm B}\rhoPar^2\nablab B}.
		\label{eq:alt_dotX_dotrhoPar}
	\end{align}
\end{subequations}

\noindent  Using the identities
\begin{subequations}
	\begin{align}
		\dot{\bm X}\cdot\nablab{\bm A}^*_{\rhoPar} =&\; \dot{\bm X}\cdot\nablab{\bm A} + \rhoPar\dot{\bm X}\cdot\nablab{\bm B},
		\\
		\nablab{\bm A}^*_{\rhoPar}\cdot\dot{\bm X} =&\; \dot{\bm X}\times(\nablab\times{\bm A}^*_{\rhoPar}) + \dot{\bm X}\cdot\nablab{\bm A}^*_{\rhoPar},
		\\
		{\bm B}^* =&\; \nablab\times{\bm A}^* = \nablab\times{\bm A}^*_{\rhoPar} {\cred\; -\; \rhoPar\hat{\bm b}\times\nablab B},
	\end{align}
\end{subequations}

\noindent Eq.~(\ref{eq:alt_dotX_dotrhoPar}) reduces to
\begin{align}
	{\bm B}\dot{\varrho}_\parallel =&\; \dot{\bm X}\times{\bm B}^* + {\bm E}^* {\cred\; -\;\rhoPar\dot{\bm X}\times(\hat{\bm b}\times\nablab B)} {\cb\; -\; \omega_{\rm B}\rhoPar^2\nablab B}, \nonumber \\
	=&\; \dot{\bm X}\times{\bm B}^* + {\bm E}^* {\cred\; -\;\rhoPar\hat{\bm b}\dot{\bm X}\cdot\nablab B},
	\label{eq:alt_eom_combined}
\end{align}

\noindent Crossing with ${\bm B}\times$ and dotting with ${\bm B}^*\cdot$ separates the equations for $\dot{\bm X}$ and $\dot{\varrho}_\parallel$:
\begin{subequations}
	\begin{align}
		B B^*_\parallel\dot{\bm X} = &\; \rhoPar\omega_{\rm B} B{\bm B}^* + {\bm E}^*\times{\bm B},
		\label{eq:alt_eom_X}
		\\
		B B^*_\parallel \dot{\varrho}_\parallel =&\; {\bm B}^*\cdot{\bm E}^* {\cred\; -\; \rhoPar B_\parallel^* \dot{\bm X}\cdot\nablab B} \nonumber \\
		=&\; {\bm B}^*\cdot{\bm E}^* - \rhoPar^2\omega_{\rm B}{\bm B}^*\cdot\nabla B - \rhoPar\nablab B\cdot({\bm E}^*\times\hat{\bm b}) \nonumber \\
		=&\; {\bm B}\cdot{\bm E}^* + \rhoPar\mu_0{\bm J}\cdot{\bm E}^* - \rhoPar^2\omega_{\rm B}{\bm B}^*\cdot\nablab B,
		\label{eq:alt_eom_rhoPar}
	\end{align}
	\label{eq:alt_eom}
\end{subequations}

\noindent where $\mu_0{\bm J} = \nablab\times{\bm B}$ is the plasma current density.

One can see that the equation for $\dot{\bm X}$ remains unchanged. Notice also that the first line of Eq.~(\ref{eq:alt_eom_rhoPar}) corresponds to the expected relation $\dot{\varrho}_\parallel = {\rm d}(u/\omega_{\rm B})/{\rm d}t = \omega_{\rm B}^{-1}(\dot{u} - u\dot{\bm X}\cdot\nablab\ln B)$ with $\dot{u} = {\bm B}^*\cdot{\bm E}^*/(B B_\parallel^*$). Finally, we can also verify energy conservation ${\rm d}_t\H = 0$:
\begin{align}
	{\rm d}_t\H =&\; \omega_{\rm B} B\rhoPar\dot{\varrho}_\parallel + \omega_{\rm B}\rhoPar^2 \dot{\bm X}\cdot\nablab B - {\cg \dot{\bm X}\cdot{\bm E}^*} \nonumber \\
	=&\; \omega_{\rm B}\rhoPar{\bm B}^*\cdot{\bm E}^*/B^*_\parallel {\cred\; -\; \omega_{\rm B}\rhoPar^2 \dot{\bm X}\cdot\nablab B} \nonumber \\
	&\; + \omega_{\rm B}\rhoPar^2 \dot{\bm X}\cdot\nablab B - {\cg \rhoPar\omega_{\rm B} {\bm B}^*\cdot{\bm E}^*/B_\parallel^*} \nonumber \\
	=&\; 0.
\end{align}

In the presence of a purely radial electric field, we have ${\bm E}^* = E_{\Psi_{\rm P}}\nablab\Psi_{\rm P} - \mu\nabla B/(Ze)$. Since ${\bm B}\cdot\nablab\Psi_{\rm P} = 0$ and ${\bm J}\cdot\nablab\Psi_{\rm P} = 0$, Eq.~(\ref{eq:alt_eom_rhoPar}) for $\rhoPar$ then reduces to
\begin{align}
	B B^*_\parallel \dot{\varrho}_\parallel =&\; -\frac{\mu}{Ze}\overbrace{({\bm B} + \rhoPar\nablab\times{\bm B})}\limits^{{\bm B}^* {\cred + \rhoPar\hat{\bm b}\times\nablab B}}\cdot\nablab B - \rhoPar^2\omega_{\rm B}{\bm B}^*\cdot\nablab B \nonumber \\
	=&\; -\left(\frac{\mu}{Ze} + \rhoPar^2\omega_{\rm B}\right){\bm B}^*\cdot\nablab B,
	\label{eq:alt_eom_rhoPar_Er}
\end{align}

\noindent After applying the normalizations in Eq.~(\ref{eq:norm}), we have $\rhoPar = \varrho_0\hat{u}/\hat{B}$, and the discretized Eq.~(\ref{eq:alt_eom_rhoPar_Er}) becomes
\begin{equation}
	\frac{\Delta\rhoPar}{\varrho_0} = -\Delta \hat{t} \left(\frac{\hat{\mu}}{\hat{B}} + \frac{\rhoPar^2}{\varrho_0^2}\right)  \frac{\hat{\bm B}^*\cdot\nablab \hat{B}}{\hat{B}^*_\parallel},
	\label{eq:alt_eom_rhoPar_Er_nrm}
\end{equation}

Notice that the electric field does not appear explicitly in Eq.~(\ref{eq:alt_eom_rhoPar_Er_nrm}) for $\rhoPar$, whereas Eq.~(\ref{eq:gc_norm_u}) for $u$ contains an electric acceleration term explicitly. This electric acceleration is, however, implicitly contained in the new term $\varrho_\parallel^2$ because $\rhoPar^2\omega_{\rm B} = \frac{2}{M\omega_{\rm B}}(\etot - Ze\Phi - \mu B)$ with $\etot$ and $\mu$ constant, so that $E_{\Psi_{\rm P}}$ will have an effect as soon as the GC departs from its initial magnetic flux surface.

\setcounter{figure}{0}
\section{From adiabatic invariants to CoM}
\label{apdx:can}

The purpose of this Appendix is to discuss in more detail some issues associated with the construction of adiabatic invariants and constants of motion (CoM) in the case of full Lorentz orbits. This will clarify our use of somewhat unconventional notation in Section~\ref{sec:intro_com} and is also closely connected to the issue of identifying a unique midplane crossing point as discussed in Section~\ref{sec:intro_fo}. Since these problems are usually resolved by invoking the abstract concept of a guiding center (GC), we begin by revisiting that model in \ref{apdx:can_gc} before proceeding to full orbits in \ref{apdx:can_full}.

\subsection{Guiding center model}
\label{apdx:can_gc}

In a perfectly stationary and axisymmetric field, unperturbed GC motion is constrained to a perfectly closed orbit surface, more generally known as an invariant torus. Each orbit surface can be identified by a triplet of exact invariants, here called constants of motion (CoM). When symmetries are imperfect but only weakly broken, these toroidal orbit surfaces become fuzzy in space and time. The place of CoM as orbit labels is then taken by adiabatic invariants, which allow to uniquely identify and track the fuzzy toroidal subspace occupied by GCs following a weakly perturbed trajectory.\footnote{As we mentioned in Section~\protect\ref{sec:intro_integrals}, one application example is the study of resonant wave-particle interactions involving so-called adiabatic chirping that can be modeled via phase-space waterbags \protect\cite{Hezaveh22}. Adiabatic invariants are also useful in helical devices \protect\cite{Cary88, Todoroki93}.}

Since the GC model treats the gyrophase $\xi$ as an ignorable symmetry coordinate, a GC's magnetic moment $\mu \equiv Ze J_1/M$ is an {\it exact} invariant by definition, so there are only two {\it adiabatic} invariants, $J_2$ and $J_3$. These are computed as canonical actions by integrating a GC's canonical momentum,
\begin{equation}
	{\bm p} \equiv Ze{\bm A} + Mu\hat{\bm b},
	\label{eq:pcan_gc}
\end{equation}

\noindent along the GC trajectory ${\bm X}_{\rm gc}(t)$ in the poloidal and toroidal angular directions:
\begin{equation}
	J_2 \equiv \oint_{\rm pol}\frac{{\rm d}{\bm X}_{\rm gc}\cdot{\bm p}}{2\pi}, \;\;\; J_3 \equiv \oint_{\rm tor}\frac{{\rm d}{\bm X}_{\rm gc}\cdot{\bm p}}{2\pi}.
	\label{eq:gc_j123}
\end{equation}

\noindent Note that the concrete definitions of the magnetic moment $\mu$ and parallel GC velocity $u$ depend on the number of terms retained in the expansion with respect to the adiabaticity ordering parameter $\varrho_{\rm L}/R_0$ in Eq.~(\ref{eq:com_b}) \cite{Littlejohn83}. This is still a subject of active research \cite{Brizard23}.

In the case of perfectly stationary and axisymmetric fields, one finds (e.g., Section 2.3 in Ref.~\cite{BrochardThesis})\vspace{-0.1cm}
\begin{align}
	J_2 \approx& \oint\frac{{\rm d}\ell_{\rm B}}{2\pi}M u + \left\{\begin{array}{ll}
		0 & : \text{trapped}, \\
		Ze\Psi_{\rm T}(P_\zeta) & : \text{passing},
	\end{array}\right\} = {\rm const}., \nonumber \\
	J_3 =& P_\zeta = {\rm const}.
	\label{eq:gc_j123_sym}
\end{align}\vspace{-0.6cm}

\noindent where $2\pi\Psi_{\rm T}$ is the toroidal magnetic flux and ${\rm d}\ell_{\rm B}$ is a length element along ${\bm B}$. The evaluation of $J_2$ is generally cumbersome, so only an approximation is shown in Eq.~(\ref{eq:gc_j123_sym}). In the time-independent stationary case, it is more convenient to use the total energy (Hamiltonian) $\H = \etot = Mu^2/2 + \mu B + Ze\Phi$ instead of $J_2$, as was done in Sections~\ref{sec:intro_com} and \ref{sec:intro_gc} to define the standard set $\{\mu,\sigma_u\etot,P_\zeta\}$ of absolute CoM.

\subsection{Full Lorentz orbits}
\label{apdx:can_full}

Except for the need to approximate $J_2$ in practical implementations, the definition of the canonical actions and their relation to adiabatic or exact invariants is straightforward in the GC model. This is not so for full Lorentz orbits $\{{\bm x}(t),{\bm v}(t)\}$. When one tries to compute the actions $2\pi\tilde{J}_\eta$ (with $\eta = 1,2,3$) by faithfully integrating the canonical momentum\vspace{-0.1cm}
\begin{equation}
	\tilde{\bm p} \equiv Ze\tilde{\bm A} + M{\bm v}, \quad \text{with} \quad \tilde{\bm A} \equiv {\bm A}({\bm x}(t)),
	\label{eq:pcan_full}\vspace{-0.1cm}
\end{equation}

\noindent along the gyrational, poloidal and toroidal angular projections $\C_\eta$ of the Lorentz orbit's actual trajectory ${\bm x}(t)$ as $\oint_{\C_\eta}{\rm d}{\bm x}\cdot\tilde{\bm p}$, one encounters two problems.

\paragraph{Problem 1: Ambiguous gyrophase $\xi$ and gyroperiod $\tau_{\rm B}$} The underlying question can be stated as follows: If our starting point ${\bm x}(0)$ is taken to correspond to $\xi = 0$, how long should we follow the trajectory ${\bm x}(t)$ for $\xi$ to reach $2\pi$? That is, how long is the gyroperiod $\tau_{\rm B}$? The answer to this question is trivial in the case of particles gyrating in a uniform magnetic field. When viewed in a moving frame of reference where $v_\parallel = 0$, the integral $\oint_{\rm gyro}{\rm d}{\bm x}\cdot\tilde{\bm p}$ is taken over a Larmor circle of radius $\varrho_{\rm L} = M v_\perp/(Ze\tilde{B}) = {\rm const}$. The particle completes this circle precisely during a period $\tau_{\rm B} = 2\pi/\tilde{\omega}_{\rm B}$, where $\tilde{\omega}_{\rm B} \equiv Ze\tilde{B}/M$. A second, slightly less trivial example is that of a particle gyrating in a helical magnetic field with straight cylindrical symmetry. Although $\varrho_{\rm L}$ and $\tilde{\omega}_{\rm B}$ now generally vary in time, the gyroperiod $\tau_{\rm B}$ ($\neq 2\pi/\tilde{\omega}_{\rm B}$!) can be uniquely defined by equating it to the period during which the magnetic field vector ${\bm B}$ oscillates in the particle's (helically rotating) frame of reference.

In a torus, however, such a preferred frame of reference with a unique gyroperiod $\tau_{\rm B}$ does not strictly exist because the mean field strength gradually changes from one gyration to the next as the particle's GC moves between the low- and high-field side of the torus (with smaller and larger $B \propto 1/R$, respectively). To illustrate the ambiguity and the resulting freedom of defining the metric of the gyrophase $\xi$ and the length of its period $\tau_{\rm B}$ in a torus, let us consider a few possible choices. For instance, one may choose to measure the gyrophase $\xi$
\begin{enumerate}
	\item[(a)] relative to the horizontal $R$ axis (where $\vartheta \approx 0,\pi$), or
	\item[(b)] relative to lines of constant $\vartheta$.
\end{enumerate}

\noindent In case (a), $\xi$ corresponds to a local version of the geometric poloidal angle $\vartheta$ measured around the instantaneous GC,\vspace{-0.3cm}
\begin{equation}
	{\bm X}_{\rm gc}(t) = {\bm x}(t) - {\bm \varrho}_{\rm L}(t), \quad \frac{Ze}{M}{\bm \varrho}_{\rm L} \equiv \frac{{\bm v}\times\tilde{\hat{\bm b}}}{\tilde{B}} \equiv \frac{{\bm v}\times\hat{\bm b}({\bm x}(t))}{B({\bm x}(t))}.
	\label{eq:can_gc}
\end{equation}

\noindent Case (b) corresponds to doing the same in a poloidally rotating frame of reference rather than the lab frame. Alternatively, one may even go as far as demanding that
\begin{enumerate}
	\item[(c)] the metric of $\xi$ and the gyroperiod $\tau_{\rm B}$ vary along the particle trajectory in precisely such a way that $\oint_{\rm gyro}{\rm d}{\bm x}\cdot\tilde{\bm p} = {\rm const}$.; i.e., exactly conserved.
\end{enumerate}

\noindent In practice, option (c) may be most easily implemented by integrating not only over one gyroperiod but many. We will return to this idea at the end of this section.

In any case, the gyroperiod $\tau_{\rm B}$ is most likely close but generally not identical to $2\pi/\tilde{\omega}_{\rm B} \neq {\rm const}$. In other words, there is no general unique mapping between the particle position ${\bm x}(t)$ and the gyrophase $\xi \in [0,2\pi)$. The gyroperiod $\tau_{\rm B}$ is ambiguous, and so is the associated action integral $\tilde{J}_1$. To highlight this ambiguity, we wrote the corresponding action integral in Section~\ref{sec:intro_com} in an unconventional way as
\begin{equation}
	\tilde{J}_1 \approx \int_{\rm gyro}\frac{{\rm d}{\bm x}\cdot\tilde{\bm p}}{2\pi},
	\label{eq:can_j1}
\end{equation}

\noindent with an approximation symbol $\approx$ instead of $\equiv$, and with an open integral $\int_{\rm gyro}$ instead of $\oint_{\rm gyro}$.

In the standard literature, this ambiguity is usually papered over by assuming that the gyrofrequency is constant. For instance, one may evaluate it at the GC position (\ref{eq:can_gc}) as $\omega_{\rm B} \equiv ZeB({\bm X}_{\rm gc})/M$, albeit this conflicts with the definition of ${\bm \varrho}_{\rm L}$ in Eq.~(\ref{eq:can_gc}), where $B$ is evaluated at the particle position ${\bm x}(t)$. Alternatively, one may take $\omega_{\rm B}$ to be the gyro-averaged mean value of $Ze\tilde{B}/M$, but this requires prior knowledge of the period $\tau_{\rm B}$ over which the averaging is performed, so one faces the proverbial chicken-and-egg problem. In any case, when $B$ is nonuniform, the assumption of constant gyrofrequency is an approximation, and this is precisely what Eq.~(\ref{eq:can_j1}) is meant to express. In this sense, the fact that $\tilde{J}_1$ for full Lorentz orbits is never an exact but only an adiabatic invariant in a nonuniform field is not only a result of its weak but inevitable residual time-dependence but also its necessarily imprecise definition.

\begin{figure}
	[tb]\vspace{-0.1cm}
	\centering
	\includegraphics[width=0.48\textwidth]{\figures/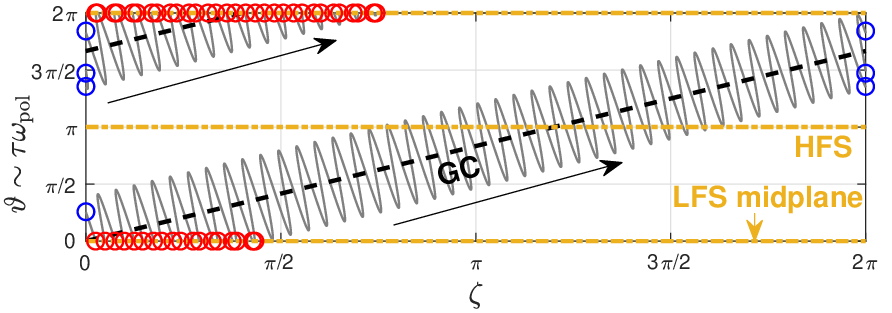}\vspace{-0.3cm}
	\caption{Projections of a Lorentz orbit (gray solid) and its GC (black dashed) in the $(\vartheta,\zeta)$-plane, showing how gyration causes repeated $2\pi$ crossings in $\vartheta$ (red) and $\zeta$ (blue) during one poloidal transit.}\vspace{-0.2cm}
	\label{fig:d01_lorentz2pi}%
\end{figure}

\paragraph{Problem 2: Ambiguous $2\pi$ crossings along $\vartheta$ and $\zeta$} In addition to gyration being an imprecisely defined concept, its presence also spoils the action integrals along the poloidal and toroidal angles. Although the geometric poloidal and toroidal angles $\vartheta \in [0,2\pi)$ and $\zeta \in [0,2\pi)$ (or any equivalent magnetic coordinates) are $2\pi$-periodic by definition (owing to the assumed toroidal geometry), the requirement of the associated action integrals to be performed along the actual particle trajectory ${\bm x}(t)$ means that is not clear when a period of $2\pi$ has been completed. The root of this problem may be said to be the superposition of two helical motions: the helically wound strong guide field ${\bm B}$ and the particle's gyration around that field. This is illustrated schematically in Fig.~\ref{fig:d01_lorentz2pi}: The helical winding of ${\bm B}$ is responsible for the tilt of the particle trajectory in the $(\vartheta,\zeta)$-plane (here, representing straight-field-line coordinates). Due to this tilt, the gyrating particles will generally cross the $0$ and $2\pi$ planes along $\vartheta$ and $\zeta$ multiple times in rapid succession at {\it variable} time intervals $0\lesssim t \lesssim \tau_{\rm B}$. This raises the following questions: Which of these crossings should be regarded as completions of the $2\pi$ periods for the action integrals defining $\tilde{J}_2$ and $\tilde{J}_3$?

To our knowledge, there exists no satisfactory answer, which presumably means that the question itself is ill-posed in the sense that $\tilde{J}_2$ and $\tilde{J}_3$ cannot be defined via integrals along actual Lorentz orbits ${\bm x}(t)$ in a tokamak. One obvious solution to this problem, which indeed seems to be the standard way to deal with this issue, is to determine the timing of a $2\pi$ crossing by monitoring the motion of the instantaneous GC as defined in Eq.~(\ref{eq:can_gc}). The canonical actions can then be defined by replacing ${\bm p}$ with $\tilde{\bm p}$ in Eq.~(\ref{eq:gc_j123}):\vspace{-0.1cm}
\begin{subequations}
	\begin{align}
		\tilde{J}_2 \equiv &\oint_{\rm pol}\frac{{\rm d}{\bm X}_{\rm gc}\cdot\tilde{\bm p}}{2\pi} \approx \int_{\rm pol}\frac{{\rm d}{\bm x}\cdot\tilde{\bm p}}{2\pi}, \\
		\tilde{J}_3 \equiv &\oint_{\rm tor}\frac{{\rm d}{\bm X}_{\rm gc}\cdot\tilde{\bm p}}{2\pi} \approx \int_{\rm tor}\frac{{\rm d}{\bm x}\cdot\tilde{\bm p}}{2\pi}.
	\end{align}
	\label{eq:full_j123}\vspace{-0.3cm}
\end{subequations}

\noindent Clearly, $\tilde{J}_2$ and $\tilde{J}_3$ defined in such a practically useful way are not {\it pure} Lorentz orbit actions (whose existence is, in fact, unclear) since they depend also on the introduction of the GC concept. We highlighted this fact in Section~\ref{sec:intro_com} and in Eq.~(\ref{eq:full_j123}) through unconventional notation, writing approximation symbols $\approx$ instead of $\equiv$, and open integrals $\int$ instead of $\oint$ for $\tilde{J}_2$ and $\tilde{J}_3$.

\paragraph{Conclusion: Necessity of the GC concept} In the literature, the canonical actions $2\pi\tilde{J}_\eta$ are usually equated to $\oint_{\C_\eta}{\rm d}{\bm x}\cdot\tilde{\bm p}$, while nonchalantly saying that the integrals are ``closed contours in phase space''. Upon closer inspection of the subsequent derivations, one finds that the authors actually work with a mixture of phase spaces: one of particles following Lorentz orbits, and one of GCs. For all three canonical actions, $\tilde{J}_1$, $\tilde{J}_2$ and $\tilde{J}_3$, the GC concept is used to resolve ambiguities in the definition of the $2\pi$ period over which the integrals are taken.

The GC is an abstract entity that is intuitively appealing and clearly very useful technically. However, it does not seem to be dictated by the equations governing the Hamiltonian dynamics of charged particles moving in prescribed ambient fields ${\bm E}$ and ${\bm B}$. In fact, the mapping ${\bm x}(t) \leftrightarrow {\bm X}_{\rm gc}(t)$ itself is ambiguous. While the forward mapping is conveniently performed using Eq.~(\ref{eq:can_gc}), the reverse is more easily carried out using\vspace{-0.1cm}
\begin{equation}
	{\bm x}(t) \approx {\bm X}_{\rm gc}(t) + \frac{M}{Ze}\frac{{\bm v}({\bm X}_{\rm gc}(t))\times\hat{\bm b}({\bm X}_{\rm gc}(t))}{B({\bm X}_{\rm gc}(t))}.
	\label{eq:can_gc_inv}\vspace{-0.1cm}
\end{equation}

\noindent Other options exist.\footnote{Theories and simulation codes that utilize gyroaveraging operations usually use more sophisticated exactly reversible mappings between particles and guiding-center-like entities (e.g., gyrocenters) in order to enforce conservation laws.}
One can say that the GC itself and its ambiguity are actively exploited in the modeling of charged particle motion in toroidal magnetic geometry.

\paragraph{Bonus: Effectively conserved magnetic moment $\tilde{\mu}$} When the ambient magnetic field is perfectly stationary and axisymmetric, and when it also satisfies the smoothness condition (\ref{eq:com_b}) sufficiently well for $\tilde{J}_1$ to oscillate effectively periodically, the above option (c) with $\oint_{\rm gyro}{\rm d}{\bm x}\cdot\tilde{\bm p} \rightarrow {\rm const}$.\ can be realized in practice with high accuracy by extending the integral from one gyroperiod to $N_{\rm pol}$ periods during which the GC completes one full poloidal transit: $\tau_{\rm B} \rightarrow \tau_{\rm pol} \approx N_{\rm pol}\tau_{\rm B}$; namely, the time interval shown in Fig.~\ref{fig:d01_lorentz2pi}. This can be written in the form\vspace{-0.0cm}
\begin{equation}
	\tilde{\mu} \equiv \oint_{\rm pol}\frac{{{\rm d}\tau}}{\tau_{\rm pol}} \int_{\rm gyro}\frac{{\rm d}{\bm x}\cdot\tilde{\bm p}}{2\pi}.
	\label{eq:can_j12}\vspace{-0.0cm}
\end{equation}

\noindent This ``bounce-averaged magnetic moment'' $\tilde{\mu}$ is expected to be practically invariant and was introduced in Section~\ref{sec:intro_com} to complete the standard set $\{\tilde{\mu},\sigma\tilde{\etot},\tilde{P}_\zeta\}$ of what we call absolute CoM, alongside $\tilde{\etot}$ that takes the place of $\tilde{J}_2$ for convenience, and $\tilde{J}_3 = \tilde{P}_\zeta = \tilde{\bm p}\cdot\partial_\zeta{\bm r}$.

\addtocontents{toc}{\setcounter{tocdepth}{1}}
\addcontentsline{toc}{section}{References}

\bibliographystyle{unsrt}
\bibliography{references}

\end{document}